\documentclass[reprint,aps,prx]{revtex4-2}

\usepackage{amsmath}
\usepackage{amssymb}
\usepackage[colorlinks=true,linkcolor=blue,citecolor=blue,urlcolor=blue]{hyperref}
\usepackage{graphicx}
\usepackage{comment}
\usepackage{epsfig}
\usepackage{epstopdf}
\usepackage[usenames,dvipsnames]{xcolor}
\usepackage{braket}
\usepackage{amsthm}
\usepackage{mathtools}
\usepackage{enumitem}
\usepackage{dsfont}
\usepackage{bbold}
\usepackage{bm}
\usepackage{eurosym}
\usepackage{mdframed}
\usepackage{cases}
\usepackage{adjustbox}
\usepackage{diagbox}
\usepackage{ragged2e}
\usepackage[symbol]{footmisc}
\usepackage{tikz-cd}

\usepackage[most]{tcolorbox}
\newtcolorbox{mybox}[2][]{
               = {yshift=-8pt},
  colback      = blue!6!white,
  colframe     = blue!1!black,
  halign       = flush left,
  fonttitle    = \bfseries\sffamily,
  colbacktitle = blue!90!black,
  title        = #2,#1,
  enhanced,
}

\newcommand{\be}{\begin{equation}}
\newcommand{\ee}{\end{equation}}
\newcommand{\ba}{\begin{aligned}}
\newcommand{\ea}{\end{aligned}}

\newcommand{\R}{\mathbb{R}}
\newcommand{\Z}{\mathbb{Z}}
\newcommand{\bc}{\begin{center}}
\newcommand{\ec}{\end{center}}
\newcommand{\beq}{\begin{equation}}
\newcommand{\eeq}{\end{equation}}
\newcommand{\beqq}{\begin{equation*}}
\newcommand{\eeqq}{\end{equation*}}
\newcommand{\beqa}{\begin{align}}
\newcommand{\eeqa}{\end{align}}
\newcommand{\barr}{\begin{array}}
\newcommand{\earr}{\end{array}}
\newcommand{\bi}{\begin{itemize}}
\newcommand{\ei}{\end{itemize}}

\newcommand{\Hi}{\mathcal{H}}

\newcommand{\Tr}{\ensuremath{\,\mathrm{Tr}}}

\newcommand{\ketbra}[2]{\ensuremath{ | #1 \rangle \langle #2 | }}

\newcommand{\thetachar}[4]{\ensuremath{ \theta \hspace{-2pt} \begin{bmatrix} #1 \\ #2 \end{bmatrix} \hspace{-3pt} ( #3, #4 ) }}

\newtheorem{lem}{Lemma}
\newtheorem{theo}{Theorem}
\newtheorem*{theo*}{Theorem}

\newtheorem*{prot*}{Protocol}

\newtheorem{coro}{Corollary}
\newtheorem{defi}{Definition}

\begin{document}

\title{Identifying quantum resources in encoded computations}

\author{Jack Davis$^1$}
\author{Nicolas Fabre$^2$}
\author{Ulysse Chabaud$^1$}
\email{ulysse.chabaud@inria.fr}

\affiliation{$^1$DIENS, \'Ecole Normale Sup\'erieure, PSL University, CNRS, INRIA, 45 rue d'Ulm, Paris 75005, France}
\affiliation{$^2$Telecom Paris, Institut Polytechnique de Paris, 19 Place Marguerite Perey, 91120 Palaiseau, France}

\date{\today}


\begin{abstract}
What is the origin of quantum computational advantage? Providing answers to this far-reaching question amounts to identifying the key properties, or \textit{quantum resources}, that distinguish quantum computers from their classical counterparts, with direct applications to the development of quantum devices. 
The advent of universal quantum computers, however, relies on error-correcting codes to
protect fragile logical quantum information by robustly encoding it into symmetric states of a quantum physical system. 
Such encodings make the task of resource identification more difficult, as what constitutes a resource from the logical and physical points of view can differ significantly.
Here we introduce a general framework which allows us to correctly identify quantum resources in encoded computations, based on phase-space techniques. 
For a given quantum code, our construction provides a Wigner function that accounts for how the symmetries of the code space are contained within the transformations of the physical space, resulting in an object capable of describing the logical content of any physical state, both within and outside the code space. 
We illustrate our general construction with the Gottesman--Kitaev--Preskill encoding of qudits with odd dimension.  The resulting Wigner function, which we call the \textit{Zak--Gross Wigner function}, is shown to correctly identify quantum resources through its phase-space negativity.  For instance, it is positive for encoded stabilizer states and negative for the bosonic vacuum.
We further prove several properties, including that its negativity provides a measure of magic for the logical content of a state, and that its marginals are modular measurement distributions associated to conjugate Zak patches.
\end{abstract}


\maketitle


\section{Introduction}

With the rapid development of quantum technologies, it is of major importance to identify the quantum resources which make quantum computers powerful. Beyond its fundamental significance, exploring this classical-to-quantum transition directly contributes to quantum architecture and algorithm design, with the aim of producing quantum devices capable of outperforming their classical counterparts.

In this context, phase-space representations are prominent tools for the understanding and advancement of quantum information science \cite{rundle2021overview}.  In addition to easing analytical computations, representing quantum states geometrically, and enabling the use of mathematical techniques from complex analysis and distribution theory, they allow for the identification and characterization of quantum resources \cite{chitambar2019quantum}---typically in relation to the presence of negativity in a particular phase-space quasi-probability distribution representing the quantum system at hand. 

The most common example of a phase-space representation is the Wigner function, first introduced in the infinite-dimensional continuous-variable (CV) setting \cite{Wigner1932} and later generalized to the finite-dimensional discrete-variable (DV) setting in multiple ways \cite{Bjork_Klimov_Sanchez_Soto_2008, Rundle_Everitt_2021}. A strong parallel exists between the original CV Wigner function and a specific DV Wigner function \cite{schwinger1960unitary, Buot_method_1974, wootters1987wigner, galetti1988}: both emerge from a conjugacy between position and momentum variables and both are uniquely singled out by their covariance 
properties \cite{Zhu_permutation_PRL_2016,Dias_Prata_2019,Schmid2022}. 
CV and DV Wigner-negative quantum states lead to statistics exhibiting quantum contextuality under CV and DV Pauli measurements, respectively \cite{howard2014contextuality, delfosse2017equivalence, haferkamp2021equivalence,Booth2021}. Moreover, the theorems of Hudson \cite{hudson1974wigner,soto1983wigner} and Gross \cite{gross2006hudson} identify pure states with positive CV and DV Wigner functions in odd dimension as Gaussian and stabilizer states, respectively. Non-stabilizerness, also known as magic, is essential for universal quantum computations \cite {bravyi2005universal} and may in fact be directly quantified by the amount of negativity in the DV Wigner function for odd-dimensional systems \cite{veitch2012negative,veitch2014resource}.  Gottesman--Knill-type theorems \cite{gottesman1998heisenberg} further demonstrate the efficient classical simulatability of Wigner-positive circuits in both the CV and DV settings \cite{Bartlett2002, Bartlett2002povm, Cormick2006, veitch2012negative, Mari2012, veitch2013cv, pashayan2015estimating, Rahimi2016sufficient, Kocia2017}.
This collection of results has led to a growing understanding that negativity of the Wigner function, or \textit{Wigner negativity} for short, is an essential resource for both CV and DV quantum computations \cite{veitch2014resource,takagi2018convex,albarelli2018resource}.

To deliver on their promise, however, quantum computations need to be protected against errors arising from unwanted interactions between the computer and its environment \cite{gottesman1997stabilizer}. This may be achieved by encoding the fragile logical information in a way that is robust against errors, which typically involves some form of redundancy, or more generally some form of symmetry. Error-correction then involves regularly projecting the intermediate quantum state of the computation onto a symmetric subspace used as a code space \cite{lidar2013quantum, roffe2019quantum}. Some of the most promising quantum error-correction schemes involve bosonic codes \cite{albert2022bosonic}, which aim to protect DV logical quantum information by encoding it into CV quantum states.  The Gottesman--Kitaev--Preskill (GKP) code \cite{Gottesman2001} is particularly prominent \cite{bourassa_blueprint_2021,sivak2023real}, and relies on so-called GKP states that have lattice-like symmetries \cite{conrad2022gottesman}.
The applications of GKP states span various fields, including quantum metrology \cite{PhysRevA.93.012315,PhysRevA.104.022208}, quantum communication \cite{rozpedek_all-photonic_2023}, and quantum computing \cite{PhysRevA.99.032344}.  While difficult to engineer, the generation of these states is quickly progressing \cite{fluhmann_encoding_2019, campagne-ibarcq_quantum_2020, sivak2023real, doi:10.1126/science.adk7560, fabre_generation_2020, PhysRevX.13.031001}. Given that error-correction is a necessary step towards full-fledged universal quantum computers, it is crucial to accurately assess quantum computational resources in the context of encoded computations, and for bosonic codes in particular. 

As previously mentioned, the negativity of the Wigner function provides a fundamental tool for such a task. Recently, however, an important mismatch between quantum computational resources and Wigner negativity has been uncovered in the context of GKP-encoded computation.
Namely, it was shown that many CV quantum states, such as stabilizer codewords, are resourceless from the point of view of the physical space, i.e.\ being DV Wigner-positive despite possessing arbitrarily high CV Wigner negativity, leading to the efficient classical simulation of quantum computations involving such states \cite{garcia2020efficient, calcluth2022efficient}. Perhaps even more surprising was the further realization that some CV Wigner-positive states (such as the vacuum, often considered as resourceless) turn out to be the missing ingredient to promote easy-to-simulate quantum computations to universal ones \cite{Baragiola_2019_allGaussianUniversality, Yamasaki_2020_costReducedGaussianUniversality, calcluth2023addingVacuum}. This mismatch was partially resolved by cleverly renormalising the CV Wigner function to allow an assessment of codeword resourcefulness using negativity \cite{Yamasaki_2020_costReducedGaussianUniversality, hahn2022quantifying, hahn2024bridgingMagic}, but a complete phase-space picture accounting for states outside the code space is missing.  The recent work \cite{calcluth2024sufficient} identifies resources for states outside the GKP code space by studying explicit encoding maps, though at the expense of abandoning phase-space descriptions and restricting to ideal GKP states.

Moreover, the mismatch between quantum computational resources and Wigner negativity goes far beyond the case of GKP codes: it is due to the fundamental difference between the physical and logical spaces, which can be observed for virtually any encoding. Namely, the Wigner negativity of a codeword, seen as a state in the physical space, generally says little about the Wigner negativity of the underlying logical state, seen as an element of the logical space (and vice-versa). This phenomenon is even more striking for states outside the code space, where quantum resourcefulness is challenging to interpret from the logical point of view. An important gap therefore remains when it comes to unambiguously identifying quantum resources in encoded computations.

Here we address this gap by providing a construction that associates to any code space a well-behaved phase-space representation for all states in the physical space, compatible with the logical structure of the code space.  Our framework generalises the seminal work of Brif and Mann~\cite{brif1999phase} by placing the symmetry of the code at the heart of the framework.  Emerging from this choice of symmetry is a Wigner-like function that describes the logical content of quantum states both within and outside the code space. 

We illustrate our construction using the odd-dimensional GKP encoding as an example.  The initial GKP code space leads us to a continuous torus as the phase space and to an appropriate toroidal Wigner function.  We also demonstrate how error correction subroutines naturally emerge and can be related to such Wigner functions.  This new quasi-probability distribution is here named the \textit{Zak--Gross Wigner function}, as we show it can be understood both as (i) a joint quasi-probability distribution for modular position and momentum measurements (i.e.\ Zak basis projections) and (ii) as a continuous collection of DV (Gross) Wigner functions, one for each displaced code space.
Moreover we show that the negativity of the Zak--Gross Wigner function is a meaningful measure of magic for all states in the physical space, encoded or otherwise.  For states in the code space it correctly identifies stabilizer codewords as non-negative and magic codewords as negative.  For states outside the code space, it captures the amount of resource in their logical content. We compute the negativity for selected examples; in particular, the vacuum state has a negative Zak--Gross Wigner function, which resolves the apparent counterintuitive fact that it provides universality to otherwise classically simulable architectures despite having a positive CV Wigner function \cite{calcluth2023addingVacuum}.
This supports the intuition that the amount of resource within a state should not change upon encoding, 
and rather it is the lens through which one is identifying the resources that should be adapted.

The rest of the paper is structured as follows. In section \ref{sec:background}, we provide preliminary material on phase-space techniques, quantum information theory, and GKP codes. In section \ref{sec:WignerQcodes} we explain how to construct a Wigner function compatible with a generic encoding of a quantum system, together with the general implications for representing encoded computations in phase space. In section \ref{sec:WignerGKP} we apply our construction to the case of GKP codes and obtain the Zak--Gross Wigner function, for which we prove important properties, provide several interpretations, and compute selected numerical examples. We conclude in section \ref{sec:concl}.

\section{Background}
\label{sec:background}

In this section, we provide the necessary preliminary material.  We review the Brif and Mann phase-space construction in~\ref{sec:BrifMan}, CV and DV Wigner functions in section~\ref{sec:CVDV_Wigner}, and GKP codes in section~\ref{sec:GKP}.


\subsection{Generic phase-space construction}
\label{sec:BrifMan}

A quantum system is naturally described by a Hilbert space $\mathcal H$, which contains its pure states, and a group $G$, which describes its dynamical symmetries.  For instance, the infinite-dimensional Hilbert space with the Heisenberg--Weyl group $H_3(\mathbb C)$ describes a quantum mechanical oscillator, a bosonic mode, and a spinless particle, while a finite-dimensional Hilbert space with the group SU(2) describes a spin-$j$ particle, a collection of indistinguishable two-level atoms, and the polarization of a monochromatic light beam. 

The phase-space formulation of quantum mechanics aims to represent the states of quantum systems by distributions over phase space rather than by density operators of a Hilbert space. This provides an equivalent theory for quantum mechanics based on phase-space representations which mirrors statistical theory on classical phase space, leading to profound insights about the differences between classical and quantum systems \cite{cahill1969ordered,cahill1969density,schroeck2013quantum}.
A generic construction of well-behaved phase-space representations for a large class of dynamical symmetry groups was derived by Brif and Mann in \cite{brif1999phase}, which we revisit from an information-theoretic perspective in section~\ref{sec:WignerQcodes}. We first provide a brief of review this construction hereafter in the case of Wigner functions and refer the reader to~\cite{brif1999phase} for more details.

Let $\pi$ be an irreducible unitary representation of a connected and simply connected, finite-dimensional Lie group $G$ acting on the Hilbert space $\mathcal H$. Choosing a fixed reference---or fiducial---state $\ket{\psi_0}\in\mathcal H$, the \textit{coherent states} of the system are defined as $\pi(g)\ket{\psi_0}$, for all $g\in G$ \cite{perelomov1972coherent}. The set of coherent states up to a phase is then determined by the homogeneous space $X=G/H$, where $H\subset G$ is the subgroup of elements that leave the reference state $\ket{\psi_0}$ invariant up to a phase factor. The subgroup $H\subset G$ is the so-called \textit{isotropy subgroup} of $\ket{\psi_0}$ (or equivalently the commutant of $\ket{\psi_0}\!\bra{\psi_0}$ in $\pi(G)$) and the homogeneous space $X=G/H$ is the \textit{phase space}, which comes with an invariant measure $d\mu$ and indexes the set of coherent states $\ket\Omega:=\pi(\Omega)\ket{\psi_0}$, for all $\Omega\in X$. Intuitively, the reference state $\ket{\psi_0}$ determines the origin of phase space and its symmetries are quotiented out in order to obtain a well-behaved phase-space representation.

Denoting the natural action of $G$ on $X$ by $g\cdot\Omega$, for $g\in G$ and $\Omega\in X$, the typical desirable properties of a phase-space (generalised) Wigner function $W$ for operators $\hat A$ acting on the Hilbert space $\mathcal H$ are known as the Stratonovich--Weyl (SW) axioms:
\begin{enumerate}[label=(\roman*)]
    \item\label{enum:SWlin} Linearity: $\hat A\mapsto W_{\hat A}$ is a linear map such that $W_{\hat A}=W_{\hat B}$ iff $\hat A=\hat B$.
    \item\label{enum:SWreal} Reality: $W_{\hat A^\dag}=(W_{\hat A})^*$.
    \item\label{enum:SWst} Standardization: $\int_Xd\mu W_{\hat A}=\Tr(\hat A)$.
    \item\label{enum:SWcov} Covariance: $W_{\pi(g)^\dag\hat A\pi(g)}(\Omega)=W_{\hat A}(g\cdot\Omega)$, for all $\Omega\in X$ and $g\in G$.
    \item\label{enum:SWtr} Traciality: $\int_Xd\mu W_{\hat A}W_{\hat B}=\Tr(\hat A\hat B)$.
\end{enumerate}

\noindent When satisfied, these properties ensure that there is a one-to-one correspondence between the usual and the phase-space formulations of quantum mechanics, and that the latter has a similar mathematical structure to that of statistical mechanics over classical phase space. 

In~\cite{brif1999phase}, it is shown how to construct a Wigner function satisfying the SW axioms starting from the choice of a reference state which provides an (over)complete system of coherent states over $\mathcal H$, i.e.,
\begin{equation}\label{eq:resolutionId}
    \int_Xd\mu(\Omega)\ket\Omega\!\bra\Omega=\hat{\mathbb I},
\end{equation}
where $\hat{\mathbb I}$ is the identity operator over $\mathcal H$.
This Wigner function takes the form
\begin{equation}\label{eq:WignerBrifMan}
    W_{\hat A}(\Omega)=\mathrm{Tr}[\hat \Delta(\Omega)\hat A],
\end{equation}
where $\hat\Delta$ is a SW kernel operator, which possess a specific mathematical structure tailored to the SW axioms that we detail in what follows. The \textit{harmonic functions} $\{Y_\nu\}_\nu$ are defined as the eigenfunctions of the Laplace--Beltrami operator and form a basis of the Hilbert space $L^2(X,\mu)$ of square-integrable functions on $X$ with invariant measure $d\mu$. Let us define the operators
\begin{equation}\label{eq:tensorOp}
    \hat D_\nu:=\tau_\nu^{-1/2}\int_Xd\mu(\Omega)Y_\nu(\Omega)\ket\Omega\!\bra\Omega,
\end{equation}
where $\tau_\nu$ are real and positive coefficients defined implicitely by the coherent state overlaps
\begin{equation}\label{eq:cohOver}
    |\langle\Omega|\Omega'\rangle|^2=\sum_\nu\tau_\nu Y_\nu^*(\Omega)Y_\nu(\Omega'),
\end{equation}
for all $\Omega,\Omega'\in X$.
These two equations determine the operators $\hat D_\nu$ up to a sign function $(-1)^{s_\nu}$, and each choice of a specific sign function leads to a SW kernel given by
\begin{equation}\label{eq:DeltaBrifMan}
    \hat\Delta(\Omega):=\sum_\nu Y_\nu(\Omega)\hat D_\nu^\dag.
\end{equation}

This construction, which relies heavily on the representation theory of the group $G$ to identify the space $X$, its invariant measure $d\mu$, and the harmonic functions $\{Y_\nu\}_\nu$ over $L^2(X,\mu)$, provides a Wigner function based on a SW kernel which satisfies all the SW axioms, assuming that the resolution of the identity in Eq.~(\ref{eq:resolutionId}) holds. 

For instance, for a quantum system described by an infinite-dimensional Hilbert space $\mathcal H$ with dynamical symmetry group $G=H_3(\mathbb C)$, the Heisenberg--Weyl group over $\mathbb C$, with the vacuum state as reference state, this construction leads to the original CV Wigner function \cite{Wigner1932}, which we review in the following section.


\subsection{CV and DV Wigner functions}
\label{sec:CVDV_Wigner}

In this section, we review important properties of the original CV Wigner function \cite{Wigner1932} in parallel with one of its DV analogues, which we refer to as the Gross Wigner function \cite{gross2006hudson}.  

Unless otherwise said, $\hbar = 1$ and $d$ refers to an odd integer. We also restrict to a single qudit/qumode, with the extension to composite systems being relatively straightforward.

The CV and DV \textit{displacement operators} are
\begin{equation}\label{eq:CVDV_displacement_operators}
    \begin{aligned}
    \hat D(x,p) &= e^{i \frac{xp}{2}} e^{ip\hat x} e^{- i x \hat p}, \qquad x,p \in \R, \\
    \hat D_d(a,b) &= \omega^{2^{-1}ab} \hat X^a \hat Z^b, \qquad a,b \in \Z_d = \Z/d\Z,
    \end{aligned}
\end{equation}
where $\hat X$ and $\hat Z$ are the generalized Pauli operators, $\omega = e^{i\frac{2\pi}{d}}$, and $2^{-1} = \frac{d+1}{2}$ is the multiplicative inverse of $2$ in $\Z_d$. Note that our definition for the DV displacement operator differs in convention from, for example, \cite{Vourdas_2004}.  This is done for later convenience and to better illustrate the similarity with the CV version. With $\bm z$ denoting $(x, p)$ or $(a, b)$ when appropriate, the displacement operators satisfy the orthogonality relations
\begin{equation}\label{eq:dis_orthogonality}
\begin{aligned}
    \Tr[\hat D^\dagger(\bm z) \hat D(\bm z')] &= 2\pi \delta(\bm z - \bm z'), \\
    \Tr[\hat D^\dagger_d (\bm z) \hat D_d(\bm z')] &= d \delta_{\bm z, \bm z'},
\end{aligned}
\end{equation}
as well as the commutation relations
\begin{equation}\label{eq:CVDV_displacement_conjugation}
\begin{aligned}
    \hat D(\bm z) \hat D(\bm z') &= e^{-i\bm z^T \sigma \bm z'} \hat D(\bm z') \hat D(\bm z), \\
    \hat D_d(\bm z) \hat D_d(\bm z') &= \omega^{-\bm z^T \sigma \bm z'} \hat D_d(\bm z') \hat D_d(\bm z),
\end{aligned}
\end{equation}
where
\begin{equation}
    \sigma = \begin{pmatrix}
        0 & 1 \\ -1 & 0
    \end{pmatrix}\!,
\end{equation}
which is the matrix representation of the standard symplectic form on $\R^2$ or $\Z_d^2$ respectively.  Those two spaces, being in one-to-one correspondence with their respective displacement operators, are the phase spaces of the systems.

The \textit{undisplaced parity operators} are defined by
\begin{equation}\label{eq:parity_def}
    \hat \Pi(0,0)\ket{x}_{\hat x} = \ket{-x}_{\hat x}, \qquad \hat \Pi_d(0,0)\ket{j} = \ket{-j},
\end{equation}
where $\{\ket{x}_{\hat x}\}_{x\in\R}$ are position basis states, $\{\ket{j}\}_{j=0,\dots,d-1}$ are computational basis states, and $-j \cong d - j$ is understood to be modulo $d$.  An important relation \cite{Leaf1968, Grossmann_1976, royer1977Wigner} is 
\begin{equation}\label{eq:LeafGrossmannRoyer}
\begin{aligned}
    \hat \Pi(0,0) &= \frac{1}{4\pi} \int d\bm z\hat D(\bm z), \quad
    \hat \Pi_d(0,0) = \frac{1}{d} \sum_{\bm z} \hat D_d(\bm z).
\end{aligned}
\end{equation}
The \textit{displaced parity operators} (cf.\ \eqref{eq:DeltaBrifMan}) are
\begin{equation}\label{eq:CVDV_dis_parity_symplectic_fourier}
\begin{aligned}
    \hat \Pi(\bm z) &= \hat D(\bm z) \hat \Pi(0,0) \hat D^\dagger(\bm z) = \frac{1}{4\pi} \int d\bm z' \, \hat D(\bm z') e^{-i \bm z^T \sigma \bm z'}, \\
    \hat \Pi_d(\bm z) &= \hat D_d(\bm z) \hat \Pi_d(0,0) \hat D_d^\dagger(\bm z) = \frac{1}{d} \sum_{\bm z'} \hat D_d(\bm z') \omega^{-\bm z^T \sigma \bm z'}.
\end{aligned}
\end{equation}
These relations, obtained by applying Eq.~\eqref{eq:CVDV_displacement_conjugation} to Eq.~\eqref{eq:LeafGrossmannRoyer}, are seen as a symplectic Fourier transform mapping the displacement basis to the parity basis.  This, together with the Hilbert--Schmidt orthogonality of the displacement operators in Eq.~\eqref{eq:dis_orthogonality}, show that the two operator bases are mutually unbiased,
\begin{equation}
\begin{aligned}\label{parity_displacement_MUB}
    \Tr[ \hat D^\dagger(\bm z) \hat \Pi(\bm z') ] &= \frac{1}{2} e^{i\sigma(\bm z, \bm z')}, \\
    \Tr[ \hat D_d^\dagger(\bm z) \hat \Pi_d(\bm z') ] &= \omega^{\sigma(\bm z, \bm z')}.
\end{aligned}
\end{equation}

The expansion coefficients of a general operator, in particular a quantum state ${\hat\rho}$, in either of these bases give rise to two quasi-probability distributions. The CV and DV \textit{characteristic functions} and \textit{Wigner functions} are respectively
\begin{equation}\label{eq:CVDV_char_wig_def}
\begin{aligned}
    \chi^\mathrm{CV}_{{\hat\rho}}(\bm z) &=\Tr[\hat D(\bm z) {\hat\rho}], \quad W^\mathrm{CV}_{\hat\rho}(\bm z) = \Tr[2\hat \Pi(\bm z) {\hat\rho}], \\
    \chi^\mathrm{DV}_{\hat\rho}(\bm z) &= \Tr[\hat D_d(\bm z) {\hat\rho}], \quad W^\mathrm{DV}_{\hat\rho}(\bm z) = \Tr[\hat \Pi_d(\bm z) {\hat\rho}],
\end{aligned}
\end{equation}
where we have dropped the subscript $d$ for the DV case. Note that the CV definition matches with \cite{cahill1969density,brif1999phase} (setting $\alpha=(x+ip)/\sqrt2$) and satisfies the SW axioms \ref{enum:SWlin}-\ref{enum:SWtr} over the phase space $\mathbb C\simeq\mathbb R^2$ with invariant measure $\frac{d^2\alpha}\pi=\frac{dxdp}{2\pi}$, while the DV definition differs from the Gross Wigner function \cite{gross2006hudson} by a factor $d$, so that it satisfies the SW axioms \ref{enum:SWlin}-\ref{enum:SWtr} over the phase space $\mathbb Z_d^2$ with invariant measure $\frac1d$ times the counting measure. Substituting \eqref{eq:CVDV_dis_parity_symplectic_fourier} in \eqref{eq:CVDV_char_wig_def}, one retrieves the original definitions of the Wigner functions.  The reconstruction formulas are respectively
\begin{equation}\label{eq:reconstruction}
\begin{aligned}
    \hat \rho &= \frac{1}{\pi} \int W^{\text{CV}}_{\hat \rho}(\bm z) \hat \Pi(\bm z) d\bm z = \frac{1}{2\pi} \int \chi^{\text{CV}}_{\hat \rho} (\bm z) \hat D(\bm z) d\bm z,\\
    \hat \rho &= \frac{1}{d} \sum_{\bm z} W^{\text{DV}}_{\hat \rho}(\bm z) \hat \Pi_d(\bm z) = \frac{1}{d} \sum_{\bm z} \chi^{\text{DV}}_{\hat \rho}(\bm z) \hat D_d(\bm z).
\end{aligned}
\end{equation}

\subsection{Encoding DV quantum information in CV: GKP codes}
\label{sec:GKP}

The GKP codes are a prominent family of qudit-to-oscillator bosonic stabilizer codes that offer protection of encoded DV quantum information against small displacements in CV phase space \cite{Gottesman2001, Albert2017PerformanceBosonic, Grimsmo2021GKP_Review, Brady2024GKP_Review}.  We fix the qudit dimension $d$ and focus on the unbiased/square code, a stabilizer code with logical operators 
\begin{equation}\label{GKP_logical_def}
    \hat{\bar{Z}} = e^{i\sqrt{\frac{2\pi}{d}} \hat x} = e^{i\ell\hat x}, \quad \hat{\bar{X}} = e^{-i\sqrt{\frac{2\pi}{d}}\hat p} = e^{-i\ell\hat p},
\end{equation}
and stabilizer generators 
\begin{equation}
    \hat{\bar{Z}}^d= e^{i\sqrt{2\pi d}\hat x} = e^{id\ell\hat x},\quad
    \hat{\bar{X}}^d=e^{-i\sqrt{2\pi d}\hat p} = e^{-id\ell\hat p},
\end{equation}
where $\ell = \sqrt{\frac{2\pi}{d}}$ is the logical step-size.  The generators commute \eqref{eq:CVDV_displacement_conjugation} and generate the abelian stabilizer group $\mathcal{S} = \langle \hat{\bar{Z}}^d, \hat{\bar{X}}^d \rangle$, the associated code space of which is the simultaneous +1 eigenspace:
\begin{equation}\label{eq:code space_def}
    \{ \ket{\psi} \in \Hi \,|\, U\ket{\psi} = \ket{\psi} \, \forall \, U \in \mathcal{S} \} \subset \mathcal{H}.
\end{equation}
Each encoded computational basis state $\ket j\mapsto\ket{\bar j}$ takes the form of a Dirac comb in the position basis with period $d\ell = \sqrt{2\pi d}$,
\begin{equation}\label{codewords_position}
    \ket{\bar j} = \frac1{\sqrt\ell}\sum_{n \in \Z} \ket{j \ell + d\ell n}_{\hat x}.
\end{equation}
More generally, we extend the map $\ket j\mapsto\ket{\bar j}$ to operators by linearity and indicate by $\hat{\bar\rho}$ the GKP-encoding of a DV density operator $\hat\rho$. Conversely, if $\hat\rho$ is a CV density operator which is the GKP-encoding of a DV state, we denote by $\hat{\underline\rho}$ the corresponding DV logical state.

Logical operators obey the generalized Clifford/Pauli commutation relation (derivable from \eqref{eq:CVDV_displacement_conjugation}):
\begin{equation}\label{encoded_Pauli_commutation}
    \hat{\bar{Z}} \hat{\bar{X}} = \omega \hat{\bar{X}} \hat{\bar{Z}},
\end{equation}
where $\omega=e^{\frac{2\pi i}d}$ is the first $d^{\text{th}}$ root of unity, and combine to give the encoded discrete displacement operator \eqref{eq:CVDV_displacement_operators}
\begin{equation}\label{GKP_encoded_displacement_def}
    \hat{\bar{D}}_d(a,b) = \omega^{2^{-1} ab} \hat{\bar{X}}^a \hat{\bar{Z}}^b,
\end{equation}
with
\begin{equation}
    \hat{\bar{X}}^a = \hat{\bar{D}}_d(a,0) \quad \text{and} \quad \hat{\bar{Z}}^b = \hat{\bar{D}}_d(0,b).
\end{equation}
In terms of the CV displacement operator \eqref{eq:CVDV_displacement_operators} this is 
\begin{equation}\label{encoded_DV_dis_via_CV_dis}
    \hat{\bar{D}}_d(a,b) = (-1)^{ab} \hat D(a \ell, b \ell).
\end{equation}
 
The projector onto the code space is 
\begin{equation}
    \hat{{P}}_{(0,0)} = \sum_{j=0}^d \ketbra{\bar j}{\bar j},
\end{equation}
and similarly for the displaced code spaces, which are orthogonal to each other,
\begin{equation}\label{eq:dispcodeproj}
    \hat{{P}}_{(s,t)} = \hat D(s,t) \hat{{P}}_{(0,0)} \hat D^\dagger(s,t), \quad s,t \in [0, \ell).
\end{equation}
Syndrome diagnosis after an arbitrary displacement error is the determination of $(s,t)$ by simultaneously measuring the stabilizers. This can be done in a number of ways; for instance, one may perform a destructive position/momentum homodyne measurement on an auxiliary GKP state that has appropriately interacted with the logical mode \cite{Gottesman2001}.  Given a general CV state $\hat\rho$, the corresponding syndrome $(s,t)$ is measured with probability
\begin{equation}\label{eq:syndrome_distribution}
    \text{Pr}[(s,t) \, | \, \hat\rho] = \Tr[\hat\rho \, \hat{{P}}_{(s,t)}],
\end{equation}
which we call the \textit{syndrome distribution} of $\hat\rho$. Upon obtaining a specific syndrome an error-correcting displacement $\hat D(-s,-t)$ may be applied, leading, up to normalisation, to the error-corrected state
\begin{equation}\label{eq:err-corr-state}
    \begin{aligned}
        \hat\rho_{(s,t)}:=&\, \hat D(-s,-t) \hat P_{(s,t)} \hat \rho \hat P_{(s,t)} \hat D^\dagger(-s,-t) \\
        =& \, \hat P_{(0,0)}\hat D(s,t)^\dag\hat\rho\hat D(s,t)\hat P_{(0,0)}.
    \end{aligned}
\end{equation}
The subset of correctable errors (i.e., those for which the above procedure does not introduce unwanted logical errors) are the displacements $\hat D(x,p)$ taking arguments within the phase space region $(-\ell/2, \ell/2)\times (-\ell/2, \ell/2)$ and its shifts by multiples of $d\ell$ in either direction.

GKP codes are closely related to the mathematical concepts of Zak transform and theta functions \cite{Pantaleoni_Zak_2024}, for which we provide additional background in Appendix~\ref{app:Zaktheta}.
Note also that there exist various ways of representing geometrically GKP states: apart from the usual CV Wigner function \cite{Wigner1932,Gottesman2001}, the modular variables formalism \cite{Aharonov1969Modular, Zak1967finiteTranslations, englert_periodic_2006, ketterer2016quantum} provides an approach for building an informationally-complete phase-space representation of GKP states on a double cylinder phase space \cite{fabre2020wigner}. However, the corresponding Wigner function does not correctly identify quantum computational resources with respect to the GKP encoding, unlike the Zak--Gross Wigner function that we introduce in this work.


\section{Wigner functions for quantum codes}
\label{sec:WignerQcodes}

Given a quantum system described by a Hilbert space $\mathcal H$ and a (connected and simply connected,
finite-dimensional topological Lie) symmetry group $G$ with irreducible unitary representation $\pi$ over $\mathcal H$, recall that
the phase-space construction due to Brif and Mann \cite{brif1999phase} picks a fixed reference state $\ket{\psi_0}\in\mathcal H$ to be the origin of phase space and builds a system of coherent states and phase-space representations out of this choice (see section \ref{sec:BrifMan}). The choice of reference state is arbitrary, such that the corresponding system of coherent states may be interpreted as a complete set of almost-classical states, and is usually motivated by physical considerations.

We propose to choose the reference state based on symmetry considerations instead: our construction starts with the choice of a code space $\mathcal C$, a specific subspace of $\mathcal H$ chosen to encode logical information. This choice gives rise to an associated symmetry described by the joint isotropy subgroup $H^{\mathcal C}\subset G$ of the code space, defined as the set of elements $h$ such that
\begin{equation}
    \pi(h)\ket\psi\!\bra\psi\pi(h)^\dag=\ket\psi\!\bra\psi,
\end{equation}
for all $\ket\psi\in\mathcal C$.
Note that there can be states outside of the code space with the same symmetry: those are elements of the ``syndrome subspace'', which can be formally defined as the bicommutant of the code space $\mathcal C$, seen as a subset of $\mathcal D(\mathcal H)$, with respect to the subgroup $H^{\mathcal C}$.

Once the symmetry associated with a code space is fixed, we pick a reference state $\ket{\psi_0}\in\mathcal C$ within the code space whose isotropy subgroup is equal to $H^{\mathcal C}$ and follow similar steps to the Brif and Mann construction \cite{brif1999phase}. The result is a Wigner function tailored to the choice of code space $\mathcal C$, over the phase space $X^{\mathcal C}=G/H^{\mathcal C}$. The coherent states are then given by $\ket\Omega:=\pi(\Omega)\ket{\psi_0}$, for all $\Omega\in X^{\mathcal C}$, and in general do not resolve the identity over the full Hilbert space $\mathcal H$. The corresponding Wigner function is then not informationally-complete, because it captures the \textit{logical content} of a quantum state with respect to the symmetry described by $H^{\mathcal C}$: different quantum states related by an element $\pi(h)$ with $h\in H^{\mathcal C}$ must have the same phase-space representation. Note that our construction generalises that of Ref.~\cite{brif1999phase} to the case where the coherent states do not necessarily resolve the identity, and we retrieve their construction as a special case when they do.

For a density operator ${\hat\rho}$ on $\mathcal H$, its Wigner function associated with the code space $\mathcal C$ is given by
\begin{equation}\label{eq:WignerGen}
    W_{\hat\rho}^{\mathcal C}(\Omega)=\mathrm{Tr}[\Delta^{\mathcal C}(\Omega){\hat\rho}],
\end{equation}
where the kernel associated with the code space $\mathcal C$ is defined as
\begin{equation}\label{eq:DeltaGen}
    \hat\Delta^{\mathcal C}(\Omega)=\sum_\nu Y_\nu(\Omega)\hat D_\nu^\dag,
\end{equation}
where $\hat D_\nu$ are the tensor operators and $Y_\nu$ the harmonic functions over $L^2(X^{\mathcal C},\mu)$, with $d\mu$ the invariant measure over $X^{\mathcal C}$ and the index $\nu$ ranges over the spectrum of the Laplace--Beltrami operator over $L^2(X^{\mathcal C},\mu)$.

Explicit expressions for the tensor operators are obtained using:
\begin{align}
    \label{eq:cohOvercode}|\langle\Omega|\Omega'\rangle|^2&=\sum_\nu\tau_\nu Y_\nu^*(\Omega)Y_\nu(\Omega')\\
    \label{eq:tensorOpcode}e^{i\varphi_\nu}\tau_\nu^{1/2}\hat D_\nu&=\int_{X^{\mathcal C}}d\mu(\Omega)Y_\nu(\Omega)\ket\Omega\!\bra\Omega,
\end{align}
where we use a generalised definition compared to \cite{brif1999phase} (see Eqs.~(\ref{eq:tensorOp}) and (\ref{eq:cohOver})) that allows for an additional choice of phase $\varphi_\nu$.
Inverting the first equation provides the expression of the real and positive coefficients $\tau_\nu$, while the second equation determines the tensor operators $\hat D_\nu$ up to a phase function. Choosing a fixed phase function then leads to an explicit expression for the kernel $\Delta^\mathcal C$ and for the Wigner function $W^\mathcal C$. Interestingly, we find in practice that the phase function may be chosen so that the resulting Wigner function is actually independent of the choice of reference state, and only depends on the symmetry of the code space. For instance, we may consider a bosonic mode with an infinite-dimensional Hilbert space and Heisenberg--Weyl symmetry group, and a trivial encoding corresponding to a $U(1)$ symmetry, when the ``syndrome space'' is the full Hilbert space. In this case, picking the vacuum state as a reference state gives back the usual CV Wigner function \cite{brif1999phase}, and we show in Appendix~\ref{app:BrifManFock} that picking any Fock state as a reference state (which also has $U(1)$ as isotropy subgroup) also yields the same Wigner function.


\section{Wigner functions for GKP codes}
\label{sec:WignerGKP}

In what follows, we illustrate our phase-space construction with a more interesting case for which the code space $\mathcal C$ is non-trivial: we still consider a bosonic mode with an infinite-dimensional Hilbert space and Heisenberg--Weyl symmetry group, but in the more subtle case of the GKP encoding.


\subsection{The Zak--Gross Wigner function}

For simplicity, we focus on the square qudit GKP code over a single bosonic mode, with infinite-dimensional Hilbert space $\mathcal H$ and symmetry group $G=H_3(\mathbb C)$. 
Recall from section~\ref{sec:GKP} that the code space in this case is $\mathcal C_\mathrm{GKP}:=\mathrm{span}\{\ket{\bar j}\}_{j=0,\dots,d-1}$, where $\ket{\bar j} = \frac1{\sqrt\ell}\sum_{n \in \Z} \ket{j \ell + d\ell n}_{\hat x}$, with $\ell=\sqrt{2\pi/d}$. 

Any element $g\in H_3(\mathbb C)$ can be parametrized as
\begin{equation}
    g:=g(\varphi,x,p),\quad\pi(g)=e^{i\varphi\hat{\mathbb I}}e^{-ix\hat p}e^{ip\hat x},
\end{equation}
for $\varphi,x,p\in\mathbb R$, i.e., as a displacement operator up to a global phase. 
The joint isotropy subgroup $H^{\mathcal C_\mathrm{GKP}}\subset H_3(\mathbb C)$ for $\mathcal C_\mathrm{GKP}$ is the group $U(1)\times\mathbb Z\times\mathbb Z$, with unitary representation
\begin{equation}
    \pi(H^{\mathcal C_\mathrm{GKP}}\!)\!:=\!\{e^{i\varphi}(e^{-id\ell\hat p})^m(e^{id\ell\hat x})^n|\varphi\in[0,2\pi),m,n\in\mathbb Z\}\!,
\end{equation}
which up to a global phase are the stabilizers of the code.
The associated phase space is $X=G/H^{\mathcal C_\mathrm{GKP}}\simeq\mathbb T^2_{d\ell}$, where 
\begin{equation}
    \mathbb{T}_{d\ell}^2:=[0,d\ell)\times [0,d\ell)=[0,\sqrt{2\pi d})\times[0,\sqrt{2\pi d})
\end{equation}
denotes the \textit{continuous} torus with invariant measure $\frac1d{dudv}$. The size of the torus is up to an arbitrary scaling and chosen for convenience.
Using the construction outlined in the previous section, we obtain a Wigner function associated with the code space $\mathcal C_\mathrm{GKP}$ over the phase space $\mathbb T^2_{d\ell}$ as follows.

\begin{defi}\label{def:ZGWigner}
Let $\hat A$ be an operator acting on the Hilbert space $\mathcal H$.
The \textit{Zak--Gross Wigner function} associated with the code space $\mathcal C_\mathrm{GKP}$ is given by
\begin{equation}\label{eq:GKP_Wigner_def}
    W_{\hat A}^{\mathcal C_\mathrm{GKP}}(u,v)=\mathrm{Tr}[\hat\Delta^{\mathcal C_\mathrm{GKP}}(u,v)\hat A],
\end{equation}
for all $(u,v)\in\mathbb{T}_{d\ell}^2$, where the kernel associated with the code space $\mathcal C_\mathrm{GKP}$ is given by
\begin{equation}\label{eq:GKP_Wigner_kernel_abstract_def}
    \hat\Delta^{\mathcal C_\mathrm{GKP}}(u,v)=\hat D (u, v)\hat \Delta^{\mathcal C_\mathrm{GKP}}(0,0)\hat D^\dagger(u, v),
\end{equation}
with
\begin{equation}\label{eq:DeltaGKP00}
    \hat\Delta^{\mathcal C_\mathrm{GKP}}(0,0)=\frac1{2\pi}\sum_{m,n\in \Z} (-1)^{mn} \hat{D}(n\ell,m\ell).
\end{equation}
\end{defi}

\noindent We provide a detailed derivation in Appendix~\ref{app:BrifManGKP}, where we also show that this definition is \textit{independent of the choice of reference state} with isotropy subgroup $H^{\mathcal C_\mathrm{GKP}}$, i.e., the Wigner function $W^{\mathcal C_\mathrm{GKP}}$ is defined purely by the choice of symmetry of the code space $\mathcal C_\mathrm{GKP}$. The name Zak--Gross Wigner function comes from profound connections with Zak bases and the Gross Wigner function, which we uncover in sections \ref{sec:phys} and \ref{sec:info}, respectively.

The full expression for the Zak--Gross Wigner is given by:
\begin{equation}\label{eq:fullexprWZG}
    W_{\hat A}^{\mathcal C_\mathrm{GKP}}\!(u,v)\!=\!\frac1{2\pi}\!\sum_{m,n\in \Z}\!\!(-1)^{mn}e^{i\ell(nv-mu)}\Tr[\hat{D}(n\ell,m\ell)\hat A],
\end{equation}
where the displacement operators are defined in Eq.~(\ref{eq:CVDV_displacement_operators}). We refer to section~\ref{sec:examples} for specific examples of Zak--Gross Wigner functions.\\

Note that using a continuous torus as a quantum phase space has been previously considered in several works \cite{de_bievre_quantization_1996, kowalski_coherent_2007, ligabo2016torus}. These constructions differ from ours, and in particular in our case dequantization is not unique in the sense that different density matrices may have the same Zak--Gross Wigner function, because this representation precisely captures the features of quantum states that are relevant with respect to a specific encoding, as we shall see in what follows.


\subsection{Mathematical description: the Stratonovich--Weyl axioms}
\label{sec:maths}

Let us define the twirling map
\begin{equation}
    \mathcal E(\hat A):= \int_{\mathbb{T}_\ell^2}dsdt\hat{P}_{(s,t)}\,\hat\rho\hat{P}_{(s,t)},
\end{equation}
where $\hat{P}_{(s,t)}$ is the projector onto a displaced code space defined in Eq.~(\ref{eq:dispcodeproj}), and where integration is performed over the smaller torus $\mathbb{T}_\ell^2$. This map removes coherence between the various displaced code spaces and yields operators that are invariant under the action of $H^{\mathcal C_\mathrm{GKP}}$.

The next result shows that the Zak--Gross Wigner function is a well-behaved phase-space representation associated with the code space $\mathcal C_\mathrm{GKP}$, in the sense that it satisfies a version of the SW axioms \ref{enum:SWlin}-\ref{enum:SWtr} tailored to the symmetry of that space:

\begin{theo}\label{th:SWGKP}
The Wigner function $W^{\mathcal C_\mathrm{GKP}}$ satisfies:
\begin{enumerate}[label=(\roman*')]
    \item\label{enum:SWlinGKP} Linearity: $\hat A\mapsto W^{\mathcal C_\mathrm{GKP}}_{\hat A}$ is a linear map such that $W^{\mathcal C_\mathrm{GKP}}_{\hat A}=W^{\mathcal C_\mathrm{GKP}}_{\hat B}$ iff $\mathcal E(\hat A)=\mathcal E(\hat B)$.
    \item\label{enum:SWrealGKP} Reality: $W^{\mathcal C_\mathrm{GKP}}_{\hat A^\dag}=(W^{\mathcal C_\mathrm{GKP}}_{\hat A})^*$.
    \item\label{enum:SWstGKP} Standardization: $\int_{\mathbb{T}_{d\ell}^2}\frac1d{dudv}\,W^{\mathcal C_\mathrm{GKP}}_{\hat A}(u,v)=\Tr[\hat A]$.
    \item\label{enum:SWcovGKP} Covariance: $W_{\pi(g)^\dag\hat A\pi(g)}(\Omega)=W_{\hat A}(g\cdot(u,v))$, for all $(u,v)\in\mathbb{T}_{d\ell}^2$ and $g\in H_3(\mathbb C)$, where the action of $H_3(\mathbb C)$ on $\mathbb{T}_{d\ell}^2$ is given by $g(\varphi,x,p)\cdot(u,v)=(u+x\mod d\ell,v+p\mod d\ell)$.
    \item\label{enum:SWtrGKP} Traciality: $\int_{\mathbb{T}_{d\ell}^2}\frac1d{dudv}\,W^{\mathcal C_\mathrm{GKP}}_{\hat A}(u,v)W^{\mathcal C_\mathrm{GKP}}_{\hat B}(u,v)=\Tr[\mathcal E(\hat A)\mathcal E(\hat B)]=\Tr[\mathcal E(\hat A)\hat B]=\Tr[\hat A\mathcal E(\hat B)]$.
\end{enumerate}
\end{theo}

\noindent We give a proof in Appendix~\ref{app:SWGKP}, together with additional properties of the SW kernel $\hat\Delta^{\mathcal C_\mathrm{GKP}}$. Along the way, we prove the following relation:
\begin{equation}\label{eq:mapE}
    \mathcal E(\hat A)=\int_{\mathbb{T}^2_{d\ell}}\frac1d{dudv}\,W^{\mathcal C_\mathrm{GKP}}_{\hat\rho}(u,v)\hat{\Delta}(u,v).
\end{equation}
Contrast this equation and \ref{enum:SWlinGKP} with Eq.~(\ref{eq:reconstruction}) and \ref{enum:SWlin}: rather than being informationally-complete, the Zak--Gross Wigner function captures the part of the state that has the same symmetry as the code space $\mathcal C_\mathrm{GKP}$, i.e., is supported on the image of the map $\mathcal E$. In spite of this, note that the traciality property \ref{enum:SWtrGKP} implies that Born statistics can be retrieved from the Zak--Gross Wigner function even when only one of either a preparation (density matrix) or a measurement (POVM element) has the same symmetry as the code space.

\subsection{Physical description: modular measurements and conjugate Zak distributions as marginals}
\label{sec:phys}

The two axes of the toroidal phase space $\mathbb T_{d\ell}^2$ are associated with a modular position and a modular momentum, each with a periodicity matching the stabilizer displacement size $d\ell=\sqrt{2\pi d}$.  The product of these two periods is an integer multiple of $2\pi$, implying that the stabilizers commute and can be simultaneously measured \cite{Busch_Lahti_1986}.  A natural question is then: where may non-classical phase-space negativity come from?  By comparison, the original Wigner function is based on a phase space where the two axes are associated with the canonically conjugate variables of position and momentum, which do not commute.

As it turns out, the non-classicality is coming from logical displacements. This is natural in the GKP construction \cite{conrad2022gottesman} and can be understood as follows. The stabilizers are indeed jointly measurable but the simultaneous eigenspaces \eqref{eq:code space_def} as a whole are invariant under logical shifts.  This results in the syndrome distribution \eqref{eq:syndrome_distribution} being defined over a cell of size $\ell^2 = \frac{2\pi}{d}$, one point for each code space.  So while the outcomes of the syndrome measurement take values in this small ``logical cell'', that is not the same thing as simultaneously measuring the position and momentum of a bosonic state modulo $\ell=\sqrt{2\pi/d}$.  Such a measurement is impossible because the product of the periods is not a multiple of $2\pi$, or equivalently because the two logical displacements do not commute \cite{ketterer2016modular}. Consider one of the logical displacements, $\hat{\bar{X}} = e^{-i\ell\hat p}$ for example; all eigenstates are of the form $\psi_p(x) = e^{i p x}\varphi_p(x)$, where $\varphi_p(x)$ is $\ell$-periodic \cite{Zak1967finiteTranslations}. The eigenvalue for this eigenstate is $e^{-ip\ell}$, and the set of all eigenvalues is therefore parameterized by the quasi-momentum interval $p\in[0, 2\pi/\ell)=[0,\sqrt{2\pi d})$. Measuring the logical displacement $\hat{\bar{X}}$ hence returns a number in this range, and similarly for $\hat{\bar{Z}}$. In general, the outcome spaces of logical and stabilizer measurements are made clear by writing:
\begin{equation}
\begin{aligned}
        \hat{\bar Z}&=e^{i\frac{2\pi}{d\ell}\hat x}, \quad \hat{\bar X}=e^{-i\frac{2\pi}{d\ell}\hat p}, \\
        \hat{\bar Z}^d&=e^{i\frac{2\pi}\ell\hat x}, \quad\hat{\bar X}^d=e^{-i\frac{2\pi}\ell\hat p}.
    \end{aligned}
\end{equation}
While one cannot simultaneously measure both logical displacements, we see that their outcomes are localized to within the ``stabilizer cell" of area $(d \ell)^2 = 2\pi d$.  This larger cell of quasi-position and quasi-momentum is our phase space and the negativity arises from the non-commutativity of the logical displacements, similar to how planar negativity arises from the non-commutativity of $\hat x$ and $\hat p$.

A more concrete way to see how the negativity arises in the Zak--Gross Wigner function is by studying its marginal distributions.  This is inspired by the CV notion that Wigner negativity can be thought of as an obstruction to the existence of a joint probability distribution that reproduces the quadrature distributions upon marginalization \cite{bertrand1987tomographic, Vogel_Risken_1989,Booth2021}.  In the present case, a similar situation occurs, and can be explicitly seen as follows.  Recall an expression for the Zak measurement (joint modular quadrature measurement) distribution of a pure CV state:
\begin{equation}\label{Zak_measurement_characteristic}
    \!|[Z_\alpha\psi](k,q)|^2\!=\!\frac1{2\pi}\!\sum_{n,m} (-1)^{nm} \chi_\psi (\alpha n, \frac{2\pi}{\alpha}m) e^{i\alpha k n} e^{-i2\pi \frac{q}{\alpha} m}\!\!,
\end{equation}
where $\chi_\psi(q,p)$ is the characteristic function of $\ket{\psi}$ \eqref{eq:CVDV_char_wig_def} and $\alpha$ defines the \textit{Zak patch} $[0,\alpha) \times [0,\frac{2\pi}{\alpha}) \subset \R^2$. This expression can be found, e.g., in \cite{janssen1988zak}, and we reproduce its derivation in Appendix \ref{appendix_gkp_dim1} for completeness.  Note that by setting the length $\alpha = \ell = \sqrt{2\pi}$ we get
\begin{equation}
    |[Z_\ell\psi](k,q)|^2 = \frac1{2\pi}\sum_{m,n} (-1)^{mn} \chi_\psi (\ell n, \ell m) e^{i\ell ( k n - q m)}.
\end{equation}
By Eqs.~\eqref{eq:CVDV_char_wig_def} and \eqref{eq:fullexprWZG}, this is the Zak--Gross Wigner function for the trivial $(d=1)$ square GKP code with the variables $(u,v)$ respectively identified with the quasi-position $q$ and quasi-momentum $k$:
\begin{equation}\label{dim1_Wigner_is_Zak}
        |[Z_\ell\psi](k,q)|^2 = W^{\mathcal C_\mathrm{GKP}}_{\ketbra{\psi}{\psi}}(q,k).
\end{equation}
It then follows from linearity that for general mixed states
\begin{equation}
    W_{{\hat\rho}}^{\mathcal C_\mathrm{GKP}}(u,v) = \langle v,u | {\hat\rho} | v,u \rangle,
\end{equation}
or, equivalently,
\begin{equation}\label{1dim_kernel_as_zak_projector}
    \hat \Delta^{\mathcal C_\mathrm{GKP}}(u,v)=\ketbra{v,u}{v,u}.
\end{equation}
This, together with the fact that logical operations are the same as stabilizers for $d=1$, means that in the trivial GKP code there is no distinction between our Wigner function, the Zak distribution, and the syndrome distribution.

For $d>1$ the situation changes and all three become distinct.  Consider the following form of marginalization.  Instead of integrating out one of the variables $(u,v)$, perform a finite sum along one of the directions with a spacing of $\ell$.  This has the effect of lowering the periodicity of one of the $(u,v)$ variables from $d\ell$ to $\ell$, resulting in a 2-dimensional marginal distribution over an area of $\ell \cdot d\ell = 2\pi$.  Such marginals correspond to the joint probability distribution for the measurements of $\hat{\bar Z}$ and $\hat{\bar X}^d$, or $\hat{\bar Z}^d$ and $\hat{\bar X}$, respectively. Moreover, they can be identified with the Zak measurement distributions over the corresponding area, now interpreted as a Zak patch:
\begin{equation}\label{eq:zak_marginalization1}
    \begin{aligned}
        &\quad \frac1d\sum_{j=0}^{d-1} W^{\mathcal C_\mathrm{GKP}}_{\ketbra{\psi}{\psi}}(u, v - j\ell) \\
        &= \frac1{2\pi} \sum_{m,n \in \Z} (-1)^{dnm} \chi_\psi(\ell dn, \ell m) e^{i\ell(dnv - mu)} \\
        &= | [Z_{d\ell}\psi](v,u) |^2,
    \end{aligned}
\end{equation}
where we used the identity $\sum_{j=0}^{d-1}\omega^{-jn} = d \delta_{dn,0}$ and $(-1)^{dnm}=(-1)^{nm}$ for odd $d$. Note that $v$ is now $\ell$-periodic.  Similarly,
\begin{equation}\label{eq:zak_marginalization2}
        \frac1d\sum_{j=0}^{d-1} W^{\mathcal C_\mathrm{GKP}}_{\ketbra{\psi}{\psi}}(u - j\ell, v) =| [Z_{\ell}\psi](v,u) |^2.
\end{equation}
These two marginals can be seen on the operator level as the Zak--Gross phase point operators \eqref{eq:GKP_Wigner_kernel_abstract_def} reducing to the Zak projector basis over the respective Zak patch.  Hence the presence of negativity in the Zak--Gross Wigner function can be seen as an obstruction to finding a joint probability distribution that reproduces the conjugate Zak distributions as marginals.  The origins of this idea of conjugate Zak transforms acting as canonically conjugate modular variables was considered by Mann, Revzen, and Zak in \cite{mann2005conjugate,Mann_Revzen_Zak_2006} but in the limited context of a $d$-dimensional system.  Here, within the wider scope provided by the GKP framework, we have extended it to the full collection of displaced code spaces in an infinite-dimensional setting and observed an information-theoretic interpretation related to (the impossibility of) simultaneously measuring logical operators in a quantum error-correcting code.

A further property of the Zak--Gross Wigner function is that if we go further and perform a ``double'' marginalization of the above kind (i.e., a finite summation along both directions), the resulting object is in fact the syndrome distribution \eqref{eq:syndrome_distribution}, i.e., the joint probability distribution for the measurements of $\hat{\bar Z}^d$ and $\hat{\bar X}^d$:
\begin{equation}\label{eq:double_marg_to_syndrome}
    \frac1d\sum_{a,b=0}^{d-1} W^{\mathcal C_\mathrm{GKP}}_{\hat \rho}(a\ell+s,b\ell+t)= \Tr[\hat\rho \, \hat{{P}}_{(s,t)}],
\end{equation}
where now $(s,t)\in\mathbb T_\ell^2$ (see Appendix \ref{appendix_double_marginalization} for details).
See Fig.~\ref{fig:pGKP} for an example of Zak--Gross Wigner function illustrating these relations.


\subsection{Information-theoretic description: GKP encoding and the Gross Wigner function}
\label{sec:info}

At this point we may start reinterpreting the Zak--Gross Wigner function as representing the logical content of quantum states in $\mathcal H$ with respect to the GKP code space $\mathcal C$. First, notice that the kernel \eqref{eq:DeltaGKP00} rewrites \eqref{encoded_DV_dis_via_CV_dis} in terms of logical operators as:
\begin{equation}
    \begin{aligned}
        \hat\Delta^{\mathcal C_\mathrm{GKP}}(0,0) &= \frac1{2\pi}\sum_{m,n \in \Z} (-1)^{mn} \hat{D}(n\ell,m\ell) \\
        &=\frac1{2\pi}\sum_{m,n \in \Z} \omega^{2^{-1}mn} \hat{\bar{X}}^n \hat{\bar{Z}}^m \\
        &=\frac1{2\pi}\sum_{m,n \in \Z} \hat{\bar{D}}_d(n, m).
    \end{aligned}
\end{equation}
In particular, we show that this operator can be interpreted as a logical parity operator:

\begin{lem}\label{lem:logicalparity}
\begin{equation}
    \hat\Delta^{\mathcal C_\mathrm{GKP}}(0,0)=\hat{\bar{\Pi}}_d(0,0),
\end{equation}
where $\hat{\bar{\Pi}}_d(0,0):=\sum_{j=0}^{d-1}\ket{-\bar j}\!\bra{\bar j}$ is the GKP-encoded parity operator over $\mathcal H$.
\end{lem}

\noindent We give a proof in Appendix~\ref{app:SWGKP} by computing the action of both operators on the position basis.

Recall from Eq.~(\ref{eq:CVDV_char_wig_def}) the expression of the DV Wigner function in terms of DV displaced parity operators as $W^\mathrm{DV}_{\hat\rho}(a,b) = \Tr[\hat \Pi_d(a,b) {\hat\rho}]$, for all $(a,b)\in\mathbb Z_d^2$. From Definition~\ref{def:ZGWigner}, this suggests that the Zak--Gross Wigner function is related to the Gross Wigner function of GKP-encoded states. The next result makes this connection precise:

\begin{theo}\label{th:collectionDV} 
Let $\hat \rho$ be a continuous-variable state and $(u,v)\in\mathbb T_{d\ell}^2$. Writing $u = s + a\ell$ and $v = t + b\ell$ for $(s,t)\in\mathbb T_\ell^2$ and $(a,b)\in\mathbb Z_d^2$, we have
\begin{equation}
    W_{\hat\rho}^{\mathcal C_\mathrm{GKP}}(u,v)=W_{\hat{\underline\rho}(s,t)}^{\mathrm{DV}}(a,b),
\end{equation}
where $\hat{\underline\rho}(s,t)$ is the finite-dimensional (sub-normalised) logical state corresponding to the error-corrected state $\hat\rho(s,t)=\hat P_{(0,0)}\hat D(s,t)^\dag\hat\rho\hat D(s,t)\hat P_{(0,0)}$ defined in Eq.~(\ref{eq:err-corr-state}).
\end{theo}
\noindent The proof of this result is based on Lemma~\ref{lem:logicalparity} and is detailed in Appendix \ref{app:SWGKP}. The result is more generally valid for any CV operator $\hat A$ such that $W_{\hat A}^{\mathcal C_\mathrm{GKP}}$ is well-defined.  

A first consequence of Theorem~\ref{th:collectionDV} is that the Zak--Gross Wigner function of a codeword $\hat{\bar \rho}$ numerically matches the Gross Wigner function of the unencoded finite-dimensional state $\hat\rho$,
\begin{equation}
    W_{\hat{\bar \rho}}^{\mathcal C_\mathrm{GKP}}(u,v) = \begin{cases}
        W_{\hat\rho}^{\mathrm{DV}}(a,b), & (u,v)=(a\ell, b\ell) \\
        0, & \text{otherwise}.
    \end{cases}
\end{equation}
This relationship can be summarized by the following schematic commutative diagram for a $d$-dimensional quantum state ${\hat\rho}$:
%
\tikzcdset{every matrix/.append style={column sep=huge}}

\begin{equation}
\label{c_diagram_code_space}
    \begin{tikzcd}
     \mathcal{D}(\mathcal{H}_d) \ni {\hat\rho} \arrow[r, "\text{encode}"] \arrow[d, "\text{DV Wigner}", shift right=0pt]
    & \hat{\bar{\rho}} \arrow[d, "\text{Zak--Gross Wigner}"] \\
    W^{\text{DV}}_{\hat\rho} (a,b) \arrow[r, "\text{embed in }\mathbb{T}^2_{d\ell}", shift right=0pt]
    &  W^{\mathcal C_\mathrm{GKP}}_{\hat{\bar{\rho}}}(a\ell,b\ell)
    \end{tikzcd}
\end{equation}
for $(a,b)\in\mathbb Z_d^2$, where $\mathcal{D}(\mathcal{H}_d)$ denotes the set of density operators over a $d$-dimensional Hilbert space $\mathcal H_d$.

Related ideas have been discussed recently in \cite{Feng_Luo_2024, hahn2024bridgingMagic}.  These works use the GKP framework to faithfully relate DV measures of non-classicality such as DV Wigner negativity and stabilizer R\'{e}nyi entropy to the non-classicality properties of CV (i.e., planar) quasiprobability distributions representing the encoded counterparts. These results also offer insight to the more general problem of understanding how a quantum resource should be characterized pre-encoding vs post-encoding.

However, such connections to CV distributions so far only apply to states within the GKP code space $\mathcal C_\mathrm{GKP}$.  By contrast, our construction assigns a well-behaved phase-space representation to all states in $\mathcal H$, both within and outside $\mathcal C_\mathrm{GKP}$.
Indeed, Theorem~\ref{th:collectionDV} implies more generally that the Zak--Gross Wigner function can in fact be interpreted as a continuous family of DV Wigner functions, one for each displaced GKP code space.  In particular, we obtain the inverse diagram of \eqref{c_diagram_code_space}, valid for a general CV state ${\hat\rho}$ and for general displacements:
\begin{equation}
    \begin{tikzcd}
    \mathcal{D(H)} \ni {\hat\rho} \arrow[r, "\text{error-correct}"] \arrow[d, "\text{Zak--Gross Wigner}"]
    & \hat\rho(s,t) \arrow[d, "\text{DV Wigner}"] \\
    W^{\mathcal C_\mathrm{GKP}}_{\hat\rho}(a\ell+s,b\ell+t) \arrow[r, "\text{restrict}"]
    &  W^{\text{DV}}_{\hat{\underline\rho}(s,t)}(a,b)
    \end{tikzcd}
\end{equation}
for $(a,b)\in\mathbb Z_d^2$ and $(s,t)\in\mathbb T_\ell^2$, where $\mathcal{D}(\mathcal{H})$ denotes the set of density operators over the infinite-dimensional Hilbert space $\mathcal H$.  

Equivalently, the restriction of the Zak--Gross Wigner function to a displaced logical lattice within the toroidal phase space, $\ell \Z_d \times \ell \Z_d \xhookrightarrow{} \mathbb{T}^2_{d\ell}$, yields the DV Wigner function of the CV state projected to that code space.  In this respect, the Zak--Gross Wigner function is a truly hybrid CV-DV phase space representation: it has finite support for states within a finite number of displaced code spaces, but continuous support in general.  It is now also clear that the conjugate Zak distributions, \eqref{eq:zak_marginalization1} and \eqref{eq:zak_marginalization2}, can be seen as collective quadrature-like marginals of a family of DV Wigner functions.

To interpret the double marginal \eqref{eq:double_marg_to_syndrome}, it should be emphasized from Theorem \ref{th:collectionDV} that the DV state that such a restricted Zak--Gross Wigner function represents is sub-normalized.  This comes from the fact that general CV states have support on more than one code space and therefore will collapse onto them with different probabilities, given by the syndrome distribution \eqref{eq:syndrome_distribution}, leading to a sub-normalization by a factor equal to this probability. For ideal codewords affected by displacement errors, this does not arise because they are always contained entirely within only one displaced code space.  Hence, the collection of DV Wigner functions comprising the Zak--Gross Wigner function are each normalized to the probability of obtaining their associated displaced code space in the syndrome measurement. The double marginal is now readily seen as calculating this probability.  We refer to Fig.~\ref{fig:pGKP} for an explicit example.

It remains to give an operational interpretation of the negativity of the Zak--Gross Wigner function.
The negative volume of the DV Wigner function of a qudit state $\hat\rho$ is defined as
\begin{equation}
    \mathcal N^{\text{DV}}_{\hat\rho}=\frac1d\sum_{a,b\in\mathbb Z_d}\left|W^{\text{DV}}_{\hat\rho}(a,b)\right|.
\end{equation}
Since the DV Wigner function is normalised, the negative volume is $1$ for density operators with non-negative DV Wigner function and greater than $1$ otherwise.
In the odd-dimensional case, it provides a measure of magic \cite{veitch2014resource}. 
We define similarly the negative volume of the Zak--Gross Wigner function of a CV state $\hat\rho$ as
\begin{equation}
    \mathcal N_{\hat\rho}^{\mathcal C_\mathrm{GKP}}:=\int_{\mathbb{T}_{d\ell}^2}\frac1ddudv\,\left|W^{\mathcal C_\mathrm{GKP}}_{\hat\rho}(u,v)\right|.
\end{equation}
As before, since the Zak--Gross Wigner function is normalised, the negative volume is $1$ for density operators with non-negative Zak--Gross Wigner function and greater than $1$ otherwise.
Recall from section \ref{sec:GKP} that GKP error correction of a state $\hat\rho$ measures a syndrome $(s,t)\in\mathbb T_\ell^2$ with probability $\text{Pr}[(s,t) \, | \, \hat\rho] =\Tr[\hat\rho\,\hat{{P}}_{(s,t)}]$, leading to the normalised post-measurement state $\hat{\rho}'(s,t):=\hat\rho(s,t)/\Tr[\hat\rho\,\hat{{P}}_{(s,t)}]$ in the code space $\mathcal C_\mathrm{GKP}$, with its corresponding DV logical state $\hat{\underline\rho}'(s,t)$ through the GKP encoding.
The next result, which is a direct consequence of Theorem~\ref{th:collectionDV} and of the linearity of the DV Wigner function, shows that the negative volume of the Zak--Gross Wigner function provides a GKP-compatible measure of magic for states both within and outside the code space.

\begin{coro}\label{coro:negGKP} 
The negative volume of the Zak--Gross Wigner function is the average negative volume of the Gross Wigner function over one round of GKP error correction:
\begin{equation}
    \mathcal N_{\hat\rho}^{\mathcal C_\mathrm{GKP}}=\int_{\mathbb T_\ell}dsdt\, \mathrm{Pr}[(s,t) \, | \, \hat\rho]\,\mathcal N^{\mathrm{DV}}_{\hat{\underline\rho}'(s,t)},
\end{equation}
for any CV state $\hat\rho$, where $\mathrm{Pr}[(s,t) \, | \, \hat\rho]$ is the syndrome distribution.
\end{coro}

This result implies that stabilizer codewords are positively represented by the Zak--Gross Wigner function, while magic codewords are negatively represented. Moreover, for states outside the code space, the negative volume of the Zak--Gross Wigner function is the average negative volume of the DV Wigner functions of the codewords obtained after error correction. Note that the negative volume of the Zak--Gross Wigner function of a CV state $\hat\rho$ is in general greater than the negative volume of the DV Wigner function of the logical state associated to the expected error-corrected state $\int_{\mathbb{T}_\ell^2}dsdt\hat{P}_{(0,0)}\hat D^\dag(s,t)\hat\rho\hat D(s,t)\hat P_{(0,0)}$, by the triangle inequality. Note also that there exist bosonic states whose projection to the GKP code space is DV-Wigner-positive yet is Zak--Gross Wigner-negative: an example is a thermal state of appropriate temperature (see section \ref{sec:examples}), or any encoded magic state displaced by a non-logical displacement.


\subsection{Examples}
\label{sec:examples}

In this section, we present explicit examples of Zak--Gross Wigner functions.


\subsubsection{Codewords}

From Sec.\ \ref{sec:info}, the Zak--Gross Wigner function of a GKP-encoded computational basis state $\ket{\bar j}$ is essentially the DV Wigner function of the unencoded state $\ket{j}$ embedded into the continuous torus $\mathbb T_{d\ell}^2$. In particular, it is a vertical sequence of $d$ delta functions starting at $v=0$ and vertically interspaced by an amount $\ell$, with fixed horizontal position $u = j\ell$. The encoded Fourier states $\ket{\bar j}_+$ similarly form a horizontal sequence of delta functions with vertical position $v =j\ell$.



\subsubsection{Coherent states}

\begin{figure}[t]
    \centering
    \includegraphics[width=\columnwidth]{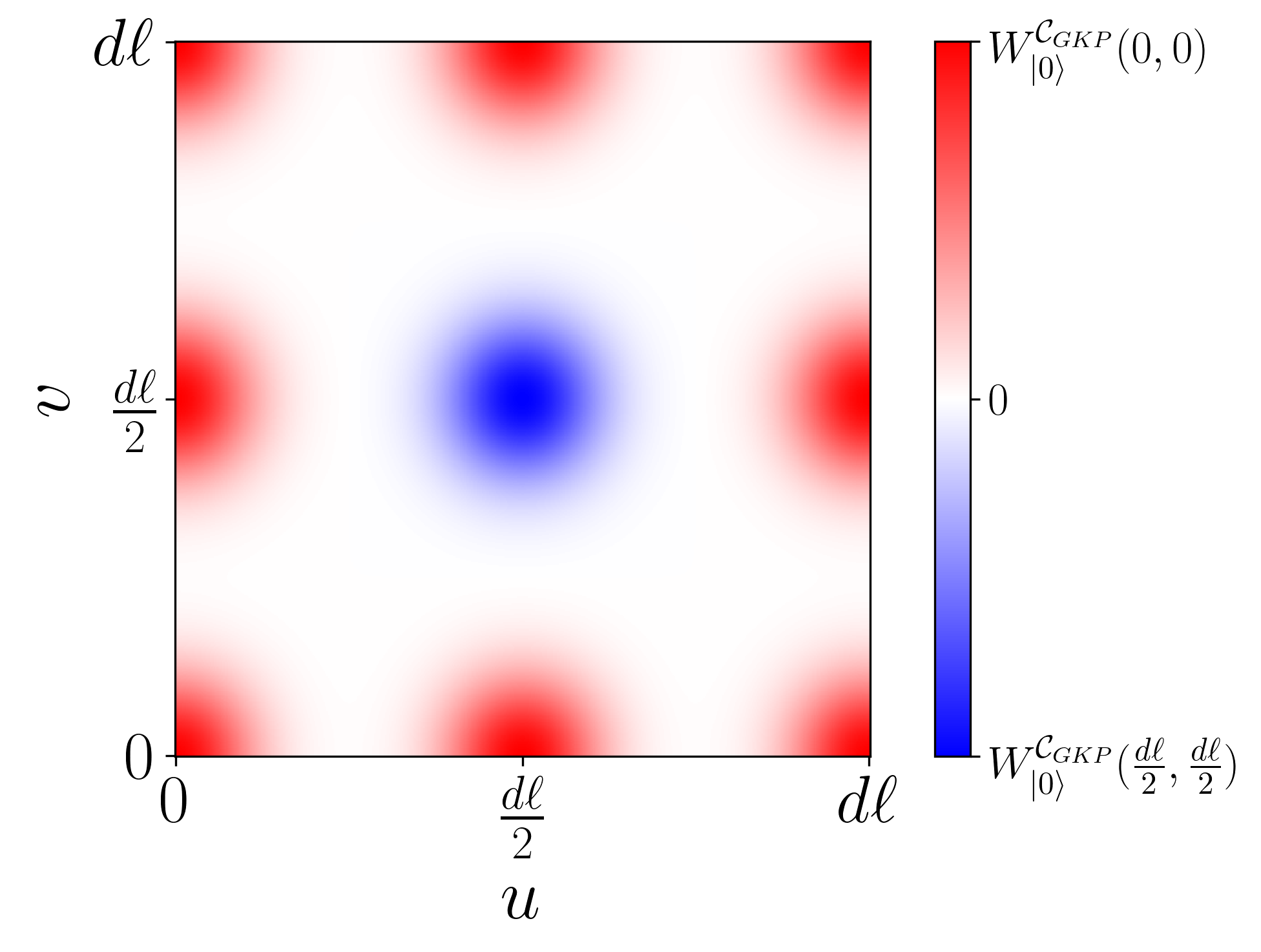}
    \caption{The Zak--Gross Wigner function of the vacuum $\ket{0}$ with logical dimension $d=13$.}
    \label{fig:vacuum_ZG}
\end{figure}

First consider the vacuum state $\ket{0}$.  With Eq.~(\ref{eq:fullexprWZG}), its Zak--Gross Wigner function is found to be
\begin{equation}\label{eq:vacuum_ZG_as_sum}
    W_{\ket{0}}^{\mathcal{C}_{\text{GKP}}}(u,v) = \frac1{2\pi} \sum_{n,m\in\mathbb Z} (-1)^{nm} e^{i\ell(nv - mu)} e^{- \frac{\pi}{2d}(n^2 + m^{2})}.
\end{equation}
Following the definition \eqref{eq:GKP_Wigner_def}, this can be found for example by computing the characteristic function $\chi_{\ket{0}}(x,p)$ then performing the weighted sum over the logical lattice.  Another approach is to proceed entirely in position basis and use the wavefunction representation $\psi_0(x) = \pi^{-1/4} e^{-x^{2}/2}$; see Appendix \ref{appendix_thermal} for details on thermal states from which the vacuum can be obtained as a special case.  Regardless of the derivation, the result can be expressed as a particular two-dimensional theta function \eqref{theta_multi} $\theta(\bm z, \bm \tau)$ restricted to real inputs in the first factor:
\begin{equation}
    W_{\ket{0}}^{\mathcal{C}_{\text{GKP}}}(u,v) = \frac1{2\pi} \theta\!\left(\left(\frac{v}{d\ell}, -\frac{u}{d\ell}\right), \bm \tau\right), \quad \bm \tau =
    \frac{1}{2}
    \begin{pmatrix}
    \frac{i}{d} & 1 \\ 1 &  \frac{i}{d}
\end{pmatrix}.
\end{equation}
The remaining coherent states follow from the covariance property under continuous displacements \ref{enum:SWcovGKP}.  In particular the coherent state $\ket{x,p} = \hat D(x,p)\ket{0}$ is
\begin{equation}\label{eq:coherent_displaced}
\begin{aligned}
    W_{\ket{x,p}}^{\mathcal{C}_{\text{GKP}}}(u,v) &= W_{\ket{0}}^{\mathcal{C}_{\text{GKP}}}(u-x,v-p) \\
    &= \frac1{2\pi} \theta\!\left(\left(\frac{v-p}{d\ell}, -\frac{u-x}{d\ell}\right), \bm \tau \right),
\end{aligned}
\end{equation}
that is represented in Fig.\ \ref{fig:vacuum_ZG}.  Unlike the original Wigner function, the Zak--Gross Wigner function takes negative values for the vacuum state. In particular, the negativity is concentrated about the point $(u,v) = (\frac{d\ell}{2}, \frac{d\ell}{2})$.

\subsubsection{Thermal states}

Associated with the harmonic oscillator Hamiltonian $\hat H = \frac{1}{2}(\hat x^2 + \hat p^2)$ are the thermal states ${\hat\rho}_\beta = e^{-\beta \hat H} / Z(\beta)$, where $\beta = 1/T$ is the inverse temperature and
\begin{equation}
    Z(\beta) = \Tr[e^{-\beta \hat H}] = \frac{1}{2 \sinh(\frac{\beta}{2})}
\end{equation}
is the partition function \cite{Feynman_1998}. The Zak--Gross Wigner function of such states is found to be
\begin{equation}\label{thermal_GKP_Wigner_full}
    W^{\mathcal C_\mathrm{GKP}}_{\hat\rho_\beta}(u,v) \!=\! \frac1{2\pi}\!\sum_{n,m\in\mathbb Z} (-1)^{nm} e^{i\ell(nv-mu)} e^{- \frac{\pi}{d}(n^2 + m^2) \langle \hat H \rangle_\beta }\!\!,
\end{equation}
where
\begin{equation}\label{thermal_expectation}
    \langle \hat{H} \rangle_\beta = \frac{1}{2} \coth(\frac{\beta}{2})
\end{equation}
is the energy; see Appendix \ref{appendix_thermal} for details.  
The vacuum \eqref{eq:vacuum_ZG_as_sum} is recovered in the limit $\beta \rightarrow 0$ where $ \langle \hat{H} \rangle_\beta \rightarrow \frac{1}{2}$.  As for coherent states, the thermal Zak--Gross Wigner function \eqref{thermal_GKP_Wigner_full} can be expressed as a two-dimensional theta function \eqref{theta_multi} restricted to real inputs in the first factor:
\begin{equation}\label{thermal_GKP_Wigner_theta}
    W_{{\hat\rho}_\beta}^{\mathcal C_\mathrm{GKP}}\!(u,v)\! =\! \frac1{2\pi}\theta\!\left(\!\left(\frac{v}{d\ell}, -\frac{u}{d\ell}\right)\!, \bm \tau \!\right)\!,\;\bm \tau\! =\! \begin{pmatrix}
        i \frac{\langle \hat{H} \rangle_\beta}{d} & \frac{1}{2} \\
        \frac{1}{2} &  i \frac{\langle \hat{H} \rangle_\beta}{d}
    \end{pmatrix}\!.
\end{equation}
The covariance of the Zak--Gross Wigner function \ref{enum:SWcovGKP} allows us to consider displaced thermal states by a relation similar to the coherent states \eqref{eq:coherent_displaced}. From \eqref{thermal_GKP_Wigner_theta}, we see that for finite logical dimension $d$ the presence of zero-point energy (i.e., $\langle \hat H \rangle_\infty \neq 0$) precludes $\tau$ from leaving the Siegel upper half-space.

\begin{figure}[t]
    \centering
    \includegraphics[width=\columnwidth]{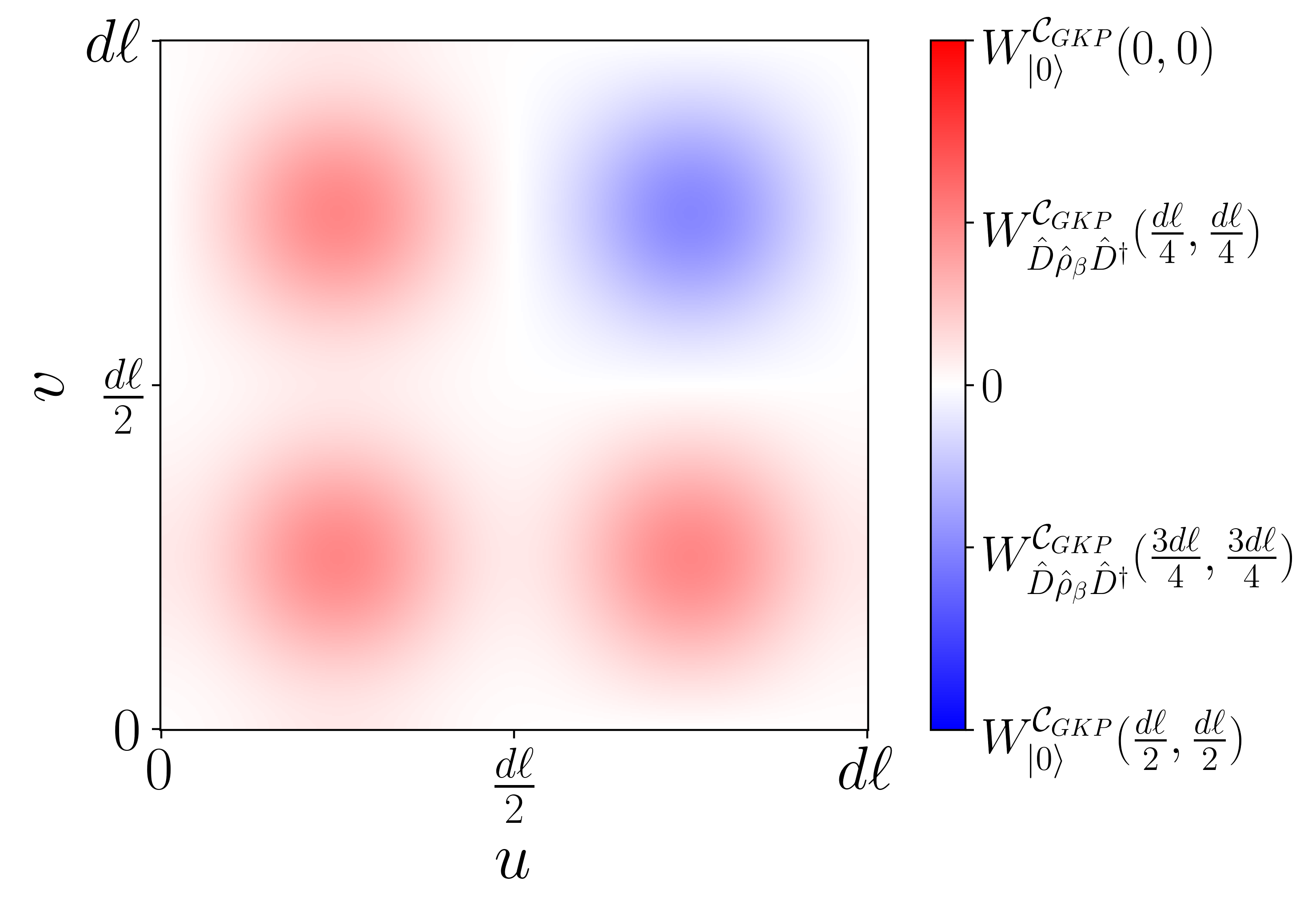}
    \caption{Zak--Gross Wigner function for $d=13$ of a thermal state at temperature $T = 1$, displaced by an amount $\hat D( \frac{d\ell}{4}, \frac{d\ell}{4})$. The colour scale is normalized to the vacuum Zak--Gross Wigner function to display the overall flattening as temperature increases.}
    \label{fig:displaced_thermal_state}
\end{figure}

See Fig.\ \ref{fig:displaced_thermal_state} for an example; this is the thermal state at temperature $T=1$ for logical dimension $d=13$ and displaced by an amount $\hat D(\frac{d\ell}{4}, \frac{d\ell}{4})$.  Comparing to the vacuum in Fig.\ \ref{fig:vacuum_ZG}, this additional displacement is done to better showcase how the relative size of the positive and negative regions increase with temperature.  We note in passing that this pattern is surprisingly similar to the CV Wigner function for stabilizer \textit{qubit} states restricted to an appropriate cell.

To characterize and interpret the presence of negativity we focus without loss of generality on undisplaced thermal states.  In every case considered the negativity concentrates about the centre of phase space $(u,v) = (\frac{d\ell}{2},\frac{d\ell}{2})$ and nowhere else;\ Fig.\ \ref{fig:vacuum_ZG} is typical up to a broadening of the characteristic regions.  This is accompanied by peaks of positivity at $(u,v) \in \{ (0,0), (0, \frac{d\ell}{2}), (\frac{d\ell}{2},0) \}$.  We conjecture this observation, which is based only the structure of theta functions, to be universal for all integers $d$, even and odd.  With that said, assuming $W_{{\hat\rho}_\beta}(\frac{d\ell}{2},\frac{d\ell}{2})$ is a global minimum for all finite $d$ and finite temperature $T = 1/\beta$, it follows that the negativity of the thermal Zak--Gross Wigner function (and therefore the presence of magic) is entirely governed by this value.  In particular, numerical evidence suggests that the following inequality must be satisfied for there to be positivity in the thermal Zak--Gross Wigner function:
\begin{equation}\label{eq:thermal_consrtaint}
    \langle \hat{H} \rangle_\beta > \frac{d}{2}.
\end{equation}
That is, the violation of \eqref{eq:thermal_consrtaint} represents the presence of negativity. 
Factoring out vacuum fluctuations, this puts a constraint on the number of thermal photons $\bar n = \langle \hat N \rangle = \langle \hat{H} \rangle_\beta - 1/2$ in the oscillator:
\begin{equation}\label{eq:thermal_photon_constraint}
    \bar n > \frac{d-1}{2}.
\end{equation}
When $d=1$ this is satisfied by any CV state, in agreement with \eqref{dim1_Wigner_is_Zak} showing that the Zak--Gross Wigner function reduces to a non-negative Zak distribution.  Eqs.\ \eqref{eq:thermal_consrtaint} and \eqref{eq:thermal_photon_constraint} can also be equivalently phrased using properties of the covariance matrix
\begin{equation}
    V = \begin{pmatrix}
        \langle \hat x \rangle^2 & \frac{1}{2}\langle \{ \hat x, \hat p \} \rangle \\
        \frac{1}{2}\langle \{ \hat x, \hat p \} \rangle & \langle \hat p \rangle^2
    \end{pmatrix} \cong \begin{pmatrix}
        \nu & 0 \\
        0 & \nu
        \end{pmatrix},
\end{equation}
where the second matrix is the Williamson normal form of $V$ and $\nu$ is the corresponding symplectic eigenvalue \cite{Simon_Mukunda_Noise_Matrix_1994, Olivares_2012}.  For thermal states this is just $\nu =  \bar n + \frac{1}{2}$, leading to the same constraint,
\begin{equation}\label{eq:uncertainty_thermal}
    \nu > \frac{d}{2}.
\end{equation}
Now when $d=1$ this is identified with the Heisenberg uncertainty relation for single-mode Gaussian states \cite{Simon_Mukunda_Noise_Matrix_1994}, which of course is satisfied by all quantum states.  But for $d>1$ we have a hierarchy of uncertainty inequalities,
\begin{equation}
    \nu = \sqrt{\det[V]} > \frac{2\pi d}{4\pi},
\end{equation}
the violation of which comes with an appealing geometric interpretation: a thermal state is Zak--Gross Wigner-negative when the phase space extent of the CV Wigner function, as measured by $\sqrt{\det[V]} = \nu$, is appropriately contained within the stabilizer cell.

\begin{figure}[ht]
    \centering
    \includegraphics[width=\columnwidth]{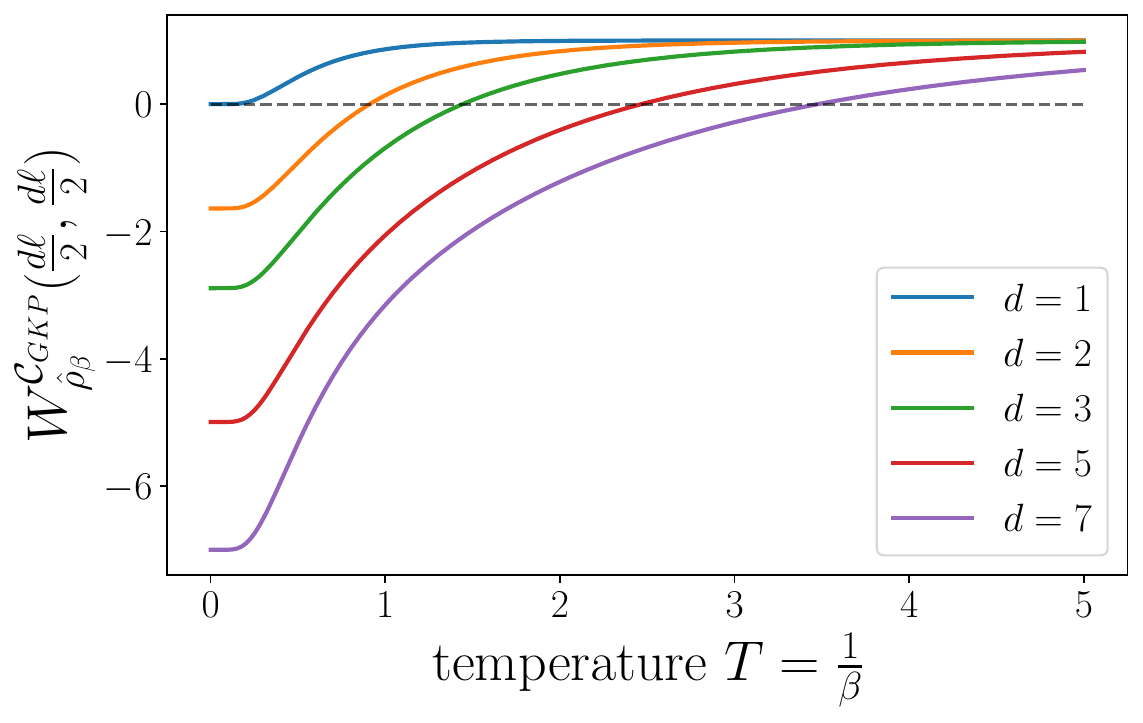}
    \caption{Minimum of the Zak--Gross Wigner function of a thermal state as a function of temperature, with $T=0$ being the vacuum.  Logical dimension $1$ is always non-negative due to the Zak--Gross Wigner function reducing to the Zak measurement distribution.  Logical dimension 2 is included for interest, though there is no underlying DV Gross Wigner function.}
    \label{fig:theta-v-beta}
\end{figure}

An immediate consequence is that the same thermal state can be Zak--Gross Wigner-negative for a high logical dimension but positive for a low logical dimension; see Figure \ref{fig:theta-v-beta}.  It also means that for a given logical dimension $d$ there is a critical temperature $T_{d}$ beyond which the associated thermal state is Wigner-positive in both the Zak--Gross sense and the usual sense.  More generally, it is likely the case that any CV state will become Wigner-positive in both senses after a sufficient degree of Gaussian blurring has been applied, e.g., by the pure loss channel.

We end this subsection on thermal states with a technical observation.  After rescaling the $\bm z$ argument appearing in the theta functions by the stabilizer length $d\ell$, the minimum value in the Wigner function (which serves as a magic witness for the error-corrected thermal state) can be written using a theta function with characteristics (see Appendix~\ref{app:Zaktheta}) evaluated at the origin,
\[ \min W_{{\hat\rho_\beta, \text{rescaled}}}^{\mathcal C_\mathrm{GKP}} \sim \thetachar{0}{(\frac{1}{2}, \frac{1}{2})}{(0,0)}{\bm \tau}. \]
This particular shift of $1/2$ is commonly seen in the study of theta functions.  Furthermore, the right hand side of the above relation, seen as a function of $\bm \tau$, is precisely what is known as a \textit{theta constant}, which is a central object in the study of modular forms \cite{Mumford_Tata_1}. The connection to quantum magic perhaps suggests that a deeper mathematical theory could bring about new tools for studying GKP resources.




\subsubsection{Physical GKP states}



Ideal GKP states are non-normalizable and so cannot be experimentally produced.  Instaed, it is crucail to consider approximate GKP codewords \cite{Gottesman2001, Terhal2016Encoding, PhysRevA.95.053819, Matsuura2020equivalenceGKPcodes}, which can be seen in the position basis as a sum of squeezed Gaussians together with an overall damping Gaussian envelope.  For the computational state $\ket{j}$ this is
\begin{equation}\label{eq:physicalGKPdef}
    \ket{\tilde j} \sim \sum_{k \in \Z} e^{-\frac{1}{2} \kappa^2 (j\ell + d\ell k)^2} \int dx e^{- \frac{1}{2} \frac{(x - j \ell - d\ell n)^2}{\sigma^2}} \ket{x}_{\hat x} ,
\end{equation}
up to some finite normalization constant.  Here $\sigma$ is the standard deviation of the individual Gaussians while $\frac{1}{\kappa}$ is the standard deviation of the envelope.  With $c_k \equiv e^{-\frac{1}{2} \kappa^2 (j\ell + d\ell k)^2}$ denoting the weights, we find that up to normalization the Zak--Gross Wigner function can be expressed as a weighted summation of two-dimensional theta functions:
\begin{equation}
\begin{aligned}
    W_{\ket{\tilde j}}^{\mathcal{C}_{\text{GKP}}}(u,v) \sim \sum_{k,k'} c_k c_{k'} e^{-\frac{\pi d}{2\sigma^2}(k' - k)^2} \theta( \bm z(u,v,k,k'), \bm \tau)
\end{aligned}
\end{equation}
where 
\begin{equation}
\begin{split}
    z_1 &= \frac{v}{d\ell}  + i \frac{k' - k}{2\sigma^2}, \\
    z_2 &= - \frac{u}{d\ell} + \frac{1}{2} (k+ k' + \frac{2}{d} j),
\end{split} \qquad
\begin{split}
    \bm \tau = \frac{1}{2}
    \begin{pmatrix}
        \frac{i}{d}\frac{1}{\sigma^2} & 1 \\ 1 & \frac{i}{d}\sigma^2
    \end{pmatrix}.
\end{split}
\end{equation}
See Appendix \ref{appendix:physical_GKP} for a derivation. 
\begin{figure}[t]
    \centering
    \includegraphics[width=\columnwidth]{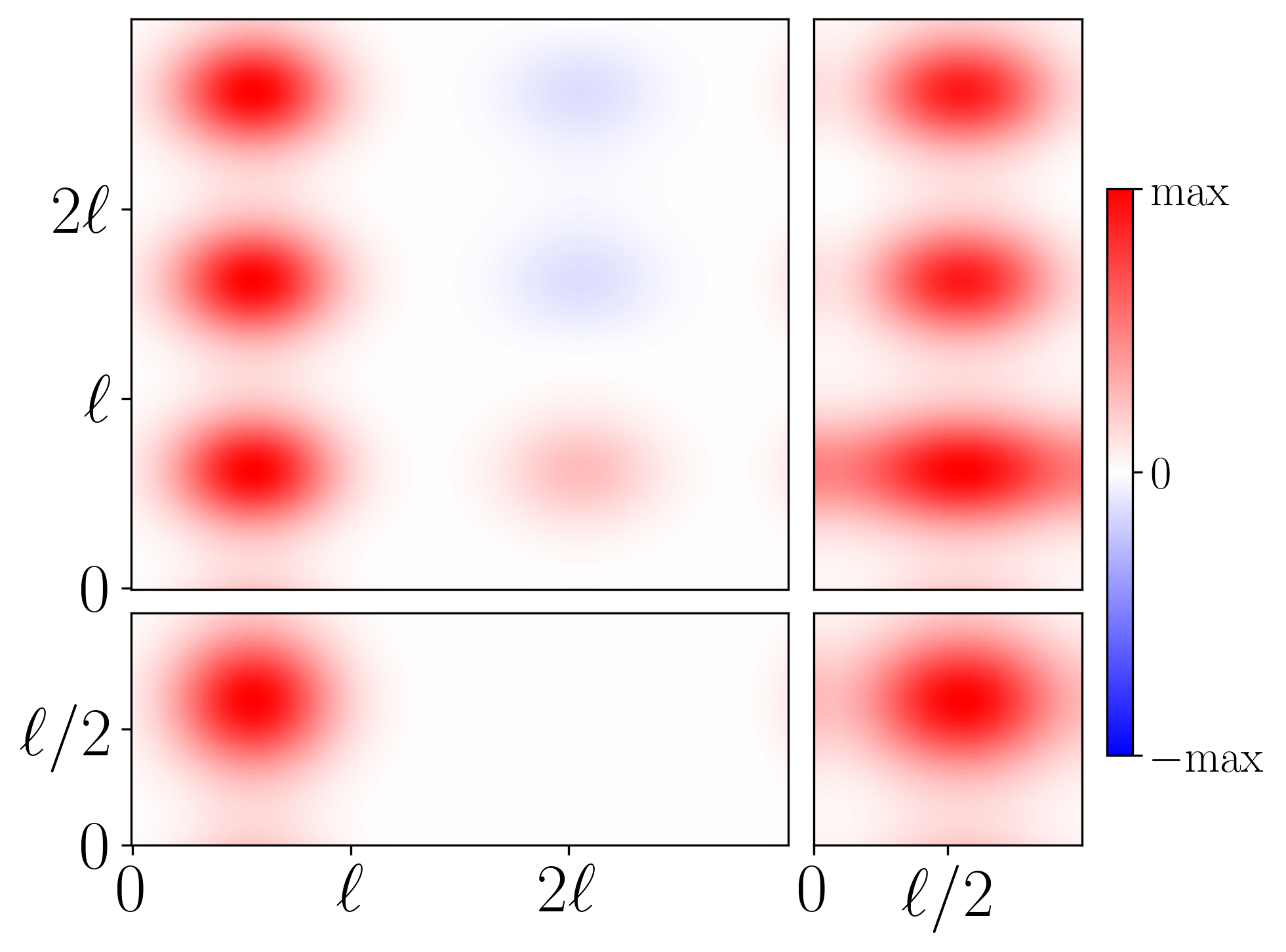}
    \caption{The Zak--Gross Wigner function of a slightly displaced approximate GKP codeword for the $d=3$ computational state $\ket{\bar0}$.  The vertical and horizontal marginals are modular/Zak measurement distributions corresponding to conjugate Zak patches.  The lower right plot is the syndrome distribution, obtained by a further marginalization of either Zak distribution.  Each plot and colour scale is self-normalized.}
    \label{fig:pGKP}
\end{figure}

We give an example in Fig.~\ref{fig:pGKP} for $d=3$, $j=0$, $\sigma = 0.51$ and $\kappa = 0.4$.  It was computed using a truncated summation of Eq.\ \eqref{eq:physicalGKPdef} to three Gaussians; zith negligible effect on the position wavefunction given the strength of $\kappa$. It has also been slightly displaced to better display its structure.  In particular, the Zak--Gross Wigner function takes on negative values on the toroidal phase space. Numerics suggest that as the state becomes more ideal through varying $\sigma$ and $\kappa$, the faint peaks along $u \approx 2\ell$ (two negative and one positive) decrease in magnitude while the three positive bright peaks along $u \approx \frac{\ell}{2}$ approach the three point-peaks expected in the ideal case.  Figure \ref{fig:pGKP} also demonstrates the general relationships between the Zak-Gross Wigner function and its non-negative marginals: the two conjugate Zak distributions and the syndrome distribution.










\section{SUmmary and Outlook}
\label{sec:concl}

We have introduced a symmetry-based construction of Wigner functions, generalizing the work of \cite{brif1999phase}.  This generalization can be thought of as considering reference spaces rather than reference states when constructing the phase-space representation. This approach is particularly well-suited in the context of quantum error-correcting codes. 
The resulting Wigner function is generally not informationally complete, but it satisfies a modified version of the Stratonovich--Weyl axioms tailored to the reference subspace.

We have applied our construction to the GKP encoding and focused on the case of odd logical dimensions where Wigner negativity in the logical space has previously been identified as a measure of magic \cite{veitch2014resource}.  The resulting Wigner function, which we named the Zak--Gross Wigner function, has several attractive properties as a phase-space representation of bosonic states over the continuous torus.  In particular, its restriction to an embedded $d \times d$ grid with spacing $\sqrt{2\pi/d}$ (i.e., a displaced logical lattice) yields the Gross Wigner function of the CV state projected to the GKP code space corresponding to the given grid. This allows for an interpretation of the Zak--Gross Wigner function as a continuous family of individual Gross Wigner functions, one for each displaced code space. Their collective DV marginals yield modular quadrature measurement distributions associated to conjugate Zak patches, while their collective double marginal yields the syndrome distribution, reflecting the property that each individual Gross Wigner function is normalized to the probability of the CV state collapsing to the corresponding code space.  Finally, we showed that the negativity volume of the Zak--Gross Wigner function provides a measure of magic of the logical content of a CV state, as it can be interpreted as the expected DV negativity over a single round of GKP error correction.

We computed the Zak--Gross Wigner function for several CV states.  As expected, for stabilizer codewords it corresponds to their positive Gross Wigner function embedded within the continuous torus.  Coherent states, on the other hand, and in particular the vacuum, are Zak--Gross Wigner-negative.  Thermal states show a structure similar to coherent states but with negativity decreasing with temperature until finally they become Wigner positive. This critical temperature is a function of the logical dimension, illustrating the general property that the same CV state may be more or less resourceful depending on the logical dimension of the encoding. Finally we have studied approximate GKP states and found that they are slightly Zak--Gross Wigner-negative.

Our results shed light on the perhaps counterintuitive results about classical simulability of GKP-encoded circuits \cite{garcia2020efficient, calcluth2022efficient, Baragiola_2019_allGaussianUniversality, Yamasaki_2020_costReducedGaussianUniversality, calcluth2023addingVacuum}. More generally, our framework contributes to the ongoing discussion of how to characterize the logical content of a state that has support on more that just the code space of a given quantum code \cite{calcluth2024sufficient}.  By employing phase-space techniques in particular, our work offers a novel perspective on the question of sufficiency of Wigner negativity as a witness for quantum resourcefulness in the context of quantum computation. That is, a fruitful way forward may be to account for the code space symmetry as it sits within the physical space.


Our work opens several directions for future research. One is to apply our general phase-space construction to other quantum codes as a means to characterize quantum resources under different encodings. This includes in particular finite-dimensional stabilizer codes, such as the surface code, as well as codes with a non-trivial symmetry group on the physical space \cite{albert2020robust}.  
Other codes of interest include alternative bosonic codes such as binomial and cat codes \cite{Albert2017PerformanceBosonic}.  We anticipate that our framework is also compatible with code concatenation, by interpreting the physical space as comprising all higher levels.

A second direction is to further harness our novel Zak--Gross Wigner function in the GKP setting. This includes deriving a Gottesman--Knill-type theorem based on the toroidal phase-space techniques developed here. Our construction has encoded Clifford operations already built-in, so the natural next step is to consider covariance under general Gaussian unitaries.  Given the results of \cite{calcluth2023addingVacuum}, we expect a strong distinction between the handling of rational vs irrational symplectic transformations.  We emphasize that our construction in principle works for even-dimensional encodings as well. We chose the odd-dimensional GKP code because of the known connections to magic in the DV setting. However another research path would be to instead focus on logical qubits and investigate if and how the result may be a continuous family of $d=2$ Wootters Wigner functions \cite{wootters1987wigner}, one for each code space. Another topic to consider is the extent to which the uncertainty relations numerically obtained in the context of unsqueezed Gaussians \eqref{eq:uncertainty_thermal} can be applied more generally and rigorously proven. This may lead to Zak--Gross negativity witnesses and bounds on magic state distillation protocols in the GKP code. Finally, from a more foundational perspective, there has been a growing interest in using modular variables as quantum tests of reality, in particular those related to violating Leggett--Garg and contextuality inequalities \cite{Asadian2014Probing, asadian2015contextuality, Fluhmann_sequential_2018}.  Given the connection between the Zak--Gross Wigner function to both the Gross Wigner function and conjugate Zak measurements, a reasonable belief is that negativity therein could be related to such modular-based tests. In particular, is Zak--Gross Wigner negativity related to contextuality for modular quadrature measurements?




Finally, a third direction is to further advance the underlying phase-space framework.  In particular, the methods of \cite{brif1999phase} require a choice of reference state (generalized here to reference space) to build the phase-space representation, together with an ansatz for the phase-space representation. That such a choice of reference state turns out to have little consequence in the case of Wigner functions suggests that the employed framework is more involved than necessary. Alternative approaches to phase-space quantization such as \cite{gazeau_integral_2022} may also be worth exploring.\\ \\

\section*{Acknowledgments}

We acknowledge inspiring discussions with R.\ I.\ Booth, F.\ Arzani, P.-E.\ Emeriau, G.\ Ferrini, A.\ Ferraro, C.\ Calcluth, J.\ Bermejo-Vega, O.\ Hahn and R.\ Takagi. U.C.\ acknowledges funding from the European Union’s Horizon Europe Framework Programme (EIC Pathfinder Challenge project Veriqub) under Grant Agreement No.\ 101114899. J.D.\ acknowledges funding from  the Plan France 2030 project NISQ2LSQ (ANR-22-PETQ-0006).


\bibliography{biblio}

\begin{thebibliography}{100}%
\makeatletter
\providecommand \@ifxundefined [1]{%
 \@ifx{#1\undefined}
}%
\providecommand \@ifnum [1]{%
 \ifnum #1\expandafter \@firstoftwo
 \else \expandafter \@secondoftwo
 \fi
}%
\providecommand \@ifx [1]{%
 \ifx #1\expandafter \@firstoftwo
 \else \expandafter \@secondoftwo
 \fi
}%
\providecommand \natexlab [1]{#1}%
\providecommand \enquote  [1]{``#1''}%
\providecommand \bibnamefont  [1]{#1}%
\providecommand \bibfnamefont [1]{#1}%
\providecommand \citenamefont [1]{#1}%
\providecommand \href@noop [0]{\@secondoftwo}%
\providecommand \href [0]{\begingroup \@sanitize@url \@href}%
\providecommand \@href[1]{\@@startlink{#1}\@@href}%
\providecommand \@@href[1]{\endgroup#1\@@endlink}%
\providecommand \@sanitize@url [0]{\catcode `\\12\catcode `\$12\catcode
  `\&12\catcode `\#12\catcode `\^12\catcode `\_12\catcode `\%12\relax}%
\providecommand \@@startlink[1]{}%
\providecommand \@@endlink[0]{}%
\providecommand \url  [0]{\begingroup\@sanitize@url \@url }%
\providecommand \@url [1]{\endgroup\@href {#1}{\urlprefix }}%
\providecommand \urlprefix  [0]{URL }%
\providecommand \Eprint [0]{\href }%
\providecommand \doibase [0]{https://doi.org/}%
\providecommand \selectlanguage [0]{\@gobble}%
\providecommand \bibinfo  [0]{\@secondoftwo}%
\providecommand \bibfield  [0]{\@secondoftwo}%
\providecommand \translation [1]{[#1]}%
\providecommand \BibitemOpen [0]{}%
\providecommand \bibitemStop [0]{}%
\providecommand \bibitemNoStop [0]{.\EOS\space}%
\providecommand \EOS [0]{\spacefactor3000\relax}%
\providecommand \BibitemShut  [1]{\csname bibitem#1\endcsname}%
\let\auto@bib@innerbib\@empty
\bibitem [{\citenamefont {Rundle}\ and\ \citenamefont
  {Everitt}(2021{\natexlab{a}})}]{rundle2021overview}%
  \BibitemOpen
  \bibfield  {author} {\bibinfo {author} {\bibfnamefont {R.~P.}\ \bibnamefont
  {Rundle}}\ and\ \bibinfo {author} {\bibfnamefont {M.~J.}\ \bibnamefont
  {Everitt}},\ }\href {https://doi.org/10.1002/qute.202100016} {\bibfield
  {journal} {\bibinfo  {journal} {Advanced Quantum Technologies}\ ,\ \bibinfo
  {pages} {2100016}} (\bibinfo {year} {2021}{\natexlab{a}})}\BibitemShut
  {NoStop}%
\bibitem [{\citenamefont {Chitambar}\ and\ \citenamefont
  {Gour}(2019)}]{chitambar2019quantum}%
  \BibitemOpen
  \bibfield  {author} {\bibinfo {author} {\bibfnamefont {E.}~\bibnamefont
  {Chitambar}}\ and\ \bibinfo {author} {\bibfnamefont {G.}~\bibnamefont
  {Gour}},\ }\href {https://doi.org/10.1103/RevModPhys.91.025001} {\bibfield
  {journal} {\bibinfo  {journal} {Rev. Mod. Phys.}\ }\textbf {\bibinfo {volume}
  {91}},\ \bibinfo {pages} {025001} (\bibinfo {year} {2019})}\BibitemShut
  {NoStop}%
\bibitem [{\citenamefont {{W}igner}(1932)}]{Wigner1932}%
  \BibitemOpen
  \bibfield  {author} {\bibinfo {author} {\bibfnamefont {E.}~\bibnamefont
  {{W}igner}},\ }\href {https://doi.org/10.1103/PhysRev.40.749} {\bibfield
  {journal} {\bibinfo  {journal} {Phys. Rev.}\ }\textbf {\bibinfo {volume}
  {40}},\ \bibinfo {pages} {749} (\bibinfo {year} {1932})}\BibitemShut
  {NoStop}%
\bibitem [{\citenamefont {Bj\"{o}rk}\ \emph {et~al.}(2008)\citenamefont
  {Bj\"{o}rk}, \citenamefont {Klimov},\ and\ \citenamefont
  {S\'{a}nchez-Soto}}]{Bjork_Klimov_Sanchez_Soto_2008}%
  \BibitemOpen
  \bibfield  {author} {\bibinfo {author} {\bibfnamefont {G.}~\bibnamefont
  {Bj\"{o}rk}}, \bibinfo {author} {\bibfnamefont {A.~B.}\ \bibnamefont
  {Klimov}},\ and\ \bibinfo {author} {\bibfnamefont {L.~L.}\ \bibnamefont
  {S\'{a}nchez-Soto}},\ }\bibinfo {title} {The discrete {Wigner} function},\
  in\ \href {https://doi.org/10.1016/S0079-6638(07)51007-3} {\emph {\bibinfo
  {booktitle} {Progress in Optics}}},\ Vol.~\bibinfo {volume} {51}\ (\bibinfo
  {publisher} {Elsevier},\ \bibinfo {year} {2008})\ p.\ \bibinfo {pages}
  {469–516}\BibitemShut {NoStop}%
\bibitem [{\citenamefont {Rundle}\ and\ \citenamefont
  {Everitt}(2021{\natexlab{b}})}]{Rundle_Everitt_2021}%
  \BibitemOpen
  \bibfield  {author} {\bibinfo {author} {\bibfnamefont {R.~P.}\ \bibnamefont
  {Rundle}}\ and\ \bibinfo {author} {\bibfnamefont {M.~J.}\ \bibnamefont
  {Everitt}},\ }\href {https://doi.org/https://doi.org/10.1002/qute.202100016}
  {\bibfield  {journal} {\bibinfo  {journal} {Adv. Quantum Technol.}\ }\textbf
  {\bibinfo {volume} {4}},\ \bibinfo {pages} {2100016} (\bibinfo {year}
  {2021}{\natexlab{b}})}\BibitemShut {NoStop}%
\bibitem [{\citenamefont {Schwinger}(1960)}]{schwinger1960unitary}%
  \BibitemOpen
  \bibfield  {author} {\bibinfo {author} {\bibfnamefont {J.}~\bibnamefont
  {Schwinger}},\ }\href {https://doi.org/10.1142/9789812795694_0026} {\bibfield
   {journal} {\bibinfo  {journal} {Proc. Natl. Acad. Sci. U.S.A.}\ }\textbf
  {\bibinfo {volume} {46}},\ \bibinfo {pages} {570} (\bibinfo {year}
  {1960})}\BibitemShut {NoStop}%
\bibitem [{\citenamefont {Buot}(1974)}]{Buot_method_1974}%
  \BibitemOpen
  \bibfield  {author} {\bibinfo {author} {\bibfnamefont {F.~A.}\ \bibnamefont
  {Buot}},\ }\href {https://doi.org/10.1103/PhysRevB.10.3700} {\bibfield
  {journal} {\bibinfo  {journal} {Phys. Rev. B}\ }\textbf {\bibinfo {volume}
  {10}},\ \bibinfo {pages} {3700} (\bibinfo {year} {1974})}\BibitemShut
  {NoStop}%
\bibitem [{\citenamefont {Wootters}(1987)}]{wootters1987wigner}%
  \BibitemOpen
  \bibfield  {author} {\bibinfo {author} {\bibfnamefont {W.~K.}\ \bibnamefont
  {Wootters}},\ }\href {https://doi.org/10.1016/0003-4916(87)90176-X}
  {\bibfield  {journal} {\bibinfo  {journal} {Ann. Phys. (N. Y.)}\ }\textbf
  {\bibinfo {volume} {176}},\ \bibinfo {pages} {1} (\bibinfo {year}
  {1987})}\BibitemShut {NoStop}%
\bibitem [{\citenamefont {Galetti}\ and\ \citenamefont {{de Toledo
  Piza}}(1988)}]{galetti1988}%
  \BibitemOpen
  \bibfield  {author} {\bibinfo {author} {\bibfnamefont {D.}~\bibnamefont
  {Galetti}}\ and\ \bibinfo {author} {\bibfnamefont {A.}~\bibnamefont {{de
  Toledo Piza}}},\ }\href
  {https://doi.org/https://doi.org/10.1016/0378-4371(88)90219-1} {\bibfield
  {journal} {\bibinfo  {journal} {Phys. A: Stat. Mech. Appl.}\ }\textbf
  {\bibinfo {volume} {149}},\ \bibinfo {pages} {267} (\bibinfo {year}
  {1988})}\BibitemShut {NoStop}%
\bibitem [{\citenamefont {Zhu}(2016)}]{Zhu_permutation_PRL_2016}%
  \BibitemOpen
  \bibfield  {author} {\bibinfo {author} {\bibfnamefont {H.}~\bibnamefont
  {Zhu}},\ }\href {https://doi.org/10.1103/PhysRevLett.116.040501} {\bibfield
  {journal} {\bibinfo  {journal} {Phys. Rev. Lett.}\ }\textbf {\bibinfo
  {volume} {116}},\ \bibinfo {pages} {040501} (\bibinfo {year}
  {2016})}\BibitemShut {NoStop}%
\bibitem [{\citenamefont {Dias}\ and\ \citenamefont
  {Prata}(2019)}]{Dias_Prata_2019}%
  \BibitemOpen
  \bibfield  {author} {\bibinfo {author} {\bibfnamefont {N.~C.}\ \bibnamefont
  {Dias}}\ and\ \bibinfo {author} {\bibfnamefont {J.~N.}\ \bibnamefont
  {Prata}},\ }\href {https://doi.org/10.4171/rmi/1056} {\bibfield  {journal}
  {\bibinfo  {journal} {Rev. Mat. Iberoam.}\ }\textbf {\bibinfo {volume}
  {35}},\ \bibinfo {pages} {317–337} (\bibinfo {year} {2019})}\BibitemShut
  {NoStop}%
\bibitem [{\citenamefont {Schmid}\ \emph {et~al.}(2022)\citenamefont {Schmid},
  \citenamefont {Du}, \citenamefont {Selby},\ and\ \citenamefont
  {Pusey}}]{Schmid2022}%
  \BibitemOpen
  \bibfield  {author} {\bibinfo {author} {\bibfnamefont {D.}~\bibnamefont
  {Schmid}}, \bibinfo {author} {\bibfnamefont {H.}~\bibnamefont {Du}}, \bibinfo
  {author} {\bibfnamefont {J.~H.}\ \bibnamefont {Selby}},\ and\ \bibinfo
  {author} {\bibfnamefont {M.~F.}\ \bibnamefont {Pusey}},\ }\href
  {https://doi.org/10.1103/PhysRevLett.129.120403} {\bibfield  {journal}
  {\bibinfo  {journal} {Phys. Rev. Lett.}\ }\textbf {\bibinfo {volume} {129}},\
  \bibinfo {pages} {120403} (\bibinfo {year} {2022})}\BibitemShut {NoStop}%
\bibitem [{\citenamefont {Howard}\ \emph {et~al.}(2014)\citenamefont {Howard},
  \citenamefont {Wallman}, \citenamefont {Veitch},\ and\ \citenamefont
  {Emerson}}]{howard2014contextuality}%
  \BibitemOpen
  \bibfield  {author} {\bibinfo {author} {\bibfnamefont {M.}~\bibnamefont
  {Howard}}, \bibinfo {author} {\bibfnamefont {J.}~\bibnamefont {Wallman}},
  \bibinfo {author} {\bibfnamefont {V.}~\bibnamefont {Veitch}},\ and\ \bibinfo
  {author} {\bibfnamefont {J.}~\bibnamefont {Emerson}},\ }\href
  {https://doi.org/10.1038/nature13460} {\bibfield  {journal} {\bibinfo
  {journal} {Nature}\ }\textbf {\bibinfo {volume} {510}},\ \bibinfo {pages}
  {351} (\bibinfo {year} {2014})}\BibitemShut {NoStop}%
\bibitem [{\citenamefont {Delfosse}\ \emph {et~al.}(2017)\citenamefont
  {Delfosse}, \citenamefont {Okay}, \citenamefont {Bermejo-Vega}, \citenamefont
  {Browne},\ and\ \citenamefont {Raussendorf}}]{delfosse2017equivalence}%
  \BibitemOpen
  \bibfield  {author} {\bibinfo {author} {\bibfnamefont {N.}~\bibnamefont
  {Delfosse}}, \bibinfo {author} {\bibfnamefont {C.}~\bibnamefont {Okay}},
  \bibinfo {author} {\bibfnamefont {J.}~\bibnamefont {Bermejo-Vega}}, \bibinfo
  {author} {\bibfnamefont {D.~E.}\ \bibnamefont {Browne}},\ and\ \bibinfo
  {author} {\bibfnamefont {R.}~\bibnamefont {Raussendorf}},\ }\href
  {https://doi.org/10.1088/1367-2630/aa8fe3} {\bibfield  {journal} {\bibinfo
  {journal} {New J. Phys.}\ }\textbf {\bibinfo {volume} {19}},\ \bibinfo
  {pages} {123024} (\bibinfo {year} {2017})}\BibitemShut {NoStop}%
\bibitem [{\citenamefont {Haferkamp}\ and\ \citenamefont
  {Bermejo-Vega}(2021)}]{haferkamp2021equivalence}%
  \BibitemOpen
  \bibfield  {author} {\bibinfo {author} {\bibfnamefont {J.}~\bibnamefont
  {Haferkamp}}\ and\ \bibinfo {author} {\bibfnamefont {J.}~\bibnamefont
  {Bermejo-Vega}},\ }\href@noop {} {} (\bibinfo {year} {2021}),\ \Eprint
  {https://arxiv.org/abs/2112.14788} {arXiv:2112.14788 [quant-ph]} \BibitemShut
  {NoStop}%
\bibitem [{\citenamefont {Booth}\ \emph {et~al.}(2022)\citenamefont {Booth},
  \citenamefont {Chabaud},\ and\ \citenamefont {Emeriau}}]{Booth2021}%
  \BibitemOpen
  \bibfield  {author} {\bibinfo {author} {\bibfnamefont {R.~I.}\ \bibnamefont
  {Booth}}, \bibinfo {author} {\bibfnamefont {U.}~\bibnamefont {Chabaud}},\
  and\ \bibinfo {author} {\bibfnamefont {P.-E.}\ \bibnamefont {Emeriau}},\
  }\href {https://doi.org/10.1103/PhysRevLett.129.230401} {\bibfield  {journal}
  {\bibinfo  {journal} {Phys. Rev. Lett.}\ }\textbf {\bibinfo {volume} {129}},\
  \bibinfo {pages} {230401} (\bibinfo {year} {2022})}\BibitemShut {NoStop}%
\bibitem [{\citenamefont {Hudson}(1974)}]{hudson1974wigner}%
  \BibitemOpen
  \bibfield  {author} {\bibinfo {author} {\bibfnamefont {R.~L.}\ \bibnamefont
  {Hudson}},\ }\href {https://doi.org/10.1016/0034-4877(74)90007-X} {\bibfield
  {journal} {\bibinfo  {journal} {Rep. Math. Phys.}\ }\textbf {\bibinfo
  {volume} {6}},\ \bibinfo {pages} {249} (\bibinfo {year} {1974})}\BibitemShut
  {NoStop}%
\bibitem [{\citenamefont {Soto}\ and\ \citenamefont
  {Claverie}(1983)}]{soto1983wigner}%
  \BibitemOpen
  \bibfield  {author} {\bibinfo {author} {\bibfnamefont {F.}~\bibnamefont
  {Soto}}\ and\ \bibinfo {author} {\bibfnamefont {P.}~\bibnamefont
  {Claverie}},\ }\href {https://doi.org/10.1063/1.525607} {\bibfield  {journal}
  {\bibinfo  {journal} {J. Math. Phys.}\ }\textbf {\bibinfo {volume} {24}},\
  \bibinfo {pages} {97} (\bibinfo {year} {1983})}\BibitemShut {NoStop}%
\bibitem [{\citenamefont {Gross}(2006)}]{gross2006hudson}%
  \BibitemOpen
  \bibfield  {author} {\bibinfo {author} {\bibfnamefont {D.}~\bibnamefont
  {Gross}},\ }\href {https://doi.org/10.1063/1.2393152} {\bibfield  {journal}
  {\bibinfo  {journal} {J. Math. Phys.}\ }\textbf {\bibinfo {volume} {47}},\
  \bibinfo {pages} {122107} (\bibinfo {year} {2006})}\BibitemShut {NoStop}%
\bibitem [{\citenamefont {Bravyi}\ and\ \citenamefont
  {Kitaev}(2005)}]{bravyi2005universal}%
  \BibitemOpen
  \bibfield  {author} {\bibinfo {author} {\bibfnamefont {S.}~\bibnamefont
  {Bravyi}}\ and\ \bibinfo {author} {\bibfnamefont {A.}~\bibnamefont
  {Kitaev}},\ }\href {https://doi.org/10.1103/PhysRevA.71.022316} {\bibfield
  {journal} {\bibinfo  {journal} {Phys. Rev. A}\ }\textbf {\bibinfo {volume}
  {71}},\ \bibinfo {pages} {022316} (\bibinfo {year} {2005})}\BibitemShut
  {NoStop}%
\bibitem [{\citenamefont {Veitch}\ \emph {et~al.}(2012)\citenamefont {Veitch},
  \citenamefont {Ferrie}, \citenamefont {Gross},\ and\ \citenamefont
  {Emerson}}]{veitch2012negative}%
  \BibitemOpen
  \bibfield  {author} {\bibinfo {author} {\bibfnamefont {V.}~\bibnamefont
  {Veitch}}, \bibinfo {author} {\bibfnamefont {C.}~\bibnamefont {Ferrie}},
  \bibinfo {author} {\bibfnamefont {D.}~\bibnamefont {Gross}},\ and\ \bibinfo
  {author} {\bibfnamefont {J.}~\bibnamefont {Emerson}},\ }\href
  {https://doi.org/10.1088/1367-2630/14/11/113011} {\bibfield  {journal}
  {\bibinfo  {journal} {New J. Phys.}\ }\textbf {\bibinfo {volume} {14}},\
  \bibinfo {pages} {113011} (\bibinfo {year} {2012})}\BibitemShut {NoStop}%
\bibitem [{\citenamefont {Veitch}\ \emph {et~al.}(2014)\citenamefont {Veitch},
  \citenamefont {Mousavian}, \citenamefont {Gottesman},\ and\ \citenamefont
  {Emerson}}]{veitch2014resource}%
  \BibitemOpen
  \bibfield  {author} {\bibinfo {author} {\bibfnamefont {V.}~\bibnamefont
  {Veitch}}, \bibinfo {author} {\bibfnamefont {S.~H.}\ \bibnamefont
  {Mousavian}}, \bibinfo {author} {\bibfnamefont {D.}~\bibnamefont
  {Gottesman}},\ and\ \bibinfo {author} {\bibfnamefont {J.}~\bibnamefont
  {Emerson}},\ }\href {https://doi.org/10.1088/1367-2630/16/1/013009}
  {\bibfield  {journal} {\bibinfo  {journal} {New J. Phys.}\ }\textbf {\bibinfo
  {volume} {16}},\ \bibinfo {pages} {013009} (\bibinfo {year}
  {2014})}\BibitemShut {NoStop}%
\bibitem [{\citenamefont {Gottesman}(1998)}]{gottesman1998heisenberg}%
  \BibitemOpen
  \bibfield  {author} {\bibinfo {author} {\bibfnamefont {D.}~\bibnamefont
  {Gottesman}},\ }\href {https://arxiv.org/abs/quant-ph/9807006} {} (\bibinfo
  {year} {1998}),\ \Eprint {https://arxiv.org/abs/quant-ph/9807006}
  {arXiv:quant-ph/9807006 [quant-ph]} \BibitemShut {NoStop}%
\bibitem [{\citenamefont {Bartlett}\ \emph {et~al.}(2002)\citenamefont
  {Bartlett}, \citenamefont {Sanders}, \citenamefont {Braunstein},\ and\
  \citenamefont {Nemoto}}]{Bartlett2002}%
  \BibitemOpen
  \bibfield  {author} {\bibinfo {author} {\bibfnamefont {S.~D.}\ \bibnamefont
  {Bartlett}}, \bibinfo {author} {\bibfnamefont {B.~C.}\ \bibnamefont
  {Sanders}}, \bibinfo {author} {\bibfnamefont {S.~L.}\ \bibnamefont
  {Braunstein}},\ and\ \bibinfo {author} {\bibfnamefont {K.}~\bibnamefont
  {Nemoto}},\ }\href {https://doi.org/10.1103/PhysRevLett.88.097904} {\bibfield
   {journal} {\bibinfo  {journal} {Phys. Rev. Lett.}\ }\textbf {\bibinfo
  {volume} {88}},\ \bibinfo {pages} {097904} (\bibinfo {year}
  {2002})}\BibitemShut {NoStop}%
\bibitem [{\citenamefont {Bartlett}\ and\ \citenamefont
  {Sanders}(2002)}]{Bartlett2002povm}%
  \BibitemOpen
  \bibfield  {author} {\bibinfo {author} {\bibfnamefont {S.~D.}\ \bibnamefont
  {Bartlett}}\ and\ \bibinfo {author} {\bibfnamefont {B.~C.}\ \bibnamefont
  {Sanders}},\ }\href {https://doi.org/10.1103/PhysRevLett.89.207903}
  {\bibfield  {journal} {\bibinfo  {journal} {Phys. Rev. Lett.}\ }\textbf
  {\bibinfo {volume} {89}},\ \bibinfo {pages} {207903} (\bibinfo {year}
  {2002})}\BibitemShut {NoStop}%
\bibitem [{\citenamefont {Cormick}\ \emph {et~al.}(2006)\citenamefont
  {Cormick}, \citenamefont {Galv\~ao}, \citenamefont {Gottesman}, \citenamefont
  {Paz},\ and\ \citenamefont {Pittenger}}]{Cormick2006}%
  \BibitemOpen
  \bibfield  {author} {\bibinfo {author} {\bibfnamefont {C.}~\bibnamefont
  {Cormick}}, \bibinfo {author} {\bibfnamefont {E.~F.}\ \bibnamefont
  {Galv\~ao}}, \bibinfo {author} {\bibfnamefont {D.}~\bibnamefont {Gottesman}},
  \bibinfo {author} {\bibfnamefont {J.~P.}\ \bibnamefont {Paz}},\ and\ \bibinfo
  {author} {\bibfnamefont {A.~O.}\ \bibnamefont {Pittenger}},\ }\href
  {https://doi.org/10.1103/PhysRevA.73.012301} {\bibfield  {journal} {\bibinfo
  {journal} {Phys. Rev. A}\ }\textbf {\bibinfo {volume} {73}},\ \bibinfo
  {pages} {012301} (\bibinfo {year} {2006})}\BibitemShut {NoStop}%
\bibitem [{\citenamefont {Mari}\ and\ \citenamefont {Eisert}(2012)}]{Mari2012}%
  \BibitemOpen
  \bibfield  {author} {\bibinfo {author} {\bibfnamefont {A.}~\bibnamefont
  {Mari}}\ and\ \bibinfo {author} {\bibfnamefont {J.}~\bibnamefont {Eisert}},\
  }\href {https://doi.org/10.1103/PhysRevLett.109.230503} {\bibfield  {journal}
  {\bibinfo  {journal} {Phys. Rev. Lett.}\ }\textbf {\bibinfo {volume} {109}},\
  \bibinfo {pages} {230503} (\bibinfo {year} {2012})}\BibitemShut {NoStop}%
\bibitem [{\citenamefont {Veitch}\ \emph {et~al.}(2013)\citenamefont {Veitch},
  \citenamefont {Wiebe}, \citenamefont {Ferrie},\ and\ \citenamefont
  {Emerson}}]{veitch2013cv}%
  \BibitemOpen
  \bibfield  {author} {\bibinfo {author} {\bibfnamefont {V.}~\bibnamefont
  {Veitch}}, \bibinfo {author} {\bibfnamefont {N.}~\bibnamefont {Wiebe}},
  \bibinfo {author} {\bibfnamefont {C.}~\bibnamefont {Ferrie}},\ and\ \bibinfo
  {author} {\bibfnamefont {J.}~\bibnamefont {Emerson}},\ }\href
  {https://doi.org/10.1088/1367-2630/15/1/013037} {\bibfield  {journal}
  {\bibinfo  {journal} {New J. Phys.}\ }\textbf {\bibinfo {volume} {15}},\
  \bibinfo {pages} {013037} (\bibinfo {year} {2013})}\BibitemShut {NoStop}%
\bibitem [{\citenamefont {Pashayan}\ \emph {et~al.}(2015)\citenamefont
  {Pashayan}, \citenamefont {Wallman},\ and\ \citenamefont
  {Bartlett}}]{pashayan2015estimating}%
  \BibitemOpen
  \bibfield  {author} {\bibinfo {author} {\bibfnamefont {H.}~\bibnamefont
  {Pashayan}}, \bibinfo {author} {\bibfnamefont {J.~J.}\ \bibnamefont
  {Wallman}},\ and\ \bibinfo {author} {\bibfnamefont {S.~D.}\ \bibnamefont
  {Bartlett}},\ }\href {https://doi.org/10.1103/PhysRevLett.115.070501}
  {\bibfield  {journal} {\bibinfo  {journal} {Phys. Rev. Lett.}\ }\textbf
  {\bibinfo {volume} {115}},\ \bibinfo {pages} {070501} (\bibinfo {year}
  {2015})}\BibitemShut {NoStop}%
\bibitem [{\citenamefont {Rahimi-Keshari}\ \emph {et~al.}(2016)\citenamefont
  {Rahimi-Keshari}, \citenamefont {Ralph},\ and\ \citenamefont
  {Caves}}]{Rahimi2016sufficient}%
  \BibitemOpen
  \bibfield  {author} {\bibinfo {author} {\bibfnamefont {S.}~\bibnamefont
  {Rahimi-Keshari}}, \bibinfo {author} {\bibfnamefont {T.~C.}\ \bibnamefont
  {Ralph}},\ and\ \bibinfo {author} {\bibfnamefont {C.~M.}\ \bibnamefont
  {Caves}},\ }\href {https://doi.org/10.1103/PhysRevX.6.021039} {\bibfield
  {journal} {\bibinfo  {journal} {Phys. Rev. X}\ }\textbf {\bibinfo {volume}
  {6}},\ \bibinfo {pages} {021039} (\bibinfo {year} {2016})}\BibitemShut
  {NoStop}%
\bibitem [{\citenamefont {Kocia}\ \emph {et~al.}(2017)\citenamefont {Kocia},
  \citenamefont {Huang},\ and\ \citenamefont {Love}}]{Kocia2017}%
  \BibitemOpen
  \bibfield  {author} {\bibinfo {author} {\bibfnamefont {L.}~\bibnamefont
  {Kocia}}, \bibinfo {author} {\bibfnamefont {Y.}~\bibnamefont {Huang}},\ and\
  \bibinfo {author} {\bibfnamefont {P.}~\bibnamefont {Love}},\ }\href
  {https://www.mdpi.com/1099-4300/19/7/353} {\bibfield  {journal} {\bibinfo
  {journal} {Entropy}\ }\textbf {\bibinfo {volume} {19}} (\bibinfo {year}
  {2017})}\BibitemShut {NoStop}%
\bibitem [{\citenamefont {Takagi}\ and\ \citenamefont
  {Zhuang}(2018)}]{takagi2018convex}%
  \BibitemOpen
  \bibfield  {author} {\bibinfo {author} {\bibfnamefont {R.}~\bibnamefont
  {Takagi}}\ and\ \bibinfo {author} {\bibfnamefont {Q.}~\bibnamefont
  {Zhuang}},\ }\href {https://doi.org/10.1103/PhysRevA.97.062337} {\bibfield
  {journal} {\bibinfo  {journal} {Phys. Rev. A}\ }\textbf {\bibinfo {volume}
  {97}},\ \bibinfo {pages} {062337} (\bibinfo {year} {2018})}\BibitemShut
  {NoStop}%
\bibitem [{\citenamefont {Albarelli}\ \emph {et~al.}(2018)\citenamefont
  {Albarelli}, \citenamefont {Genoni}, \citenamefont {Paris},\ and\
  \citenamefont {Ferraro}}]{albarelli2018resource}%
  \BibitemOpen
  \bibfield  {author} {\bibinfo {author} {\bibfnamefont {F.}~\bibnamefont
  {Albarelli}}, \bibinfo {author} {\bibfnamefont {M.~G.}\ \bibnamefont
  {Genoni}}, \bibinfo {author} {\bibfnamefont {M.~G.}\ \bibnamefont {Paris}},\
  and\ \bibinfo {author} {\bibfnamefont {A.}~\bibnamefont {Ferraro}},\ }\href
  {https://doi.org/10.1103/PhysRevA.98.052350} {\bibfield  {journal} {\bibinfo
  {journal} {Phys. Rev. A}\ }\textbf {\bibinfo {volume} {98}},\ \bibinfo
  {pages} {052350} (\bibinfo {year} {2018})}\BibitemShut {NoStop}%
\bibitem [{\citenamefont {Gottesman}(1997)}]{gottesman1997stabilizer}%
  \BibitemOpen
  \bibfield  {author} {\bibinfo {author} {\bibfnamefont {D.}~\bibnamefont
  {Gottesman}},\ }\href@noop {} {\emph {\bibinfo {title} {Stabilizer codes and
  quantum error correction}}}\ (\bibinfo  {publisher} {California Institute of
  Technology},\ \bibinfo {year} {1997})\BibitemShut {NoStop}%
\bibitem [{\citenamefont {Lidar}\ and\ \citenamefont
  {Brun}(2013)}]{lidar2013quantum}%
  \BibitemOpen
  \bibfield  {author} {\bibinfo {author} {\bibfnamefont {D.~A.}\ \bibnamefont
  {Lidar}}\ and\ \bibinfo {author} {\bibfnamefont {T.~A.}\ \bibnamefont
  {Brun}},\ }\href {https://doi.org/https://doi.org/10.1017/CBO9781139034807}
  {\emph {\bibinfo {title} {Quantum error correction}}}\ (\bibinfo  {publisher}
  {Cambridge university press},\ \bibinfo {year} {2013})\BibitemShut {NoStop}%
\bibitem [{\citenamefont {Roffe}(2019)}]{roffe2019quantum}%
  \BibitemOpen
  \bibfield  {author} {\bibinfo {author} {\bibfnamefont {J.}~\bibnamefont
  {Roffe}},\ }\href
  {https://doi.org/https://doi.org/10.1080/00107514.2019.1667078} {\bibfield
  {journal} {\bibinfo  {journal} {Contemporary Physics}\ }\textbf {\bibinfo
  {volume} {60}},\ \bibinfo {pages} {226} (\bibinfo {year} {2019})}\BibitemShut
  {NoStop}%
\bibitem [{\citenamefont {Albert}(2022)}]{albert2022bosonic}%
  \BibitemOpen
  \bibfield  {author} {\bibinfo {author} {\bibfnamefont {V.~V.}\ \bibnamefont
  {Albert}},\ }\href {https://arxiv.org/abs/2211.05714} {} (\bibinfo {year}
  {2022}),\ \Eprint {https://arxiv.org/abs/2211.05714} {arXiv:2211.05714
  [quant-ph]} \BibitemShut {NoStop}%
\bibitem [{\citenamefont {Gottesman}\ \emph {et~al.}(2001)\citenamefont
  {Gottesman}, \citenamefont {Kitaev},\ and\ \citenamefont
  {Preskill}}]{Gottesman2001}%
  \BibitemOpen
  \bibfield  {author} {\bibinfo {author} {\bibfnamefont {D.}~\bibnamefont
  {Gottesman}}, \bibinfo {author} {\bibfnamefont {A.}~\bibnamefont {Kitaev}},\
  and\ \bibinfo {author} {\bibfnamefont {J.}~\bibnamefont {Preskill}},\ }\href
  {https://doi.org/10.1103/PhysRevA.64.012310} {\bibfield  {journal} {\bibinfo
  {journal} {Phys. Rev. A}\ }\textbf {\bibinfo {volume} {64}},\ \bibinfo
  {pages} {012310} (\bibinfo {year} {2001})}\BibitemShut {NoStop}%
\bibitem [{\citenamefont {Bourassa}\ \emph {et~al.}(2021)\citenamefont
  {Bourassa}, \citenamefont {Alexander}, \citenamefont {Vasmer}, \citenamefont
  {Patil}, \citenamefont {Tzitrin}, \citenamefont {Matsuura}, \citenamefont
  {Su}, \citenamefont {Baragiola}, \citenamefont {Guha}, \citenamefont
  {Dauphinais}, \citenamefont {Sabapathy}, \citenamefont {Menicucci},\ and\
  \citenamefont {Dhand}}]{bourassa_blueprint_2021}%
  \BibitemOpen
  \bibfield  {author} {\bibinfo {author} {\bibfnamefont {J.~E.}\ \bibnamefont
  {Bourassa}}, \bibinfo {author} {\bibfnamefont {R.~N.}\ \bibnamefont
  {Alexander}}, \bibinfo {author} {\bibfnamefont {M.}~\bibnamefont {Vasmer}},
  \bibinfo {author} {\bibfnamefont {A.}~\bibnamefont {Patil}}, \bibinfo
  {author} {\bibfnamefont {I.}~\bibnamefont {Tzitrin}}, \bibinfo {author}
  {\bibfnamefont {T.}~\bibnamefont {Matsuura}}, \bibinfo {author}
  {\bibfnamefont {D.}~\bibnamefont {Su}}, \bibinfo {author} {\bibfnamefont
  {B.~Q.}\ \bibnamefont {Baragiola}}, \bibinfo {author} {\bibfnamefont
  {S.}~\bibnamefont {Guha}}, \bibinfo {author} {\bibfnamefont {G.}~\bibnamefont
  {Dauphinais}}, \bibinfo {author} {\bibfnamefont {K.~K.}\ \bibnamefont
  {Sabapathy}}, \bibinfo {author} {\bibfnamefont {N.~C.}\ \bibnamefont
  {Menicucci}},\ and\ \bibinfo {author} {\bibfnamefont {I.}~\bibnamefont
  {Dhand}},\ }\href {https://doi.org/10.22331/q-2021-02-04-392} {\bibfield
  {journal} {\bibinfo  {journal} {{Quantum}}\ }\textbf {\bibinfo {volume}
  {5}},\ \bibinfo {pages} {392} (\bibinfo {year} {2021})}\BibitemShut {NoStop}%
\bibitem [{\citenamefont {Sivak}\ \emph {et~al.}(2023)\citenamefont {Sivak},
  \citenamefont {Eickbusch}, \citenamefont {Royer}, \citenamefont {Singh},
  \citenamefont {Tsioutsios}, \citenamefont {Ganjam}, \citenamefont {Miano},
  \citenamefont {Brock}, \citenamefont {Ding}, \citenamefont {Frunzio},
  \citenamefont {Girvin}, \citenamefont {Schoelkopf},\ and\ \citenamefont
  {Devoret}}]{sivak2023real}%
  \BibitemOpen
  \bibfield  {author} {\bibinfo {author} {\bibfnamefont {V.~V.}\ \bibnamefont
  {Sivak}}, \bibinfo {author} {\bibfnamefont {A.}~\bibnamefont {Eickbusch}},
  \bibinfo {author} {\bibfnamefont {B.}~\bibnamefont {Royer}}, \bibinfo
  {author} {\bibfnamefont {S.}~\bibnamefont {Singh}}, \bibinfo {author}
  {\bibfnamefont {I.}~\bibnamefont {Tsioutsios}}, \bibinfo {author}
  {\bibfnamefont {S.}~\bibnamefont {Ganjam}}, \bibinfo {author} {\bibfnamefont
  {A.}~\bibnamefont {Miano}}, \bibinfo {author} {\bibfnamefont {B.~L.}\
  \bibnamefont {Brock}}, \bibinfo {author} {\bibfnamefont {A.~Z.}\ \bibnamefont
  {Ding}}, \bibinfo {author} {\bibfnamefont {L.}~\bibnamefont {Frunzio}},
  \bibinfo {author} {\bibfnamefont {S.~M.}\ \bibnamefont {Girvin}}, \bibinfo
  {author} {\bibfnamefont {R.~J.}\ \bibnamefont {Schoelkopf}},\ and\ \bibinfo
  {author} {\bibfnamefont {M.~H.}\ \bibnamefont {Devoret}},\ }\href
  {https://doi.org/10.1038/s41586-023-05782-6} {\bibfield  {journal} {\bibinfo
  {journal} {Nature}\ }\textbf {\bibinfo {volume} {616}},\ \bibinfo {pages}
  {50–55} (\bibinfo {year} {2023})}\BibitemShut {NoStop}%
\bibitem [{\citenamefont {Conrad}\ \emph {et~al.}(2022)\citenamefont {Conrad},
  \citenamefont {Eisert},\ and\ \citenamefont {Arzani}}]{conrad2022gottesman}%
  \BibitemOpen
  \bibfield  {author} {\bibinfo {author} {\bibfnamefont {J.}~\bibnamefont
  {Conrad}}, \bibinfo {author} {\bibfnamefont {J.}~\bibnamefont {Eisert}},\
  and\ \bibinfo {author} {\bibfnamefont {F.}~\bibnamefont {Arzani}},\ }\href
  {https://doi.org/10.22331/q-2022-02-10-648} {\bibfield  {journal} {\bibinfo
  {journal} {{Quantum}}\ }\textbf {\bibinfo {volume} {6}},\ \bibinfo {pages}
  {648} (\bibinfo {year} {2022})}\BibitemShut {NoStop}%
\bibitem [{\citenamefont {Terhal}\ and\ \citenamefont
  {Weigand}(2016{\natexlab{a}})}]{PhysRevA.93.012315}%
  \BibitemOpen
  \bibfield  {author} {\bibinfo {author} {\bibfnamefont {B.~M.}\ \bibnamefont
  {Terhal}}\ and\ \bibinfo {author} {\bibfnamefont {D.}~\bibnamefont
  {Weigand}},\ }\href {https://doi.org/10.1103/PhysRevA.93.012315} {\bibfield
  {journal} {\bibinfo  {journal} {Phys. Rev. A}\ }\textbf {\bibinfo {volume}
  {93}},\ \bibinfo {pages} {012315} (\bibinfo {year}
  {2016}{\natexlab{a}})}\BibitemShut {NoStop}%
\bibitem [{\citenamefont {Fabre}\ and\ \citenamefont
  {Felicetti}(2021)}]{PhysRevA.104.022208}%
  \BibitemOpen
  \bibfield  {author} {\bibinfo {author} {\bibfnamefont {N.}~\bibnamefont
  {Fabre}}\ and\ \bibinfo {author} {\bibfnamefont {S.}~\bibnamefont
  {Felicetti}},\ }\href {https://doi.org/10.1103/PhysRevA.104.022208}
  {\bibfield  {journal} {\bibinfo  {journal} {Phys. Rev. A}\ }\textbf {\bibinfo
  {volume} {104}},\ \bibinfo {pages} {022208} (\bibinfo {year}
  {2021})}\BibitemShut {NoStop}%
\bibitem [{\citenamefont {Rozpedek}\ \emph {et~al.}(2023)\citenamefont
  {Rozpedek}, \citenamefont {Seshadreesan}, \citenamefont {Polakos},
  \citenamefont {Jiang},\ and\ \citenamefont
  {Guha}}]{rozpedek_all-photonic_2023}%
  \BibitemOpen
  \bibfield  {author} {\bibinfo {author} {\bibfnamefont {F.}~\bibnamefont
  {Rozpedek}}, \bibinfo {author} {\bibfnamefont {K.~P.}\ \bibnamefont
  {Seshadreesan}}, \bibinfo {author} {\bibfnamefont {P.}~\bibnamefont
  {Polakos}}, \bibinfo {author} {\bibfnamefont {L.}~\bibnamefont {Jiang}},\
  and\ \bibinfo {author} {\bibfnamefont {S.}~\bibnamefont {Guha}},\ }\href
  {https://doi.org/10.1103/PhysRevResearch.5.043056} {\bibfield  {journal}
  {\bibinfo  {journal} {Phys. Rev. Research}\ }\textbf {\bibinfo {volume}
  {5}},\ \bibinfo {pages} {043056} (\bibinfo {year} {2023})}\BibitemShut
  {NoStop}%
\bibitem [{\citenamefont {Vuillot}\ \emph {et~al.}(2019)\citenamefont
  {Vuillot}, \citenamefont {Asasi}, \citenamefont {Wang}, \citenamefont
  {Pryadko},\ and\ \citenamefont {Terhal}}]{PhysRevA.99.032344}%
  \BibitemOpen
  \bibfield  {author} {\bibinfo {author} {\bibfnamefont {C.}~\bibnamefont
  {Vuillot}}, \bibinfo {author} {\bibfnamefont {H.}~\bibnamefont {Asasi}},
  \bibinfo {author} {\bibfnamefont {Y.}~\bibnamefont {Wang}}, \bibinfo {author}
  {\bibfnamefont {L.~P.}\ \bibnamefont {Pryadko}},\ and\ \bibinfo {author}
  {\bibfnamefont {B.~M.}\ \bibnamefont {Terhal}},\ }\href
  {https://doi.org/10.1103/PhysRevA.99.032344} {\bibfield  {journal} {\bibinfo
  {journal} {Phys. Rev. A}\ }\textbf {\bibinfo {volume} {99}},\ \bibinfo
  {pages} {032344} (\bibinfo {year} {2019})}\BibitemShut {NoStop}%
\bibitem [{\citenamefont {Flühmann}\ \emph {et~al.}(2019)\citenamefont
  {Flühmann}, \citenamefont {Nguyen}, \citenamefont {Marinelli}, \citenamefont
  {Negnevitsky}, \citenamefont {Mehta},\ and\ \citenamefont
  {Home}}]{fluhmann_encoding_2019}%
  \BibitemOpen
  \bibfield  {author} {\bibinfo {author} {\bibfnamefont {C.}~\bibnamefont
  {Flühmann}}, \bibinfo {author} {\bibfnamefont {T.~L.}\ \bibnamefont
  {Nguyen}}, \bibinfo {author} {\bibfnamefont {M.}~\bibnamefont {Marinelli}},
  \bibinfo {author} {\bibfnamefont {V.}~\bibnamefont {Negnevitsky}}, \bibinfo
  {author} {\bibfnamefont {K.}~\bibnamefont {Mehta}},\ and\ \bibinfo {author}
  {\bibfnamefont {J.~P.}\ \bibnamefont {Home}},\ }\href
  {https://doi.org/10.1038/s41586-019-0960-6} {\bibfield  {journal} {\bibinfo
  {journal} {Nature}\ }\textbf {\bibinfo {volume} {566}},\ \bibinfo {pages}
  {513} (\bibinfo {year} {2019})}\BibitemShut {NoStop}%
\bibitem [{\citenamefont {Campagne-Ibarcq}\ \emph {et~al.}(2020)\citenamefont
  {Campagne-Ibarcq}, \citenamefont {Eickbusch}, \citenamefont {Touzard},
  \citenamefont {Zalys-Geller}, \citenamefont {Frattini}, \citenamefont
  {Sivak}, \citenamefont {Reinhold}, \citenamefont {Puri}, \citenamefont
  {Shankar}, \citenamefont {Schoelkopf}, \citenamefont {Frunzio}, \citenamefont
  {Mirrahimi},\ and\ \citenamefont {Devoret}}]{campagne-ibarcq_quantum_2020}%
  \BibitemOpen
  \bibfield  {author} {\bibinfo {author} {\bibfnamefont {P.}~\bibnamefont
  {Campagne-Ibarcq}}, \bibinfo {author} {\bibfnamefont {A.}~\bibnamefont
  {Eickbusch}}, \bibinfo {author} {\bibfnamefont {S.}~\bibnamefont {Touzard}},
  \bibinfo {author} {\bibfnamefont {E.}~\bibnamefont {Zalys-Geller}}, \bibinfo
  {author} {\bibfnamefont {N.~E.}\ \bibnamefont {Frattini}}, \bibinfo {author}
  {\bibfnamefont {V.~V.}\ \bibnamefont {Sivak}}, \bibinfo {author}
  {\bibfnamefont {P.}~\bibnamefont {Reinhold}}, \bibinfo {author}
  {\bibfnamefont {S.}~\bibnamefont {Puri}}, \bibinfo {author} {\bibfnamefont
  {S.}~\bibnamefont {Shankar}}, \bibinfo {author} {\bibfnamefont {R.~J.}\
  \bibnamefont {Schoelkopf}}, \bibinfo {author} {\bibfnamefont
  {L.}~\bibnamefont {Frunzio}}, \bibinfo {author} {\bibfnamefont
  {M.}~\bibnamefont {Mirrahimi}},\ and\ \bibinfo {author} {\bibfnamefont
  {M.~H.}\ \bibnamefont {Devoret}},\ }\href
  {https://doi.org/10.1038/s41586-020-2603-3} {\bibfield  {journal} {\bibinfo
  {journal} {Nature}\ }\textbf {\bibinfo {volume} {584}},\ \bibinfo {pages}
  {368–372} (\bibinfo {year} {2020})}\BibitemShut {NoStop}%
\bibitem [{\citenamefont {Konno}\ \emph {et~al.}(2024)\citenamefont {Konno},
  \citenamefont {Asavanant}, \citenamefont {Hanamura}, \citenamefont
  {Nagayoshi}, \citenamefont {Fukui}, \citenamefont {Sakaguchi}, \citenamefont
  {Ide}, \citenamefont {China}, \citenamefont {Yabuno}, \citenamefont {Miki},
  \citenamefont {Terai}, \citenamefont {Takase}, \citenamefont {Endo},
  \citenamefont {Marek}, \citenamefont {Filip}, \citenamefont {van Loock},\
  and\ \citenamefont {Furusawa}}]{doi:10.1126/science.adk7560}%
  \BibitemOpen
  \bibfield  {author} {\bibinfo {author} {\bibfnamefont {S.}~\bibnamefont
  {Konno}}, \bibinfo {author} {\bibfnamefont {W.}~\bibnamefont {Asavanant}},
  \bibinfo {author} {\bibfnamefont {F.}~\bibnamefont {Hanamura}}, \bibinfo
  {author} {\bibfnamefont {H.}~\bibnamefont {Nagayoshi}}, \bibinfo {author}
  {\bibfnamefont {K.}~\bibnamefont {Fukui}}, \bibinfo {author} {\bibfnamefont
  {A.}~\bibnamefont {Sakaguchi}}, \bibinfo {author} {\bibfnamefont
  {R.}~\bibnamefont {Ide}}, \bibinfo {author} {\bibfnamefont {F.}~\bibnamefont
  {China}}, \bibinfo {author} {\bibfnamefont {M.}~\bibnamefont {Yabuno}},
  \bibinfo {author} {\bibfnamefont {S.}~\bibnamefont {Miki}}, \bibinfo {author}
  {\bibfnamefont {H.}~\bibnamefont {Terai}}, \bibinfo {author} {\bibfnamefont
  {K.}~\bibnamefont {Takase}}, \bibinfo {author} {\bibfnamefont
  {M.}~\bibnamefont {Endo}}, \bibinfo {author} {\bibfnamefont {P.}~\bibnamefont
  {Marek}}, \bibinfo {author} {\bibfnamefont {R.}~\bibnamefont {Filip}},
  \bibinfo {author} {\bibfnamefont {P.}~\bibnamefont {van Loock}},\ and\
  \bibinfo {author} {\bibfnamefont {A.}~\bibnamefont {Furusawa}},\ }\href
  {https://doi.org/10.1126/science.adk7560} {\bibfield  {journal} {\bibinfo
  {journal} {Science}\ }\textbf {\bibinfo {volume} {383}},\ \bibinfo {pages}
  {289} (\bibinfo {year} {2024})}\BibitemShut {NoStop}%
\bibitem [{\citenamefont {Fabre}\ \emph
  {et~al.}(2020{\natexlab{a}})\citenamefont {Fabre}, \citenamefont {Maltese},
  \citenamefont {Appas}, \citenamefont {Felicetti}, \citenamefont {Ketterer},
  \citenamefont {Keller}, \citenamefont {Coudreau}, \citenamefont {Baboux},
  \citenamefont {Amanti}, \citenamefont {Ducci},\ and\ \citenamefont
  {Milman}}]{fabre_generation_2020}%
  \BibitemOpen
  \bibfield  {author} {\bibinfo {author} {\bibfnamefont {N.}~\bibnamefont
  {Fabre}}, \bibinfo {author} {\bibfnamefont {G.}~\bibnamefont {Maltese}},
  \bibinfo {author} {\bibfnamefont {F.}~\bibnamefont {Appas}}, \bibinfo
  {author} {\bibfnamefont {S.}~\bibnamefont {Felicetti}}, \bibinfo {author}
  {\bibfnamefont {A.}~\bibnamefont {Ketterer}}, \bibinfo {author}
  {\bibfnamefont {A.}~\bibnamefont {Keller}}, \bibinfo {author} {\bibfnamefont
  {T.}~\bibnamefont {Coudreau}}, \bibinfo {author} {\bibfnamefont
  {F.}~\bibnamefont {Baboux}}, \bibinfo {author} {\bibfnamefont {M.~I.}\
  \bibnamefont {Amanti}}, \bibinfo {author} {\bibfnamefont {S.}~\bibnamefont
  {Ducci}},\ and\ \bibinfo {author} {\bibfnamefont {P.}~\bibnamefont
  {Milman}},\ }\href {https://doi.org/10.1103/PhysRevA.102.012607} {\bibfield
  {journal} {\bibinfo  {journal} {Phys. Rev. A}\ }\textbf {\bibinfo {volume}
  {102}},\ \bibinfo {pages} {012607} (\bibinfo {year}
  {2020}{\natexlab{a}})}\BibitemShut {NoStop}%
\bibitem [{\citenamefont {Dahan}\ \emph {et~al.}(2023)\citenamefont {Dahan},
  \citenamefont {Baranes}, \citenamefont {Gorlach}, \citenamefont {Ruimy},
  \citenamefont {Rivera},\ and\ \citenamefont {Kaminer}}]{PhysRevX.13.031001}%
  \BibitemOpen
  \bibfield  {author} {\bibinfo {author} {\bibfnamefont {R.}~\bibnamefont
  {Dahan}}, \bibinfo {author} {\bibfnamefont {G.}~\bibnamefont {Baranes}},
  \bibinfo {author} {\bibfnamefont {A.}~\bibnamefont {Gorlach}}, \bibinfo
  {author} {\bibfnamefont {R.}~\bibnamefont {Ruimy}}, \bibinfo {author}
  {\bibfnamefont {N.}~\bibnamefont {Rivera}},\ and\ \bibinfo {author}
  {\bibfnamefont {I.}~\bibnamefont {Kaminer}},\ }\href
  {https://doi.org/10.1103/PhysRevX.13.031001} {\bibfield  {journal} {\bibinfo
  {journal} {Phys. Rev. X}\ }\textbf {\bibinfo {volume} {13}},\ \bibinfo
  {pages} {031001} (\bibinfo {year} {2023})}\BibitemShut {NoStop}%
\bibitem [{\citenamefont {Garc\'{\i}a-\'Alvarez}\ \emph
  {et~al.}(2020)\citenamefont {Garc\'{\i}a-\'Alvarez}, \citenamefont
  {Calcluth}, \citenamefont {Ferraro},\ and\ \citenamefont
  {Ferrini}}]{garcia2020efficient}%
  \BibitemOpen
  \bibfield  {author} {\bibinfo {author} {\bibfnamefont {L.}~\bibnamefont
  {Garc\'{\i}a-\'Alvarez}}, \bibinfo {author} {\bibfnamefont {C.}~\bibnamefont
  {Calcluth}}, \bibinfo {author} {\bibfnamefont {A.}~\bibnamefont {Ferraro}},\
  and\ \bibinfo {author} {\bibfnamefont {G.}~\bibnamefont {Ferrini}},\ }\href
  {https://doi.org/10.1103/PhysRevResearch.2.043322} {\bibfield  {journal}
  {\bibinfo  {journal} {Phys. Rev. Res.}\ }\textbf {\bibinfo {volume} {2}},\
  \bibinfo {pages} {043322} (\bibinfo {year} {2020})}\BibitemShut {NoStop}%
\bibitem [{\citenamefont {Calcluth}\ \emph {et~al.}(2022)\citenamefont
  {Calcluth}, \citenamefont {Ferraro},\ and\ \citenamefont
  {Ferrini}}]{calcluth2022efficient}%
  \BibitemOpen
  \bibfield  {author} {\bibinfo {author} {\bibfnamefont {C.}~\bibnamefont
  {Calcluth}}, \bibinfo {author} {\bibfnamefont {A.}~\bibnamefont {Ferraro}},\
  and\ \bibinfo {author} {\bibfnamefont {G.}~\bibnamefont {Ferrini}},\ }\href
  {https://doi.org/10.22331/q-2022-12-01-867} {\bibfield  {journal} {\bibinfo
  {journal} {{Quantum}}\ }\textbf {\bibinfo {volume} {6}},\ \bibinfo {pages}
  {867} (\bibinfo {year} {2022})}\BibitemShut {NoStop}%
\bibitem [{\citenamefont {Baragiola}\ \emph {et~al.}(2019)\citenamefont
  {Baragiola}, \citenamefont {Pantaleoni}, \citenamefont {Alexander},
  \citenamefont {Karanjai},\ and\ \citenamefont
  {Menicucci}}]{Baragiola_2019_allGaussianUniversality}%
  \BibitemOpen
  \bibfield  {author} {\bibinfo {author} {\bibfnamefont {B.~Q.}\ \bibnamefont
  {Baragiola}}, \bibinfo {author} {\bibfnamefont {G.}~\bibnamefont
  {Pantaleoni}}, \bibinfo {author} {\bibfnamefont {R.~N.}\ \bibnamefont
  {Alexander}}, \bibinfo {author} {\bibfnamefont {A.}~\bibnamefont
  {Karanjai}},\ and\ \bibinfo {author} {\bibfnamefont {N.~C.}\ \bibnamefont
  {Menicucci}},\ }\href {https://doi.org/10.1103/PhysRevLett.123.200502}
  {\bibfield  {journal} {\bibinfo  {journal} {Phys. Rev. Lett.}\ }\textbf
  {\bibinfo {volume} {123}},\ \bibinfo {pages} {200502} (\bibinfo {year}
  {2019})}\BibitemShut {NoStop}%
\bibitem [{\citenamefont {Yamasaki}\ \emph {et~al.}(2020)\citenamefont
  {Yamasaki}, \citenamefont {Matsuura},\ and\ \citenamefont
  {Koashi}}]{Yamasaki_2020_costReducedGaussianUniversality}%
  \BibitemOpen
  \bibfield  {author} {\bibinfo {author} {\bibfnamefont {H.}~\bibnamefont
  {Yamasaki}}, \bibinfo {author} {\bibfnamefont {T.}~\bibnamefont {Matsuura}},\
  and\ \bibinfo {author} {\bibfnamefont {M.}~\bibnamefont {Koashi}},\ }\href
  {https://doi.org/10.1103/PhysRevResearch.2.023270} {\bibfield  {journal}
  {\bibinfo  {journal} {Phys. Rev. Res.}\ }\textbf {\bibinfo {volume} {2}},\
  \bibinfo {pages} {023270} (\bibinfo {year} {2020})}\BibitemShut {NoStop}%
\bibitem [{\citenamefont {Calcluth}\ \emph {et~al.}(2023)\citenamefont
  {Calcluth}, \citenamefont {Ferraro},\ and\ \citenamefont
  {Ferrini}}]{calcluth2023addingVacuum}%
  \BibitemOpen
  \bibfield  {author} {\bibinfo {author} {\bibfnamefont {C.}~\bibnamefont
  {Calcluth}}, \bibinfo {author} {\bibfnamefont {A.}~\bibnamefont {Ferraro}},\
  and\ \bibinfo {author} {\bibfnamefont {G.}~\bibnamefont {Ferrini}},\ }\href
  {https://doi.org/10.1103/PhysRevA.107.062414} {\bibfield  {journal} {\bibinfo
   {journal} {Phys. Rev. A}\ }\textbf {\bibinfo {volume} {107}},\ \bibinfo
  {pages} {062414} (\bibinfo {year} {2023})}\BibitemShut {NoStop}%
\bibitem [{\citenamefont {Hahn}\ \emph {et~al.}(2022)\citenamefont {Hahn},
  \citenamefont {Ferraro}, \citenamefont {Hultquist}, \citenamefont {Ferrini},\
  and\ \citenamefont {Garc\'{\i}a-\'Alvarez}}]{hahn2022quantifying}%
  \BibitemOpen
  \bibfield  {author} {\bibinfo {author} {\bibfnamefont {O.}~\bibnamefont
  {Hahn}}, \bibinfo {author} {\bibfnamefont {A.}~\bibnamefont {Ferraro}},
  \bibinfo {author} {\bibfnamefont {L.}~\bibnamefont {Hultquist}}, \bibinfo
  {author} {\bibfnamefont {G.}~\bibnamefont {Ferrini}},\ and\ \bibinfo {author}
  {\bibfnamefont {L.}~\bibnamefont {Garc\'{\i}a-\'Alvarez}},\ }\href
  {https://doi.org/10.1103/PhysRevLett.128.210502} {\bibfield  {journal}
  {\bibinfo  {journal} {Phys. Rev. Lett.}\ }\textbf {\bibinfo {volume} {128}},\
  \bibinfo {pages} {210502} (\bibinfo {year} {2022})}\BibitemShut {NoStop}%
\bibitem [{\citenamefont {Hahn}\ \emph {et~al.}(2024)\citenamefont {Hahn},
  \citenamefont {Ferrini},\ and\ \citenamefont
  {Takagi}}]{hahn2024bridgingMagic}%
  \BibitemOpen
  \bibfield  {author} {\bibinfo {author} {\bibfnamefont {O.}~\bibnamefont
  {Hahn}}, \bibinfo {author} {\bibfnamefont {G.}~\bibnamefont {Ferrini}},\ and\
  \bibinfo {author} {\bibfnamefont {R.}~\bibnamefont {Takagi}},\ }\href@noop {}
  {} (\bibinfo {year} {2024}),\ \Eprint {https://arxiv.org/abs/2406.06418}
  {arXiv:2406.06418 [quant-ph]} \BibitemShut {NoStop}%
\bibitem [{\citenamefont {Calcluth}\ \emph {et~al.}(2024)\citenamefont
  {Calcluth}, \citenamefont {Reichel}, \citenamefont {Ferraro},\ and\
  \citenamefont {Ferrini}}]{calcluth2024sufficient}%
  \BibitemOpen
  \bibfield  {author} {\bibinfo {author} {\bibfnamefont {C.}~\bibnamefont
  {Calcluth}}, \bibinfo {author} {\bibfnamefont {N.}~\bibnamefont {Reichel}},
  \bibinfo {author} {\bibfnamefont {A.}~\bibnamefont {Ferraro}},\ and\ \bibinfo
  {author} {\bibfnamefont {G.}~\bibnamefont {Ferrini}},\ }\href
  {https://doi.org/10.1103/PRXQuantum.5.020337} {\bibfield  {journal} {\bibinfo
   {journal} {PRX Quantum}\ }\textbf {\bibinfo {volume} {5}},\ \bibinfo {pages}
  {020337} (\bibinfo {year} {2024})}\BibitemShut {NoStop}%
\bibitem [{\citenamefont {Brif}\ and\ \citenamefont
  {Mann}(1999)}]{brif1999phase}%
  \BibitemOpen
  \bibfield  {author} {\bibinfo {author} {\bibfnamefont {C.}~\bibnamefont
  {Brif}}\ and\ \bibinfo {author} {\bibfnamefont {A.}~\bibnamefont {Mann}},\
  }\href {https://doi.org/10.1103/PhysRevA.59.971} {\bibfield  {journal}
  {\bibinfo  {journal} {Phys. Rev. A}\ }\textbf {\bibinfo {volume} {59}},\
  \bibinfo {pages} {971} (\bibinfo {year} {1999})}\BibitemShut {NoStop}%
\bibitem [{\citenamefont {Cahill}\ and\ \citenamefont
  {Glauber}(1969{\natexlab{a}})}]{cahill1969ordered}%
  \BibitemOpen
  \bibfield  {author} {\bibinfo {author} {\bibfnamefont {K.~E.}\ \bibnamefont
  {Cahill}}\ and\ \bibinfo {author} {\bibfnamefont {R.~J.}\ \bibnamefont
  {Glauber}},\ }\href {https://doi.org/10.1103/PhysRev.177.1857} {\bibfield
  {journal} {\bibinfo  {journal} {Phys. Rev.}\ }\textbf {\bibinfo {volume}
  {177}},\ \bibinfo {pages} {1857} (\bibinfo {year}
  {1969}{\natexlab{a}})}\BibitemShut {NoStop}%
\bibitem [{\citenamefont {Cahill}\ and\ \citenamefont
  {Glauber}(1969{\natexlab{b}})}]{cahill1969density}%
  \BibitemOpen
  \bibfield  {author} {\bibinfo {author} {\bibfnamefont {K.~E.}\ \bibnamefont
  {Cahill}}\ and\ \bibinfo {author} {\bibfnamefont {R.~J.}\ \bibnamefont
  {Glauber}},\ }\href {https://doi.org/10.1103/PhysRev.177.1882} {\bibfield
  {journal} {\bibinfo  {journal} {Phys. Rev.}\ }\textbf {\bibinfo {volume}
  {177}},\ \bibinfo {pages} {1882} (\bibinfo {year}
  {1969}{\natexlab{b}})}\BibitemShut {NoStop}%
\bibitem [{\citenamefont {Schroeck~Jr}(2013)}]{schroeck2013quantum}%
  \BibitemOpen
  \bibfield  {author} {\bibinfo {author} {\bibfnamefont {F.~E.}\ \bibnamefont
  {Schroeck~Jr}},\ }\href
  {https://doi.org/https://doi.org/10.1007/978-94-017-2830-0} {\emph {\bibinfo
  {title} {Quantum mechanics on phase space}}},\ Vol.~\bibinfo {volume} {74}\
  (\bibinfo  {publisher} {Springer Science \& Business Media},\ \bibinfo {year}
  {2013})\BibitemShut {NoStop}%
\bibitem [{\citenamefont {Perelomov}(1972)}]{perelomov1972coherent}%
  \BibitemOpen
  \bibfield  {author} {\bibinfo {author} {\bibfnamefont {A.~M.}\ \bibnamefont
  {Perelomov}},\ }\href {https://doi.org/10.1007/BF01645091} {\bibfield
  {journal} {\bibinfo  {journal} {Commun. Math. Phys.}\ }\textbf {\bibinfo
  {volume} {26}},\ \bibinfo {pages} {222–236} (\bibinfo {year}
  {1972})}\BibitemShut {NoStop}%
\bibitem [{\citenamefont {Vourdas}(2004)}]{Vourdas_2004}%
  \BibitemOpen
  \bibfield  {author} {\bibinfo {author} {\bibfnamefont {A.}~\bibnamefont
  {Vourdas}},\ }\href {https://doi.org/10.1088/0034-4885/67/3/R03} {\bibfield
  {journal} {\bibinfo  {journal} {Rep. Prog. Phys.}\ }\textbf {\bibinfo
  {volume} {67}},\ \bibinfo {pages} {267} (\bibinfo {year} {2004})}\BibitemShut
  {NoStop}%
\bibitem [{\citenamefont {Leaf}(1968)}]{Leaf1968}%
  \BibitemOpen
  \bibfield  {author} {\bibinfo {author} {\bibfnamefont {B.}~\bibnamefont
  {Leaf}},\ }\href {https://doi.org/10.1063/1.1664478} {\bibfield  {journal}
  {\bibinfo  {journal} {J. Math. Phys.}\ }\textbf {\bibinfo {volume} {9}},\
  \bibinfo {pages} {65} (\bibinfo {year} {1968})}\BibitemShut {NoStop}%
\bibitem [{\citenamefont {Grossmann}(1976)}]{Grossmann_1976}%
  \BibitemOpen
  \bibfield  {author} {\bibinfo {author} {\bibfnamefont {A.}~\bibnamefont
  {Grossmann}},\ }\href {https://doi.org/10.1007/BF01617867} {\bibfield
  {journal} {\bibinfo  {journal} {Commun. Math. Phys.}\ }\textbf {\bibinfo
  {volume} {48}},\ \bibinfo {pages} {191–194} (\bibinfo {year}
  {1976})}\BibitemShut {NoStop}%
\bibitem [{\citenamefont {Royer}(1977)}]{royer1977Wigner}%
  \BibitemOpen
  \bibfield  {author} {\bibinfo {author} {\bibfnamefont {A.}~\bibnamefont
  {Royer}},\ }\href {https://doi.org/10.1103/PhysRevA.15.449} {\bibfield
  {journal} {\bibinfo  {journal} {Phys. Rev. A}\ }\textbf {\bibinfo {volume}
  {15}},\ \bibinfo {pages} {449} (\bibinfo {year} {1977})}\BibitemShut
  {NoStop}%
\bibitem [{\citenamefont {Albert}\ \emph {et~al.}(2018)\citenamefont {Albert},
  \citenamefont {Noh}, \citenamefont {Duivenvoorden}, \citenamefont {Young},
  \citenamefont {Brierley}, \citenamefont {Reinhold}, \citenamefont {Vuillot},
  \citenamefont {Li}, \citenamefont {Shen}, \citenamefont {Girvin},
  \citenamefont {Terhal},\ and\ \citenamefont
  {Jiang}}]{Albert2017PerformanceBosonic}%
  \BibitemOpen
  \bibfield  {author} {\bibinfo {author} {\bibfnamefont {V.~V.}\ \bibnamefont
  {Albert}}, \bibinfo {author} {\bibfnamefont {K.}~\bibnamefont {Noh}},
  \bibinfo {author} {\bibfnamefont {K.}~\bibnamefont {Duivenvoorden}}, \bibinfo
  {author} {\bibfnamefont {D.~J.}\ \bibnamefont {Young}}, \bibinfo {author}
  {\bibfnamefont {R.~T.}\ \bibnamefont {Brierley}}, \bibinfo {author}
  {\bibfnamefont {P.}~\bibnamefont {Reinhold}}, \bibinfo {author}
  {\bibfnamefont {C.}~\bibnamefont {Vuillot}}, \bibinfo {author} {\bibfnamefont
  {L.}~\bibnamefont {Li}}, \bibinfo {author} {\bibfnamefont {C.}~\bibnamefont
  {Shen}}, \bibinfo {author} {\bibfnamefont {S.~M.}\ \bibnamefont {Girvin}},
  \bibinfo {author} {\bibfnamefont {B.~M.}\ \bibnamefont {Terhal}},\ and\
  \bibinfo {author} {\bibfnamefont {L.}~\bibnamefont {Jiang}},\ }\href
  {https://doi.org/10.1103/PhysRevA.97.032346} {\bibfield  {journal} {\bibinfo
  {journal} {Phys. Rev. A}\ }\textbf {\bibinfo {volume} {97}},\ \bibinfo
  {pages} {032346} (\bibinfo {year} {2018})}\BibitemShut {NoStop}%
\bibitem [{\citenamefont {Grimsmo}\ and\ \citenamefont
  {Puri}(2021)}]{Grimsmo2021GKP_Review}%
  \BibitemOpen
  \bibfield  {author} {\bibinfo {author} {\bibfnamefont {A.~L.}\ \bibnamefont
  {Grimsmo}}\ and\ \bibinfo {author} {\bibfnamefont {S.}~\bibnamefont {Puri}},\
  }\href {https://doi.org/10.1103/PRXQuantum.2.020101} {\bibfield  {journal}
  {\bibinfo  {journal} {PRX Quantum}\ }\textbf {\bibinfo {volume} {2}},\
  \bibinfo {pages} {020101} (\bibinfo {year} {2021})}\BibitemShut {NoStop}%
\bibitem [{\citenamefont {Brady}\ \emph {et~al.}(2024)\citenamefont {Brady},
  \citenamefont {Eickbusch}, \citenamefont {Singh}, \citenamefont {Wu},\ and\
  \citenamefont {Zhuang}}]{Brady2024GKP_Review}%
  \BibitemOpen
  \bibfield  {author} {\bibinfo {author} {\bibfnamefont {A.~J.}\ \bibnamefont
  {Brady}}, \bibinfo {author} {\bibfnamefont {A.}~\bibnamefont {Eickbusch}},
  \bibinfo {author} {\bibfnamefont {S.}~\bibnamefont {Singh}}, \bibinfo
  {author} {\bibfnamefont {J.}~\bibnamefont {Wu}},\ and\ \bibinfo {author}
  {\bibfnamefont {Q.}~\bibnamefont {Zhuang}},\ }\href
  {https://doi.org/https://doi.org/10.1016/j.pquantelec.2023.100496} {\bibfield
   {journal} {\bibinfo  {journal} {Prog. Quantum Electron.}\ }\textbf {\bibinfo
  {volume} {93}},\ \bibinfo {pages} {100496} (\bibinfo {year}
  {2024})}\BibitemShut {NoStop}%
\bibitem [{\citenamefont {Pantaleoni}\ \emph {et~al.}(2023)\citenamefont
  {Pantaleoni}, \citenamefont {Baragiola},\ and\ \citenamefont
  {Menicucci}}]{Pantaleoni_Zak_2024}%
  \BibitemOpen
  \bibfield  {author} {\bibinfo {author} {\bibfnamefont {G.}~\bibnamefont
  {Pantaleoni}}, \bibinfo {author} {\bibfnamefont {B.~Q.}\ \bibnamefont
  {Baragiola}},\ and\ \bibinfo {author} {\bibfnamefont {N.~C.}\ \bibnamefont
  {Menicucci}},\ }\href {https://doi.org/10.1103/PhysRevA.107.062611}
  {\bibfield  {journal} {\bibinfo  {journal} {Phys. Rev. A}\ }\textbf {\bibinfo
  {volume} {107}},\ \bibinfo {pages} {062611} (\bibinfo {year}
  {2023})}\BibitemShut {NoStop}%
\bibitem [{\citenamefont {Aharonov}\ \emph {et~al.}(1969)\citenamefont
  {Aharonov}, \citenamefont {Pendleton},\ and\ \citenamefont
  {Petersen}}]{Aharonov1969Modular}%
  \BibitemOpen
  \bibfield  {author} {\bibinfo {author} {\bibfnamefont {Y.}~\bibnamefont
  {Aharonov}}, \bibinfo {author} {\bibfnamefont {H.}~\bibnamefont
  {Pendleton}},\ and\ \bibinfo {author} {\bibfnamefont {A.}~\bibnamefont
  {Petersen}},\ }\href {https://doi.org/10.1007/BF00670008} {\bibfield
  {journal} {\bibinfo  {journal} {Int. J. Theor. Phys.}\ }\textbf {\bibinfo
  {volume} {2}},\ \bibinfo {pages} {213–230} (\bibinfo {year}
  {1969})}\BibitemShut {NoStop}%
\bibitem [{\citenamefont {Zak}(1967)}]{Zak1967finiteTranslations}%
  \BibitemOpen
  \bibfield  {author} {\bibinfo {author} {\bibfnamefont {J.}~\bibnamefont
  {Zak}},\ }\href {https://doi.org/10.1103/PhysRevLett.19.1385} {\bibfield
  {journal} {\bibinfo  {journal} {Phys. Rev. Lett.}\ }\textbf {\bibinfo
  {volume} {19}},\ \bibinfo {pages} {1385} (\bibinfo {year}
  {1967})}\BibitemShut {NoStop}%
\bibitem [{\citenamefont {Englert}\ \emph {et~al.}(2006)\citenamefont
  {Englert}, \citenamefont {Lee}, \citenamefont {Mann},\ and\ \citenamefont
  {Revzen}}]{englert_periodic_2006}%
  \BibitemOpen
  \bibfield  {author} {\bibinfo {author} {\bibfnamefont {B.-G.}\ \bibnamefont
  {Englert}}, \bibinfo {author} {\bibfnamefont {K.~L.}\ \bibnamefont {Lee}},
  \bibinfo {author} {\bibfnamefont {A.}~\bibnamefont {Mann}},\ and\ \bibinfo
  {author} {\bibfnamefont {M.}~\bibnamefont {Revzen}},\ }\href
  {https://doi.org/10.1088/0305-4470/39/7/011} {\bibfield  {journal} {\bibinfo
  {journal} {J. Phys. A: Math. Gen.}\ }\textbf {\bibinfo {volume} {39}},\
  \bibinfo {pages} {1669} (\bibinfo {year} {2006})}\BibitemShut {NoStop}%
\bibitem [{\citenamefont {Ketterer}\ \emph {et~al.}(2016)\citenamefont
  {Ketterer}, \citenamefont {Keller}, \citenamefont {Walborn}, \citenamefont
  {Coudreau},\ and\ \citenamefont {Milman}}]{ketterer2016quantum}%
  \BibitemOpen
  \bibfield  {author} {\bibinfo {author} {\bibfnamefont {A.}~\bibnamefont
  {Ketterer}}, \bibinfo {author} {\bibfnamefont {A.}~\bibnamefont {Keller}},
  \bibinfo {author} {\bibfnamefont {S.~P.}\ \bibnamefont {Walborn}}, \bibinfo
  {author} {\bibfnamefont {T.}~\bibnamefont {Coudreau}},\ and\ \bibinfo
  {author} {\bibfnamefont {P.}~\bibnamefont {Milman}},\ }\href
  {https://doi.org/10.1103/PhysRevA.94.022325} {\bibfield  {journal} {\bibinfo
  {journal} {Phys. Rev. A}\ }\textbf {\bibinfo {volume} {94}},\ \bibinfo
  {pages} {022325} (\bibinfo {year} {2016})}\BibitemShut {NoStop}%
\bibitem [{\citenamefont {Fabre}\ \emph
  {et~al.}(2020{\natexlab{b}})\citenamefont {Fabre}, \citenamefont {Keller},\
  and\ \citenamefont {Milman}}]{fabre2020wigner}%
  \BibitemOpen
  \bibfield  {author} {\bibinfo {author} {\bibfnamefont {N.}~\bibnamefont
  {Fabre}}, \bibinfo {author} {\bibfnamefont {A.}~\bibnamefont {Keller}},\ and\
  \bibinfo {author} {\bibfnamefont {P.}~\bibnamefont {Milman}},\ }\href
  {https://doi.org/10.1103/PhysRevA.102.022411} {\bibfield  {journal} {\bibinfo
   {journal} {Phys. Rev. A}\ }\textbf {\bibinfo {volume} {102}},\ \bibinfo
  {pages} {022411} (\bibinfo {year} {2020}{\natexlab{b}})}\BibitemShut
  {NoStop}%
\bibitem [{\citenamefont {De~Bi\`{e}vre}\ \emph {et~al.}(1996)\citenamefont
  {De~Bi\`{e}vre}, \citenamefont {Esposti},\ and\ \citenamefont
  {Giachetti}}]{de_bievre_quantization_1996}%
  \BibitemOpen
  \bibfield  {author} {\bibinfo {author} {\bibfnamefont {S.}~\bibnamefont
  {De~Bi\`{e}vre}}, \bibinfo {author} {\bibfnamefont {M.~D.}\ \bibnamefont
  {Esposti}},\ and\ \bibinfo {author} {\bibfnamefont {R.}~\bibnamefont
  {Giachetti}},\ }\href {https://doi.org/10.1007/BF02099363} {\bibfield
  {journal} {\bibinfo  {journal} {Commun. Math. Phys.}\ }\textbf {\bibinfo
  {volume} {176}},\ \bibinfo {pages} {73} (\bibinfo {year} {1996})}\BibitemShut
  {NoStop}%
\bibitem [{\citenamefont {Kowalski}\ and\ \citenamefont
  {Rembieli\ifmmode~\acute{n}\else
  \'{n}\fi{}ski}(2007)}]{kowalski_coherent_2007}%
  \BibitemOpen
  \bibfield  {author} {\bibinfo {author} {\bibfnamefont {K.}~\bibnamefont
  {Kowalski}}\ and\ \bibinfo {author} {\bibfnamefont {J.}~\bibnamefont
  {Rembieli\ifmmode~\acute{n}\else \'{n}\fi{}ski}},\ }\href
  {https://doi.org/10.1103/PhysRevA.75.052102} {\bibfield  {journal} {\bibinfo
  {journal} {Phys. Rev. A}\ }\textbf {\bibinfo {volume} {75}},\ \bibinfo
  {pages} {052102} (\bibinfo {year} {2007})}\BibitemShut {NoStop}%
\bibitem [{\citenamefont {Ligab{\`o}}(2016)}]{ligabo2016torus}%
  \BibitemOpen
  \bibfield  {author} {\bibinfo {author} {\bibfnamefont {M.}~\bibnamefont
  {Ligab{\`o}}},\ }\href {https://doi.org/10.1063/1.4961325} {\bibfield
  {journal} {\bibinfo  {journal} {J. Math. Phys.}\ }\textbf {\bibinfo {volume}
  {57}},\ \bibinfo {pages} {082110} (\bibinfo {year} {2016})}\BibitemShut
  {NoStop}%
\bibitem [{\citenamefont {Busch}\ and\ \citenamefont
  {Lahti}(1986)}]{Busch_Lahti_1986}%
  \BibitemOpen
  \bibfield  {author} {\bibinfo {author} {\bibfnamefont {P.}~\bibnamefont
  {Busch}}\ and\ \bibinfo {author} {\bibfnamefont {P.~J.}\ \bibnamefont
  {Lahti}},\ }\href {https://doi.org/10.1016/0375-9601(86)90549-9} {\bibfield
  {journal} {\bibinfo  {journal} {Phys. Lett. A}\ }\textbf {\bibinfo {volume}
  {115}},\ \bibinfo {pages} {259–264} (\bibinfo {year} {1986})}\BibitemShut
  {NoStop}%
\bibitem [{\citenamefont {Ketterer}(2016)}]{ketterer2016modular}%
  \BibitemOpen
  \bibfield  {author} {\bibinfo {author} {\bibfnamefont {A.}~\bibnamefont
  {Ketterer}},\ }\emph {\bibinfo {title} {Modular variables in quantum
  information}},\ \href {https://theses.hal.science/tel-01502539} {Ph.D.
  thesis},\ \bibinfo  {school} {Universit{\'e} Paris 7, Sorbonne Paris
  Cit{\'e}} (\bibinfo {year} {2016})\BibitemShut {NoStop}%
\bibitem [{\citenamefont {Bertrand}\ and\ \citenamefont
  {Bertrand}(1987)}]{bertrand1987tomographic}%
  \BibitemOpen
  \bibfield  {author} {\bibinfo {author} {\bibfnamefont {J.}~\bibnamefont
  {Bertrand}}\ and\ \bibinfo {author} {\bibfnamefont {P.}~\bibnamefont
  {Bertrand}},\ }\href {https://doi.org/10.1007/BF00733376} {\bibfield
  {journal} {\bibinfo  {journal} {Found. Phys.}\ }\textbf {\bibinfo {volume}
  {17}},\ \bibinfo {pages} {397} (\bibinfo {year} {1987})}\BibitemShut
  {NoStop}%
\bibitem [{\citenamefont {Vogel}\ and\ \citenamefont
  {Risken}(1989)}]{Vogel_Risken_1989}%
  \BibitemOpen
  \bibfield  {author} {\bibinfo {author} {\bibfnamefont {K.}~\bibnamefont
  {Vogel}}\ and\ \bibinfo {author} {\bibfnamefont {H.}~\bibnamefont {Risken}},\
  }\href {https://doi.org/10.1103/PhysRevA.40.2847} {\bibfield  {journal}
  {\bibinfo  {journal} {Phys. Rev. A}\ }\textbf {\bibinfo {volume} {40}},\
  \bibinfo {pages} {2847} (\bibinfo {year} {1989})}\BibitemShut {NoStop}%
\bibitem [{\citenamefont {Janssen}(1988)}]{janssen1988zak}%
  \BibitemOpen
  \bibfield  {author} {\bibinfo {author} {\bibfnamefont {A.~J. E.~M.}\
  \bibnamefont {Janssen}},\ }\href@noop {} {\bibfield  {journal} {\bibinfo
  {journal} {Philips J. Res.}\ }\textbf {\bibinfo {volume} {43}},\ \bibinfo
  {pages} {23} (\bibinfo {year} {1988})}\BibitemShut {NoStop}%
\bibitem [{\citenamefont {Mann}\ \emph {et~al.}(2005)\citenamefont {Mann},
  \citenamefont {Revzen},\ and\ \citenamefont {Zak}}]{mann2005conjugate}%
  \BibitemOpen
  \bibfield  {author} {\bibinfo {author} {\bibfnamefont {A.}~\bibnamefont
  {Mann}}, \bibinfo {author} {\bibfnamefont {M.}~\bibnamefont {Revzen}},\ and\
  \bibinfo {author} {\bibfnamefont {J.}~\bibnamefont {Zak}},\ }\href
  {https://doi.org/10.1088/0305-4470/38/21/L03} {\bibfield  {journal} {\bibinfo
   {journal} {J. Phys. A: Math. Gen.}\ }\textbf {\bibinfo {volume} {38}},\
  \bibinfo {pages} {L389} (\bibinfo {year} {2005})}\BibitemShut {NoStop}%
\bibitem [{\citenamefont {Mann}\ \emph {et~al.}(2006)\citenamefont {Mann},
  \citenamefont {Revzen},\ and\ \citenamefont {Zak}}]{Mann_Revzen_Zak_2006}%
  \BibitemOpen
  \bibfield  {author} {\bibinfo {author} {\bibfnamefont {A.}~\bibnamefont
  {Mann}}, \bibinfo {author} {\bibfnamefont {M.}~\bibnamefont {Revzen}},\ and\
  \bibinfo {author} {\bibfnamefont {J.}~\bibnamefont {Zak}},\ }\href
  {https://doi.org/10.1142/S0219749906001670} {\bibfield  {journal} {\bibinfo
  {journal} {Int. J. Quantum Inf.}\ }\textbf {\bibinfo {volume} {04}},\
  \bibinfo {pages} {173–180} (\bibinfo {year} {2006})}\BibitemShut {NoStop}%
\bibitem [{\citenamefont {Feng}\ and\ \citenamefont
  {Luo}(2024)}]{Feng_Luo_2024}%
  \BibitemOpen
  \bibfield  {author} {\bibinfo {author} {\bibfnamefont {L.}~\bibnamefont
  {Feng}}\ and\ \bibinfo {author} {\bibfnamefont {S.}~\bibnamefont {Luo}},\
  }\href {https://doi.org/10.1007/s10773-024-05549-3} {\bibfield  {journal}
  {\bibinfo  {journal} {Int. J. Theor. Phys.}\ }\textbf {\bibinfo {volume}
  {63}},\ \bibinfo {pages} {40} (\bibinfo {year} {2024})}\BibitemShut {NoStop}%
\bibitem [{\citenamefont {Feynman}(1998)}]{Feynman_1998}%
  \BibitemOpen
  \bibfield  {author} {\bibinfo {author} {\bibfnamefont {R.~P.}\ \bibnamefont
  {Feynman}},\ }\href {https://doi.org/10.1201/9780429493034} {\emph {\bibinfo
  {title} {Statistical Mechanics: A Set Of Lectures}}},\ \bibinfo {edition}
  {1st}\ ed.\ (\bibinfo  {publisher} {CRC Press},\ \bibinfo {address} {Boca
  Raton},\ \bibinfo {year} {1998})\BibitemShut {NoStop}%
\bibitem [{\citenamefont {Simon}\ \emph {et~al.}(1994)\citenamefont {Simon},
  \citenamefont {Mukunda},\ and\ \citenamefont
  {Dutta}}]{Simon_Mukunda_Noise_Matrix_1994}%
  \BibitemOpen
  \bibfield  {author} {\bibinfo {author} {\bibfnamefont {R.}~\bibnamefont
  {Simon}}, \bibinfo {author} {\bibfnamefont {N.}~\bibnamefont {Mukunda}},\
  and\ \bibinfo {author} {\bibfnamefont {B.}~\bibnamefont {Dutta}},\ }\href
  {https://doi.org/10.1103/PhysRevA.49.1567} {\bibfield  {journal} {\bibinfo
  {journal} {Phys. Rev. A}\ }\textbf {\bibinfo {volume} {49}},\ \bibinfo
  {pages} {1567} (\bibinfo {year} {1994})}\BibitemShut {NoStop}%
\bibitem [{\citenamefont {Olivares}(2012)}]{Olivares_2012}%
  \BibitemOpen
  \bibfield  {author} {\bibinfo {author} {\bibfnamefont {S.}~\bibnamefont
  {Olivares}},\ }\href {https://doi.org/10.1140/epjst/e2012-01532-4} {\bibfield
   {journal} {\bibinfo  {journal} {Eur. Phys. J.: Spec. Top.}\ }\textbf
  {\bibinfo {volume} {203}},\ \bibinfo {pages} {3–24} (\bibinfo {year}
  {2012})}\BibitemShut {NoStop}%
\bibitem [{\citenamefont {Mumford}(2007)}]{Mumford_Tata_1}%
  \BibitemOpen
  \bibfield  {author} {\bibinfo {author} {\bibfnamefont {D.}~\bibnamefont
  {Mumford}},\ }\href {https://doi.org/10.1007/978-0-8176-4577-9} {\emph
  {\bibinfo {title} {{Tata Lectures on Theta 1}}}},\ Modern Birkh\"{a}user
  Classics\ (\bibinfo  {publisher} {Birkh\"{a}user Boston},\ \bibinfo {address}
  {Boston, MA},\ \bibinfo {year} {2007})\BibitemShut {NoStop}%
\bibitem [{\citenamefont {Terhal}\ and\ \citenamefont
  {Weigand}(2016{\natexlab{b}})}]{Terhal2016Encoding}%
  \BibitemOpen
  \bibfield  {author} {\bibinfo {author} {\bibfnamefont {B.~M.}\ \bibnamefont
  {Terhal}}\ and\ \bibinfo {author} {\bibfnamefont {D.}~\bibnamefont
  {Weigand}},\ }\href {https://doi.org/10.1103/PhysRevA.93.012315} {\bibfield
  {journal} {\bibinfo  {journal} {Phys. Rev. A}\ }\textbf {\bibinfo {volume}
  {93}},\ \bibinfo {pages} {012315} (\bibinfo {year}
  {2016}{\natexlab{b}})}\BibitemShut {NoStop}%
\bibitem [{\citenamefont {Motes}\ \emph {et~al.}(2017)\citenamefont {Motes},
  \citenamefont {Baragiola}, \citenamefont {Gilchrist},\ and\ \citenamefont
  {Menicucci}}]{PhysRevA.95.053819}%
  \BibitemOpen
  \bibfield  {author} {\bibinfo {author} {\bibfnamefont {K.~R.}\ \bibnamefont
  {Motes}}, \bibinfo {author} {\bibfnamefont {B.~Q.}\ \bibnamefont
  {Baragiola}}, \bibinfo {author} {\bibfnamefont {A.}~\bibnamefont
  {Gilchrist}},\ and\ \bibinfo {author} {\bibfnamefont {N.~C.}\ \bibnamefont
  {Menicucci}},\ }\href {https://doi.org/10.1103/PhysRevA.95.053819} {\bibfield
   {journal} {\bibinfo  {journal} {Phys. Rev. A}\ }\textbf {\bibinfo {volume}
  {95}},\ \bibinfo {pages} {053819} (\bibinfo {year} {2017})}\BibitemShut
  {NoStop}%
\bibitem [{\citenamefont {Matsuura}\ \emph {et~al.}(2020)\citenamefont
  {Matsuura}, \citenamefont {Yamasaki},\ and\ \citenamefont
  {Koashi}}]{Matsuura2020equivalenceGKPcodes}%
  \BibitemOpen
  \bibfield  {author} {\bibinfo {author} {\bibfnamefont {T.}~\bibnamefont
  {Matsuura}}, \bibinfo {author} {\bibfnamefont {H.}~\bibnamefont {Yamasaki}},\
  and\ \bibinfo {author} {\bibfnamefont {M.}~\bibnamefont {Koashi}},\ }\href
  {https://doi.org/10.1103/PhysRevA.102.032408} {\bibfield  {journal} {\bibinfo
   {journal} {Phys. Rev. A}\ }\textbf {\bibinfo {volume} {102}},\ \bibinfo
  {pages} {032408} (\bibinfo {year} {2020})}\BibitemShut {NoStop}%
\bibitem [{\citenamefont {Albert}\ \emph {et~al.}(2020)\citenamefont {Albert},
  \citenamefont {Covey},\ and\ \citenamefont {Preskill}}]{albert2020robust}%
  \BibitemOpen
  \bibfield  {author} {\bibinfo {author} {\bibfnamefont {V.~V.}\ \bibnamefont
  {Albert}}, \bibinfo {author} {\bibfnamefont {J.~P.}\ \bibnamefont {Covey}},\
  and\ \bibinfo {author} {\bibfnamefont {J.}~\bibnamefont {Preskill}},\ }\href
  {https://doi.org/10.1103/PhysRevX.10.031050} {\bibfield  {journal} {\bibinfo
  {journal} {Phys. Rev. X}\ }\textbf {\bibinfo {volume} {10}},\ \bibinfo
  {pages} {031050} (\bibinfo {year} {2020})}\BibitemShut {NoStop}%
\bibitem [{\citenamefont {Asadian}\ \emph {et~al.}(2014)\citenamefont
  {Asadian}, \citenamefont {Brukner},\ and\ \citenamefont
  {Rabl}}]{Asadian2014Probing}%
  \BibitemOpen
  \bibfield  {author} {\bibinfo {author} {\bibfnamefont {A.}~\bibnamefont
  {Asadian}}, \bibinfo {author} {\bibfnamefont {C.}~\bibnamefont {Brukner}},\
  and\ \bibinfo {author} {\bibfnamefont {P.}~\bibnamefont {Rabl}},\ }\href
  {https://doi.org/10.1103/PhysRevLett.112.190402} {\bibfield  {journal}
  {\bibinfo  {journal} {Phys. Rev. Lett.}\ }\textbf {\bibinfo {volume} {112}},\
  \bibinfo {pages} {190402} (\bibinfo {year} {2014})}\BibitemShut {NoStop}%
\bibitem [{\citenamefont {Asadian}\ \emph {et~al.}(2015)\citenamefont
  {Asadian}, \citenamefont {Budroni}, \citenamefont {Steinhoff}, \citenamefont
  {Rabl},\ and\ \citenamefont {G{\"u}hne}}]{asadian2015contextuality}%
  \BibitemOpen
  \bibfield  {author} {\bibinfo {author} {\bibfnamefont {A.}~\bibnamefont
  {Asadian}}, \bibinfo {author} {\bibfnamefont {C.}~\bibnamefont {Budroni}},
  \bibinfo {author} {\bibfnamefont {F.~E.}\ \bibnamefont {Steinhoff}}, \bibinfo
  {author} {\bibfnamefont {P.}~\bibnamefont {Rabl}},\ and\ \bibinfo {author}
  {\bibfnamefont {O.}~\bibnamefont {G{\"u}hne}},\ }\href
  {https://doi.org/10.1103/PhysRevLett.114.250403} {\bibfield  {journal}
  {\bibinfo  {journal} {Phys. Rev. Lett.}\ }\textbf {\bibinfo {volume} {114}},\
  \bibinfo {pages} {250403} (\bibinfo {year} {2015})}\BibitemShut {NoStop}%
\bibitem [{\citenamefont {Fl\"uhmann}\ \emph {et~al.}(2018)\citenamefont
  {Fl\"uhmann}, \citenamefont {Negnevitsky}, \citenamefont {Marinelli},\ and\
  \citenamefont {Home}}]{Fluhmann_sequential_2018}%
  \BibitemOpen
  \bibfield  {author} {\bibinfo {author} {\bibfnamefont {C.}~\bibnamefont
  {Fl\"uhmann}}, \bibinfo {author} {\bibfnamefont {V.}~\bibnamefont
  {Negnevitsky}}, \bibinfo {author} {\bibfnamefont {M.}~\bibnamefont
  {Marinelli}},\ and\ \bibinfo {author} {\bibfnamefont {J.~P.}\ \bibnamefont
  {Home}},\ }\href {https://doi.org/10.1103/PhysRevX.8.021001} {\bibfield
  {journal} {\bibinfo  {journal} {Phys. Rev. X}\ }\textbf {\bibinfo {volume}
  {8}},\ \bibinfo {pages} {021001} (\bibinfo {year} {2018})}\BibitemShut
  {NoStop}%
\bibitem [{\citenamefont {Gazeau}\ and\ \citenamefont
  {Murenzi}(2022)}]{gazeau_integral_2022}%
  \BibitemOpen
  \bibfield  {author} {\bibinfo {author} {\bibfnamefont {J.-P.}\ \bibnamefont
  {Gazeau}}\ and\ \bibinfo {author} {\bibfnamefont {R.}~\bibnamefont
  {Murenzi}},\ }\href {https://doi.org/10.3390/quantum4040026} {\bibfield
  {journal} {\bibinfo  {journal} {Quantum Rep.}\ }\textbf {\bibinfo {volume}
  {4}},\ \bibinfo {pages} {362} (\bibinfo {year} {2022})}\BibitemShut {NoStop}%
\bibitem [{\citenamefont {Zak}(1968)}]{Zak1968dyamics}%
  \BibitemOpen
  \bibfield  {author} {\bibinfo {author} {\bibfnamefont {J.}~\bibnamefont
  {Zak}},\ }\href {https://doi.org/10.1103/PhysRev.168.686} {\bibfield
  {journal} {\bibinfo  {journal} {Phys. Rev.}\ }\textbf {\bibinfo {volume}
  {168}},\ \bibinfo {pages} {686} (\bibinfo {year} {1968})}\BibitemShut
  {NoStop}%
\end{thebibliography}%


\widetext
\appendix

\section{The Zak picture and theta functions}
\label{app:Zaktheta}

In this section we provide additional background on the Zak representation of quantum mechanics and theta functions.  The \textit{Zak transform} of a function $\psi\in L^2(\mathbb{R})$ is 
\begin{equation}
    [Z_\alpha\psi](k,q) = \sqrt{\frac{\alpha}{2\pi}} \sum_{n\in\Z} e^{ikn\alpha} \psi(q - n\alpha),
\end{equation}
where $\alpha$ is a characteristic length \cite{Zak1967finiteTranslations, Zak1968dyamics, janssen1988zak} (see also \cite{Pantaleoni_Zak_2024} for a recent treatment). The image of the Zak transform consists of quasiperiodic functions on $\R^2$, equipped with the usual $L^2$-inner product but restricted to any rectangle of area $2\pi$ (i.e., the Planck constant $h = 2\pi \hbar$ with units restored).  With respect to this inner product the Zak transform is a unitary isomorphism between the Schr\"{o}dinger representation and the \textit{Zak representation} of the Heisenberg--Weyl group:
\begin{equation}\label{zak_inner_iso}
    \int_0^{\alpha} dq \int_0^{\frac{2\pi}{\alpha}} dk [Z_\alpha\psi]^*(k,q) [Z_\alpha\phi](k,q) = \int_{\R} dx \psi^*(x) \phi(x),
\end{equation}
where $\psi,\phi \in L^2(\R)$ and we chose the rectangle $[0,\alpha) \times [0,2\pi/\alpha)$ without loss of generality.  Because of this, only the values of $Z\psi$ from any rectangle of area $2\pi$ are needed to completely specify a pure state.  The output Hilbert space is denoted $L^2(\mathbb{T}^2)$ for a torus $\mathbb{T}^2$, but it is understood that $Z_\alpha\psi$ is only quasiperiodic,
\begin{equation}\label{Zak_quasi-periodicity}
    \begin{aligned}
        [Z_\alpha\psi](k - \frac{2\pi}{\alpha},q) &= [Z_\alpha f](k,q) \\
        [Z_\alpha\psi](k, q - \alpha) &= e^{-ik\alpha}[Z_\alpha f](k,q),
    \end{aligned}
\end{equation}
and so strictly speaking is a function on $\R^2$.  The image of the position basis under the Zak transform constitutes the \textit{Zak basis}
\begin{equation}
    \ket{k,q} = \sqrt{\frac{\alpha}{2\pi}} \sum_{n\in\Z} e^{ikn\alpha} \ket{q + n\alpha}_{\hat x},
\end{equation}
which, as a consequence of \eqref{zak_inner_iso}, is complete in the full infinite-dimensional Hilbert space $\mathcal{H}$,
\begin{equation}
    \int_{0}^\alpha dq \int_0^{\frac{2\pi}{\alpha}} dk\ketbra{k,q}{k,q} = \hat{\mathbb{I}}.
\end{equation}
The \textit{Zak measurement} is the PVM of projectors onto the Zak basis, $\{\ketbra{k,q}{k,q}\}$.  This is strongly related to modular variables \cite{Aharonov1969Modular, ketterer2016quantum, Pantaleoni_Zak_2024}, and is also called a \textit{modular measurement}.  Outcomes are obtained with a probability determined by the standard Born rule, 
\begin{equation}\label{Zak_modular_measurement}
    \text{Pr}[(k,q) | {\hat\rho}] = \langle k,q | {\hat\rho} | k,q \rangle.
\end{equation}
Note that the asymmetry in the quasiperiodicity relations \eqref{Zak_quasi-periodicity} is due to a phase convention initiated by Zak that in some sense ``prefers'' position over momentum.  A symmetric version is available, though we will not use it because much of our Zak-related analysis lives on the measurement level where such phases play less of a role. 




The single-variable \textit{theta function} and the single-variable \textit{theta function with characteristics} are
\begin{equation}\label{theta_single}
    \theta(z,\tau) = \sum_{n\in\Z} e^{i2\pi n z}e^{i\pi n^2 \tau} \qquad \text{and} \qquad \thetachar{v_1}{v_2}{z}{\tau} = \sum_{n \in \Z} e^{i2\pi (n + v_1)(z + v_2)} e^{i\pi (n + v_1)^2},
\end{equation}
where the characteristics $v_1, v_2 \in \mathbb{Q}$, $z \in \mathbb{C}$, and $\tau$ is an element of the upper half-space of complex numbers, i.e., $\text{Im}[\tau] > 0$ \cite{Mumford_Tata_1}.  The $N$-dimensional generalizations are
\begin{equation}\label{theta_multi}
    \theta(\bm z, \bm \tau) = \sum_{\bm n \in \Z^N} e^{i2\pi \bm n^T \bm z } e^{i\pi \bm n^T \bm \tau \bm n} \qquad \text{and} \qquad \thetachar{\bm v_1}{\bm v_2}{\bm z}{\bm \tau} = \sum_{\bm n \in \Z^N} e^{i2\pi (\bm n + \bm v_1)^T (\bm z + \bm v_2 ) } e^{i\pi (\bm n + \bm v_1)^T \bm \tau (\bm n + \bm v_1 )}
\end{equation}
where now $\bm v_1, \bm v_2 \in \mathbb{Q}^N$, $\bm z \in \mathbb{C}^N$, and $\bm \tau$ is an element of the Siegel upper half-space, i.e., an $N\times N$ symmetric matrix with strictly positive imaginary part \cite{Mumford_Tata_1}.  In the single-variable case these functions are often called Jacobi theta functions and in the multi-variable case they are often called Riemann or Siegel theta functions.  We will tend to interpret them as functions of $\bm z$ with $\bm \tau$ a tunable parameter. 





The key property that connects all of the above is that, up to particular parameter values and phase conventions, the Zak transform of a Gaussian is a theta function. Since the Zak transform is a unitary isomorphism (an intertwiner between Heisenberg-Weyl unitary irreps to be precise), the role of (certain) theta functions in the Zak picture is equivalent to the role of Gaussians in the Schr\"odinger picture.


\section{CV Wigner function from Fock reference state}
\label{app:BrifManFock}

We consider a quantum system described by an infinite-dimensional Hilbert space $\mathcal H$ with dynamical symmetry group $G=H_3(\mathbb C)$, the Heisenberg--Weyl group over $\mathbb C$. The Hilbert space has an infinite countable Fock basis $\{\ket n\}_{n\in\mathbb N}$. For a fixed $n\in\mathbb N$, we follow the phase-space construction from~\cite{brif1999phase} with reference state $\ket n$ and show that the resulting Wigner function is independent of the choice of $n$.

Any element $g\in G$ can be parametrized as
\begin{equation}
    \pi(g)=e^{i\varphi\hat{\mathbb I}}e^{\gamma\hat a^\dag-\gamma^*\hat a},
\end{equation}
for $\varphi\in\mathbb R$ and $\gamma\in\mathbb C$, i.e., as a displacement operator up to a global phase. Now:

\begin{itemize}
    \item The isotropy subgroup of the state $\ket n$ is $U(1)$ and thus the phase space is given by $H_3(\mathbb C)/U(1)\simeq\mathbb C$.
    \item The coherent states are $\hat D(\alpha)\ket n$ for all $\alpha\in\mathbb C$. 
    \item The invariant measure over this space is $d\mu(\alpha)=\frac{d^2\alpha}\pi=\frac{d\Re(\alpha)d\Im(\alpha)}\pi$.
    \item The harmonic functions over $L^2(\mathbb C,\mu)$ are given by $Y_\nu(\alpha)=e^{\nu\alpha^*-\nu^*\alpha}$, for all $\nu,\alpha\in\mathbb C$. The Plancherel measure over the spectrum of the Laplace--Beltrami operator is then given by $\frac{d^2\nu}\pi=\frac{d\Re(\nu)d\Im(\nu)}\pi$.
\end{itemize}

\noindent With this preliminary material in place, we are ready to determine the real and positive coefficients $\tau_\nu$ and the tensor operators $\hat D_\nu$ via Eqs.~(\ref{eq:cohOvercode}) and (\ref{eq:tensorOpcode}), which in this case read
\begin{align}
    |\langle n|\hat D^\dag(\beta)\hat D(\alpha)|n\rangle|^2&=\int_{\nu\in\mathbb C}\frac{d^2\nu}\pi\tau_\nu e^{\nu^*(\alpha-\beta)-\nu(\alpha^*-\beta^*)}\\
    \label{eq:DnuFock}e^{i\varphi_\nu}\tau_\nu^{-1/2}\hat D_\nu&=\int_{\alpha\in\mathbb C}\frac{d^2\alpha}\pi e^{\nu\alpha^*-\nu^*\alpha}\hat D(\alpha)\ket n\!\bra n\hat D^\dag(\alpha).
\end{align}
Setting $\beta=0$ in the first equation yields$|\bra n\hat D(\alpha)\ket n|^2=e^{-|\alpha|^2}[L_n(|\alpha|^2)]^2=\int_{\mathbb C}\frac{d^2\nu}\pi\tau_\nu e^{\nu^*\alpha-\nu\alpha^*}$. Taking the inverse Fourier transform yields
\begin{equation}
    \tau_\nu=\left[\frac1{n!}L_n(|\nu|^2)\right]^2e^{-|\nu|^2},
\end{equation}
for all $\nu\in\mathbb C$.
With Eq.~(\ref{eq:DnuFock}), this implies $e^{i\varphi_\nu}\hat D_\nu=\frac{n!}{L_n(|\nu|^2)}e^{|\nu|^2/2}\int_{\mathbb C}\frac{d^2\alpha}\pi e^{\nu\alpha^*-\nu^*\alpha}\hat D(\alpha)\ket n\!\bra n\hat D^\dag(\alpha)$, where $e^{i\varphi_\nu}$ is a phase function which we set to be identically $1$.
Let us compute:
\begin{equation}
    \begin{aligned}
        L_n(|\nu|^2)\hat D_\nu&=n!e^{|\nu|^2/2}\int_{\mathbb C}\frac{d^2\alpha}\pi e^{\nu\alpha^*-\nu^*\alpha}\hat D(\alpha)\ket n\!\bra n\hat D^\dag(\alpha)\\
        &=e^{|\nu|^2/2}\int_{\mathbb C}\frac{d^2\alpha}\pi e^{\nu\alpha^*-\nu^*\alpha}\hat D(\alpha)\hat a^{\dag n}\ket0\!\bra0\hat a^n\hat D^\dag(\alpha)\\
        &=e^{|\nu|^2/2}\int_{\mathbb C}\frac{d^2\alpha}\pi e^{\nu\alpha^*-\nu^*\alpha}(\hat a^\dag-\alpha^*)^n\hat D(\alpha)\ket0\!\bra0\hat D^\dag(\alpha)(\hat a-\alpha)^n\\
        &=\sum_{k,l=0}^n\binom nk\binom nle^{|\nu|^2/2}\int_{\mathbb C}\frac{d^2\alpha}\pi e^{\nu\alpha^*-\nu^*\alpha}\hat a^{\dag k}(-\alpha^*)^{n-k}\hat D(\alpha)\ket0\!\bra0\hat D^\dag(\alpha)\hat a^l(-\alpha)^{n-l}\\
        &=\sum_{k,l=0}^n\binom nk\binom nl(-1)^{k+l}\hat a^{\dag k}\left(e^{|\nu|^2/2}\int_{\mathbb C}\frac{d^2\alpha}\pi e^{\nu\alpha^*-\nu^*\alpha}\alpha^{n-l}\ket\alpha\!\bra\alpha\alpha^{*n-k}\right)\hat a^l\\
        &=\sum_{k,l=0}^n\binom nk\binom nl(-1)^{k+l}\hat a^{\dag k}\hat a^{n-l}\left(e^{|\nu|^2/2}\int_{\mathbb C}\frac{d^2\alpha}\pi e^{\nu\alpha^*-\nu^*\alpha}\ket\alpha\!\bra\alpha\right)\hat a^{\dag n-k}\hat a^l\\
        &=\sum_{k,l=0}^n\binom nk\binom nl(-1)^{k+l}\hat a^{\dag k}\hat a^{n-l}\hat D(\nu)\hat a^{\dag n-k}\hat a^l\\
        &=\sum_{k,l=0}^n\binom nk\binom nl(-1)^{k+l}\hat a^{\dag k}\hat a^{n-l}(\hat a^\dag-\nu^*)^{n-k}(\hat a-\nu)^l\hat D(\nu),
    \end{aligned}
\end{equation}
where we have used the expansion of the displacement operator in terms of Glauber coherent states $\hat D(\nu)=e^{|\nu|^2/2}\int_{\mathbb C}\frac{d^2\alpha}\pi e^{\nu\alpha^*-\nu^*\alpha}\ket\alpha\!\bra\alpha$.
At this point, we make use of the commutation relation
\begin{equation}
    \hat X^i\hat Y^j=\sum_{m=0}^{\min i,j}m!\binom im\binom jm\hat Y^{j-m}\hat X^{i-m},
\end{equation}
valid for all operators $\hat X$ and $\hat Y$ such that $[\hat X,\hat Y]=\hat {\mathbb I}$. For $\hat X=\hat a$, $\hat Y=\hat a^\dag-\nu^*$, $i=n-l$, and $j=n-k$ this gives
\begin{equation}
    \begin{aligned}
        L_n(|\nu|^2)\hat D_\nu&=\sum_{k,l=0}^n\sum_{m=0}^{\min n-k,n-l}m!\binom{n-l}m\binom{n-k}m\binom nk\binom nl(-1)^{k+l}\\
        &\times \hat a^{\dag k}(\hat a^\dag-\nu^*)^{n-k-m}\hat a^{n-l-m}(\hat a-\nu)^l\hat D(\nu)\\
        &=\sum_{k,l=0}^n\sum_{m=0}^{\min n-k,n-l}\sum_{u=0}^{n-k-m}\sum_{v=0}^{n-l-m}\binom{n-k-m}u\binom{n-l-m}v\\
        &\times m!\binom{n-l}m\binom{n-k}m\binom nk\binom nl(-1)^{k+l+u+v}\nu^{*u}\nu^v\hat a^{\dag n-m-u}\hat a^{n-m-v}\hat D(\nu)\\
        &=\sum_{p=0}^n\sum_{u,v=0}^p\sum_{k=0}^{p-u}\sum_{l=0}^{p-v}\binom{p-k}u\binom{p-l}v\\
        &\times (n-p)!\binom{n-l}{n-p}\binom{n-k}{n-p}\binom nk\binom nl(-1)^{k+l+u+v}\nu^{*u}\nu^v\hat a^{\dag p-u}\hat a^{p-v}\hat D(\nu)\\
        &=\sum_{p=0}^n\sum_{u,v=0}^p\frac{(-1)^{u+v}}{p!}\binom np\binom pu\binom pv\nu^v\nu^{*u}\hat a^{\dag p-u}\hat a^{p-v}\\
        &\times\left(\sum_{k=0}^{p-u}\binom{p-u}k(-1)^k\right)\left(\sum_{l=0}^{p-v}\binom{p-v}l(-1)^l\right)\hat D(\nu).
    \end{aligned}
\end{equation}
where we have set $p=n-m$. Due to the alternating signs in the sums of index $k$ and $l$, all terms vanish in the above expression except when $u=v=p$. We finally obtain:
\begin{equation}
    \begin{aligned}                
        L_n(|\nu|^2)\hat D_\nu&=\left(\sum_{p=0}^n\frac1{p!}\binom np|\nu|^{2p}\right)\hat D(\nu)\\
        &=L_n(|\nu|^2)\hat D(\nu),
    \end{aligned}
\end{equation}
so that $D_\nu=\hat D(\nu)$ is a displacement operator.
Finally, we obtain
\begin{equation}
    \hat\Delta(\alpha)=\int_{\mathbb C}\frac{d^2\nu}\pi e^{\nu\alpha^*-\nu^*\alpha}\hat D^\dag(\nu),
\end{equation}
from Eq.~(\ref{eq:DeltaBrifMan}). In particular, this kernel and the corresponding Wigner function $W_{\hat\rho}(\alpha)=\Tr[\hat\Delta(\alpha)\hat\rho]$ do not depend on the choice of $n$, and matches with $W_{\hat\rho}^\mathrm{CV}(\alpha)$ in Eq.~(\ref{eq:CVDV_char_wig_def}) (setting $\alpha=\frac{x+ip}{\sqrt2}$).


\section{Wigner function from GKP code space}
\label{app:BrifManGKP}

In this section, we show that the general phase-space construction in Section \ref{sec:WignerQcodes} leads to Definition~\ref{def:ZGWigner} of the Zak--Gross Wigner function in the case of the GKP code space, independently of the choice of reference state.

As in the previous section, we consider a quantum system described by an infinite-dimensional Hilbert space $\mathcal H$ with dynamical symmetry group $G=H_3(\mathbb C)$, the Heisenberg--Weyl group over $\mathbb C$. 
Recall that any element $g\in G$ can be parametrized as
\begin{equation}
    \pi(g)=e^{i\varphi\hat{\mathbb I}}e^{i\frac{xp}2\hat{\mathbb I}}e^{ip\hat x}e^{-ix\hat p}=e^{i\varphi}\hat D(x,p),
\end{equation}
for $\varphi,x,p\in\mathbb R$, i.e., as a displacement operator up to a global phase. Recall also the logical Pauli operators for GKP qudits in terms of displacement operators:
\begin{equation}
    \begin{aligned}
        \hat{\bar Z}&=e^{i\sqrt{\frac{2\pi}d}\hat x}=e^{i\ell\hat x}=\hat D\left(0,\ell\right),\\
        \hat{\bar X}&=e^{-i\sqrt{\frac{2\pi}d}\hat p}=e^{-i\ell\hat p}=\hat D\left(\ell,0\right).
    \end{aligned}
\end{equation}
We now  choose a reference state $\ket{\psi_0}\in\mathcal C_\mathrm{GKP}$ with the same isotropy subgroup $H^{\mathcal C_\mathrm{GKP}}=U(1)\times\mathbb Z\times\mathbb Z$ as the code space.
\begin{itemize}
\item In this case, the phase space is the continuous torus: $X(H_3,\ket{\mathrm{GKP}})=H_3/(U(1)\times\mathbb Z\times\mathbb Z)\simeq\mathbb T^2_{d\ell}=[0,d\ell)\times[0,d\ell)$.
    \item The coherent states are the displaced GKP states $\ket{u,v}_{\psi_0}:=\hat D(u,v)\ket{\psi_0}$ for $(u,v)\in\mathbb T^2_{d\ell}$.
    \item The invariant measure over $\mathbb T^2_{d\ell}$ is $d\mu(u,v)=\frac1ddudv$.
    \item The harmonic functions over $L^2(X,\mu)$ are given by $Y_{mn}(u,v)=\frac1{\sqrt{2\pi}}e^{i\ell(nv-mu)}$, for all $(m,n)\in\mathbb Z^2$ and all $(u,v)\in\mathbb T^2_{d\ell}$. They satisfy $\int_{\mathbb T^2_{d\ell}}\frac1ddudv\,Y_{mn}(u,v)Y^*_{m'n'}(u,v)=\delta_{mm'}\delta_{nn'}$, the (Kronecker) delta function over the spectrum of the Laplace--Beltrami operator, and $\frac1d\sum_{m,n\in\mathbb Z}Y_{mn}(u,v)Y^*_{mn}(u',v')=\delta(u-u')\delta(v-v')$, with the Plancherel measure being $\frac1d$ times the counting measure over $\mathbb Z^2$.
\end{itemize}
With this preliminary material in place, we are ready to determine the real and positive coefficients $\tau_\nu$ and the tensor operators $\hat D_\nu$ via Eqs.~(\ref{eq:cohOvercode}) and (\ref{eq:tensorOpcode}) for all $\nu=(m,n)\in\mathbb Z^2$, which in this case read
\begin{align}
    \label{eq:taunuGKP}|\langle\psi_0|\hat D^\dag(u,v)\hat D(u',v')|\psi_0\rangle|^2&=\frac1 d\sum_{m,n\in\mathbb Z}\tau_{mn}\frac1{2\pi}e^{i\ell[n(v'-v)-m(u'-u)]}\\
    \label{eq:DnuGKP}\sqrt{2\pi}e^{i\varphi_{mn}}\tau_{mn}^{1/2}\hat D_{mn}&=\int_{\mathbb T^2_{d\ell}}\frac1ddudv\,e^{i\ell(nv-mu)}\hat D(u,v)\ket{\psi_0}\!\bra{\psi_0}\hat D^\dag(u,v).
\end{align}
The reference state can be written as a superposition of GKP qudit basis states: 
\begin{equation}
    \ket{\psi_0}=\sum_{j=0}^{d-1}c_j|\bar j\rangle,
\end{equation}
where $|\bar j\rangle=\frac1{\sqrt\ell}\sum_{n\in\mathbb Z}\ket{\ell(nd+j)}_{\hat x}$. We will specify conditions on the amplitudes $c_j$ later on.
Setting $u'=v'=0$ in Eq.~(\ref{eq:taunuGKP}) yields $|\langle\psi_0|\hat D^\dag(u,v)|\psi_0\rangle|^2=|\langle\psi_0|\hat D(u,v)|\psi_0\rangle|^2=\frac1{2\pi d}\sum_{m,n\in\mathbb Z}\tau_{mn}e^{i\ell(nv-mu)}$. Since $\langle\psi_0|\hat D(u,v)|\psi_0\rangle=0$ whenever $u\notin\ell\mathbb Z$ or $v\notin\ell\mathbb Z$, taking the inverse Fourier transform yields
\begin{equation}
    \begin{aligned}
        \tau_{mn}&=\int_{\mathbb T_{d\ell}^2}dudv\,|\langle\psi_0|\hat D(u,v)|\psi_0\rangle|^2e^{i\ell(nv-mu)}\\
        &=\ell^2\sum_{k,l=0}^{d-1}|\langle\psi_0|\hat D(k\ell,l\ell)|\psi_0\rangle|^2e^{i\ell^2(nl-mk)}\\
        &=\frac{2\pi}{d}\sum_{k,l=0}^{d-1}|\langle\psi_0|\hat D(k\ell,l\ell)|\psi_0\rangle|^2\omega^{nl-mk},
    \end{aligned}
\end{equation}
for all $(m,n)\in\mathbb Z^2$, where we used $\delta(\frac u\ell-k)=\ell\delta(u-k\ell)$ and $\delta(\frac v\ell-l)=\ell\delta(v-kl)$ in the second line, and where we have set $\omega=e^{\frac{2\pi i}d}$ in the last line. Now $|\langle\psi_0|\hat D(k\ell,l\ell)|\psi_0\rangle|^2=|\sum_{j=0}^{d-1}c_{j\oplus_dk}^*c_j\omega^{lj}|^2$, where $\oplus_d$ denotes addition modulo $d$, so
\begin{equation}
    \tau_{mn}=\frac{2\pi}{d}\sum_{k,l,j,j'=0}^{d-1}c_{j'\oplus_dk}c_{j'}^*c_{j\oplus_dk}^*c_j\omega^{l(j-j')}\omega^{nl-mk}.
\end{equation}
Summing over $l$, the non-vanishing terms satisfy $n+j-j'=0\mod d$ and we get
\begin{equation}\label{eq:exprtauGKP}
    \begin{aligned}
        \tau_{mn}&=2\pi\sum_{k,j=0}^{d-1}c_{j\oplus_dn\oplus_dk}c_{j\oplus_dn}^*c_{j\oplus_dk}^*c_j\omega^{-mk}\\
        &=2\pi\sum_{j,j'=0}^{d-1}c_{j'\oplus_dn}c_{j\oplus_dn}^*c_{j'}^*c_j\omega^{-m(j'-j)}\\
        &=2\pi\left|\sum_{j=0}^{d-1}c_jc_{j\oplus_dn}^*\omega^{mj}\right|^2\\
        &=2\pi|\langle\psi_0|\hat D(n\ell,m\ell)|\psi_0\rangle|^2,
    \end{aligned}
\end{equation}
where we have set $j'=j\oplus_dk$ in the second line. In what follows, we assume that the amplitudes $c_j$ are such that this quantity is non-vanishing.

With the explicit form of the coefficients $\tau_{mn}$ determined for all $m,n\in\mathbb Z$, we now compute Eq.~(\ref{eq:DnuGKP}):
\begin{equation}
    \begin{aligned}                        
        \sqrt{2\pi}e^{i\varphi_{mn}}\tau_{mn}^{1/2}\hat D_{mn}&=\int_{\mathbb T^2_{d\ell}}\frac1ddudv\,e^{i\ell(nv-mu)}\hat D(u,v)\ket{\psi_0}\!\bra{\psi_0}\hat D^\dag(u,v)\\
        &=\sum_{j,j'=0}^{d-1}c_jc_{j'}^*\sum_{k,k'\in\mathbb Z}\int_{\mathbb T^2_{d\ell}}\frac1{d\ell}dudv\,e^{i\ell(nv-mu)}e^{-iu\hat p}e^{iv\hat x}\,\!_{\hat x}\!\ket{\ell(kd+j)}\!\bra{\ell(k'd+j')}_{\hat x}e^{-iv\hat x}e^{iu\hat p}\\
        &=\frac1{d\ell}\sum_{j,j'=0}^{d-1}c_jc_{j'}^*\sum_{k,k'\in\mathbb Z}\int_{\mathbb T_{d\ell}}du e^{-i\ell mu}\left(\int_{\mathbb T_{d\ell}}\!\!\!dve^{i\ell v(n+kd+j-k'd-j')}\right)\!_{\hat x}\!\ket{\ell(kd+j)+u}\!\bra{\ell(k'd+j')+u}_{\hat x}\\
        &=\sum_{j,j'=0}^{d-1}c_jc_{j'}^*\sum_{k,k'\in\mathbb Z}\int_{\mathbb T_{d\ell}}du e^{-i\ell mu}\delta_{n+kd+j-k'd-j'}\,\!_{\hat x}\!\ket{\ell(kd+j)+u}\!\bra{\ell(k'd+j')+u}_{\hat x}.
    \end{aligned}
\end{equation}
The Kronecker delta implies $k'd=n+kd+j-j'$ and $j'=j\oplus_dn$. Hence,
\begin{equation}
    \begin{aligned}                        
        \sqrt{2\pi}e^{i\varphi_{mn}}\tau_{mn}^{1/2}\hat D_{mn}&=\sum_{j=0}^{d-1}c_jc_{j\oplus_dn}^*\sum_{k\in\mathbb Z}\int_{\mathbb T_{d\ell}}du e^{-i\ell mu}\,\!_{\hat x}\!\ket{\ell(kd+j)+u}\!\bra{\ell(n+kd+j)+u}_{\hat x}\\
        &=\hat{\bar Z}^{-m}\left(\sum_{j=0}^{d-1}c_jc_{j\oplus_dn}^*\omega^{mj}\sum_{k\in\mathbb Z}\int_{\mathbb T_{d\ell}}du\,\!_{\hat x}\!\ket{\ell(kd+j)+u}\!\bra{\ell(kd+j)+u}_{\hat x}\right)\hat{\bar X}^{-n},
    \end{aligned}
\end{equation}
where we used $e^{i\ell^2mkd}e^{i\ell^2mj}=\omega^{mj}$ in the last line.
With the change of variable $u'=u+kd\ell$ we obtain
\begin{equation}\label{eq:cell_resolution}
    \begin{aligned}                        
        \sum_{k\in\mathbb Z}\int_{\mathbb T_{d\ell}}du\,\!_{\hat x}\!\ket{\ell(kd+j)+u}\!\bra{\ell(kd+j)+u}_{\hat x}&=\sum_{k\in\mathbb Z}\int_{kd\ell}^{(k+1)d\ell}du'\,\!_{\hat x}\!\ket{u'+\ell j}\!\bra{u'+\ell j}_{\hat x}\\
        &=\int_{u'\in\mathbb R}du'\,\!_{\hat x}\!\ket{u'+\ell j}\!\bra{u'+\ell j}_{\hat x}\\
        &=\int_{u''\in\mathbb R}du''\,\!_{\hat x}\!\ket{u''}\!\bra{u''}_{\hat x}\\
        &=\hat{\mathbb I}.
    \end{aligned}
\end{equation}
We thus have
\begin{equation}                     
    \begin{aligned} 
            \sqrt{2\pi}e^{i\varphi_{mn}}\tau_{mn}^{1/2}\hat D_{mn}&=\left(\sum_{j=0}^{d-1}c_jc_{j\oplus_dn}^*\omega^{mj}\right)(\hat{\bar X}^n\hat{\bar Z}^m)^\dag\\
            &=\langle\psi_0|\hat D(n\ell,m\ell)|\psi_0\rangle(\omega^{2^{-1}mn}\hat{\bar X}^n\hat{\bar Z}^m)^\dag.
    \end{aligned}
\end{equation}
We now set $\varphi_{mn}=\mathrm{arg}(\langle\psi_0|\hat D(n\ell,m\ell)|\psi_0\rangle)$. Note that other choices for the phase function leading to Wigner functions that are independent of the choice of reference state are possible. For instance, setting $\varphi_{mn}=\pi mn+\mathrm{arg}(\langle\psi_0|\hat D(n\ell,m\ell)|\psi_0\rangle)$ leads to a Wigner function compatible with the DV toroidal Wigner function from \cite{ligabo2016torus}, whereas our choice leads to a Wigner function compatible with the Gross Wigner function \cite{gross2006hudson}. With Eq.~(\ref{eq:exprtauGKP}) this finally implies
\begin{equation}
    \begin{aligned}
        \hat D_{mn}^\dag&=\frac1{2\pi}\omega^{2^{-1}mn}\hat{\bar X}^n\hat{\bar Z}^m\\
        &=\frac1{2\pi}(-1)^{kl}\hat{D}(n\ell,m\ell).
    \end{aligned}
\end{equation}
The corresponding SW kernel is given by
\begin{align}
    \hat\Delta^{\mathcal C_\mathrm{GKP}}(u,v)&=\frac1{2\pi}\sum_{m,n\in\mathbb Z} e^{i\ell(nv-mu)}\hat D_{mn}^\dag\\
    &=\frac1{2\pi}\sum_{m,n\in\mathbb Z}e^{i\ell(nv-mu)}\omega^{2^{-1}mn}\hat{\bar X}^n\hat{\bar Z}^m\\
    &=\frac1{2\pi}\sum_{m,n\in\mathbb Z}e^{i\ell(nv-mu)}(-1)^{mn}\hat{D}(n\ell,m\ell)\\
    &=\hat D(u,v)\hat\Delta^{\mathcal C_\mathrm{GKP}}(0,0)\hat D^\dag(u,v),
\end{align}
for all $(u,v)\in\mathbb T^2_{d\ell}$, with $\hat\Delta^{\mathcal C_\mathrm{GKP}}(0,0)=\frac1{2\pi}\sum_{m,n\in\mathbb Z}(-1)^{mn}\hat{D}(n\ell,m\ell)$. 
The result is independent of the choice of reference state $\ket{\psi_0}$ with isotropy subgroup $H^{\mathcal C_\mathrm{GKP}}$.


\section{Properties of the Zak--Gross Wigner function}
\label{app:SWGKP}

In this section, we prove the properties of the Zak--Gross Wigner function given in the main text.
We start by proving Lemma~\ref{lem:logicalparity} in section~\ref{app:prooflogicalparity}, then Theorem~\ref{th:collectionDV} in section~\ref{app:proofcollectionDV} and finally the Stratonovich--Weyl axioms from Theorem~\ref{th:SWGKP} in section~\ref{app:proofSWGKP}.


\subsection{Proof of Lemma~\ref{lem:logicalparity}: Zak--Gross kernel as a logical parity operator}
\label{app:prooflogicalparity}

Recall that the kernel associated with the code space $\mathcal C_\mathrm{GKP}$ is given by
\begin{equation}\label{app:GKP_Wigner_kernel_abstract_def}
    \hat\Delta^{\mathcal C_\mathrm{GKP}}(u,v)=\hat D (u, v)\hat \Delta^{\mathcal C_\mathrm{GKP}}(0,0)\hat D^\dagger(u, v),
\end{equation}
for all $(u,v)\in\mathbb{T}_{d\ell}^2$, with
\begin{equation}
    \begin{aligned}
        \hat\Delta^{\mathcal C_\mathrm{GKP}}(0,0)&=\frac1{{2\pi}}\sum_{m,n \in \Z} (-1)^{mn} \hat{D}(\ell n, \ell m)\\
        &= \frac1{{2\pi}}\sum_{m,n \in \Z} \omega^{2^{-1}mn} \hat{\bar{X}}^n \hat{\bar{Z}}^m,
    \end{aligned}
\end{equation}
In order to relate $\hat\Delta^{\mathcal C_\mathrm{GKP}}(0,0)$ to the parity operator on the code space $\mathcal C_\mathrm{GKP}$, we compute its action on the position basis:
\begin{equation}
    \begin{aligned}
        \hat\Delta^{\mathcal C_\mathrm{GKP}}(0,0)\ket{x}_{\hat x} &=\frac1{{2\pi}}\sum_{m,n \in \Z} \omega^{2^{-1}mn} \hat{\bar{X}}^n \hat{\bar{Z}}^m \ket{x}_{\hat x} \\
        &=\frac1{{2\pi}}\sum_{m,n \in \Z} \omega^{2^{-1}mn} e^{-in\ell \hat p} e^{im\ell \hat x} \ket{x}_{\hat x} \\
        &=\frac1{{2\pi}}\sum_{m,n \in \Z} \omega^{2^{-1}mn} e^{im\ell x} \ket{x + \ell n}_{\hat x} 
    \end{aligned}
\end{equation}
where $\ell = \sqrt{\frac{2\pi}{d}}$. Split the summations according to
\begin{equation}
    \begin{split}
        m &= b + dm' \\
        n &= a + dn'
    \end{split}
    \quad \text{where} \quad
    \begin{split}
        a, b \in \Z_d, \quad m',n' \in \Z.
    \end{split}
\end{equation}
Because $2^{-1} = \left( \frac{d+1}{2} \right) \in \Z$ the phase factor reduces from $\omega^{2^{-1}nm}$ to $\omega^{2^{-1}ab}$, yielding
\begin{equation}
    \frac1{{2\pi}}\sum_{a,b \in \Z_d} \sum_{m',n' \in \Z} \omega^{2^{-1}ab} e^{i(b + dm')\ell x} \ket{x + \ell a + d\ell n'}_{\hat x}.
\end{equation}
Isolate the $m'$-sum and apply Poisson summation:
\begin{equation}
    \sum_{m'\in \Z} e^{idm'\ell x} = \sum_{m'} \delta\left(\frac{d\ell}{2\pi}x - m'\right) = \sum_{m'\in \Z} \delta\left(\frac{x}{\ell} - m' \right)=\ell\sum_{m'\in \Z}\delta(x - m'\ell ).
\end{equation}
To get a non-vanishing result it must be that $x \in \ell\Z \subset \R$.  Let this be so and denote $x = x_L \ell + x_B d \ell$ for $x_L \in \Z_d$ the ``logical index''and $x_B \in \Z$ the ``bin index''.  Now we have
\begin{equation}
     \frac\ell{{2\pi}}\sum_{a,b \in \Z_d} \sum_{n'\in \Z} \omega^{2^{-1}ab} e^{ib\ell^2 (x_L + dx_B)} \ket{\ell (a + x_L) + d\ell (n' + x_B)}_{\hat x},
\end{equation}
but it is clear that $x_B$ is immaterial due to $e^{ib\ell^2 dx_B} = 1$ and the ability to shift the $n'$ sum. So drop $x_B$ and isolate the $b$ sum to get a sum-of-roots: 
\begin{equation}
    \sum_{b=0}^{d-1} \omega^{(2^{-1}a + x_L)b} = 
    \begin{cases}
    d, \quad 2^{-1}a\oplus_dx_L = 0, \\
    0, \quad \text{otherwise},
    \end{cases}
\end{equation}
where $\oplus_d$ denotes addition modulo $d$. Solve to get $a=-2x_L$ and finally obtain, with $\frac{d\ell}{{2\pi}}=\frac1\ell$,
\begin{equation}
    \hat\Delta^{\mathcal C_\mathrm{GKP}}(0,0)\ket{x}_{\hat x} = \begin{cases}
        0, & x \notin \ell \Z \\
        \frac1\ell\sum_{n'\in\Z}\ket{\ell(dn'-x_L)}_{\hat x}, & x \in \ell \Z,
    \end{cases},
\end{equation}
where in the second case $x = x_L \ell + d\ell x_B$ for some $x_L\in\Z_d$ and $x_B \in\Z$. In other words, the operator $\frac1{\ell}\hat\Delta^{\mathcal C_\mathrm{GKP}}(0,0)$ takes a position eigenstate supported on the logical position lattice $\ell \Z \subset \R$, finds the associated logical index $x_L$, then encodes the reflected state. This transformation on position states is equivalent to the one obtained by the encoded DV parity operator when expressed in the encoded computational basis:
\begin{equation}
    \hat{\bar{\Pi}}_d(0,0) = \sum_{j=0}^{d-1} \ket{-\bar j}\!\bra{\bar j},
\end{equation}
where $|\bar j\rangle=\frac1{\sqrt\ell}\sum_{n\in\mathbb Z}\ket{\ell(nd+j)}_{\hat x}$ are codewords \eqref{codewords_position}. Indeed,
\begin{equation}
    \begin{aligned}
        \hat{\bar{\Pi}}_d(0,0)\ket{x}_{\hat x}&=\sum_{j=0}^{d-1}|-\bar j\rangle\langle\bar j|x\rangle_{\hat x}\\
        &=\frac1{\sqrt\ell}\sum_{j=0}^{d-1}|-\bar j\rangle\sum_{n\in\mathbb Z}\,_{\hat x}\langle\ell(nd+j)|x\rangle_{\hat x}.
    \end{aligned}
\end{equation}
The inner product of position eigenstates gives $0$ if $x\notin\ell\Z$. Otherwise, we write as before $x = x_L \ell + d\ell x_B$ for some $x_L\in\Z_d$ and $x_B\in\Z$, and the only nonvanishing term above satisfies $n=x_B$ and $j=x_L$, leading to
\begin{equation}
    \hat{\bar{\Pi}}_d(0,0)\ket{x}_{\hat x} = \begin{cases}
        0, & x \notin \ell \Z \\
        \frac1\ell\sum_{n\in\Z}\ket{\ell(dn-x_L)}_{\hat x}, & x \in \ell \Z,
    \end{cases},
\end{equation}
where in the second case $x = x_L \ell + d\ell x_B$ for some $x_L\in\Z_d$ and $x_B \in\Z$. With $\ell=\sqrt{\frac{2\pi}d}$, this proves that $\hat\Delta^{\mathcal C_\mathrm{GKP}}(0,0)=\hat{\bar{\Pi}}_d(0,0)$ over $\mathcal H$ and concludes the proof of Lemma~\ref{lem:logicalparity}.


\subsection{Proof of Theorem~\ref{th:collectionDV}: relating the Zak--Gross and the Gross Wigner functions}
\label{app:proofcollectionDV}

A direct consequence of Lemma~\ref{lem:logicalparity} is
\begin{equation}\label{eq:kernelprojcode}
    \hat P_{(0,0)}\hat\Delta^{\mathcal C_\mathrm{GKP}}(0,0)\hat P_{(0,0)}=\hat\Delta^{\mathcal C_\mathrm{GKP}}(0,0),
\end{equation}
where $\hat P_{(0,0)}=\sum_{j=0}^{d-1}\ket{\bar j}\!\bra{\bar j}$ is the projector onto the code space $\mathcal C_\mathrm{GKP}$.

Given a density operator $\hat\rho$ over $\mathcal H$, recall the definition of the (unnormalised) error-corrected state $\hat\rho(s,t)=\hat P_{(0,0)}\hat D^\dag(s,t)\hat\rho\hat D(s,t)\hat P_{(0,0)}$, for all $(s,t)\in\mathbb T_\ell^2$, where $\ell=\sqrt{\frac{2\pi}d}$.

Let $(u,v)\in\mathbb T_{d\ell}^2$. Writing $u=a\ell+s$ and $v=b\ell+t$, for $(a,b)\in\mathbb Z_d^2$ and $(s,t)\in\mathbb T_\ell^2$ we have
\begin{equation}
    \begin{aligned}
        W_{\hat\rho}^{\mathcal C_\mathrm{GKP}}(u,v)&=\Tr[\hat D(u,v)\hat\Delta^{\mathcal C_\mathrm{GKP}}(0,0)\hat D^\dag(u,v)\hat\rho]\\
        &=\Tr[\hat D(a\ell+s,b\ell+t)\hat\Delta^{\mathcal C_\mathrm{GKP}}(0,0)\hat D^\dag(a\ell+s,b\ell+t)\hat\rho]\\
        &=\Tr[\hat D(s,t)\hat D(a\ell,b\ell)\hat\Delta^{\mathcal C_\mathrm{GKP}}(0,0)\hat D^\dag(a\ell,b\ell)\hat D^\dag(s,t)\hat\rho]\\
        &=\Tr[\hat D(a\ell,b\ell)\hat P_{(0,0)}\hat\Delta^{\mathcal C_\mathrm{GKP}}(0,0)\hat P_{(0,0)}\hat D^\dag(a\ell,b\ell)\hat D^\dag(s,t)\hat\rho\hat D(s,t)]\\
        &=\Tr[\hat D(a\ell,b\ell)\hat\Delta^{\mathcal C_\mathrm{GKP}}(0,0)\hat D^\dag(a\ell,b\ell)\hat P_{(0,0)}\hat D^\dag(s,t)\hat\rho\hat D(s,t)\hat P_{(0,0)}]\\
        &=\Tr[\hat D(a\ell,b\ell)\hat{\bar{\Pi}}_d(0,0)\hat D^\dag(a\ell,b\ell)\hat\rho(s,t)]\\
        &=\Tr[\hat{\bar{\Pi}}_d(a,b)\hat\rho(s,t)]\\
        &=\Tr[\hat{\Pi}_d(a,b)\hat{\underline\rho}(s,t)]\\
        &=W_{\hat{\underline\rho}(s,t)}^{\mathrm{DV}}(a,b),
    \end{aligned}
\end{equation}
where we have used Eq.~(\ref{eq:kernelprojcode}) in the 4th line, Lemma~\ref{lem:logicalparity} in the 5th line, and identified the codeword $\hat\rho(s,t)$ with its corresponding DV logical state $\hat{\underline\rho}(s,t)$ in the 8th line, together with the definition of the DV Wigner function \eqref{eq:CVDV_char_wig_def}.


\subsection{Proof of Theorem \ref{th:SWGKP}: Stratonovich--Weyl axioms for the Zak--Gross Wigner function}
\label{app:proofSWGKP}

We now turn to the proof of the SW axioms \ref{enum:SWlinGKP}-\ref{enum:SWtrGKP} for the Zak--Gross Wigner function associated with the code space $\mathcal C_\mathrm{GKP}$. We recall the expression of the twirling map
\begin{equation}
    \mathcal E(\hat A)= \int_{\mathbb{T}_\ell^2}dsdt\hat{P}_{(s,t)}\hat\rho\hat{P}_{(s,t)},
\end{equation}
where $\hat{P}_{(s,t)}$ is the projector onto a displaced code space defined in Eq.~(\ref{eq:dispcodeproj}).

\medskip

\ref{enum:SWlinGKP}: 
The linearity part is trivial. 
Let $\hat A$ and $\hat B$ be operators over $\mathcal H$ such that $W_{\hat A}^{\mathcal C_\mathrm{GKP}}(u,v)=W_{\hat B}^{\mathcal C_\mathrm{GKP}}(u,v)$ for all $(u,v)\in\mathbb T_{d\ell}^2$. By Theorem~\ref{th:collectionDV} and the fact that the Gross Wigner function is one-to-one, this implies
$\hat A(s,t)=\hat B(s,t)$ for all $(s,t)\in\mathbb T_\ell^2$, and thus $\mathcal E(\hat A)=\mathcal E(\hat B)$.
Reciprocally, assume that $\mathcal E(\hat A)=\mathcal E(\hat B)$. For all $(s,t)\in\mathbb T_\ell^2$ we have $\hat{P}_{(s,t)}\mathcal E(\hat A)\hat{P}_{(s,t)}=\hat{P}_{(s,t)}\mathcal E(\hat B)\hat{P}_{(s,t)}$. Since $\hat{P}_{(s,t)}$ are orthogonal projectors, $\hat{P}_{(s,t)}\mathcal E(\hat A)\hat{P}_{(s,t)}=\hat{P}_{(s,t)}\hat A\hat{P}_{(s,t)}$ and similarly for $\hat B$. Hence, $\hat{P}_{(s,t)}\hat A\hat{P}_{(s,t)}=\hat{P}_{(s,t)}\hat B\hat{P}_{(s,t)}$ and thus $\hat A(s,t)=\hat B(s,t)$ for all $(s,t)\in\mathbb T_\ell^2$. With Theorem~\ref{th:collectionDV} this implies $W_{\hat A}^{\mathcal C_\mathrm{GKP}}(u,v)=W_{\hat B}^{\mathcal C_\mathrm{GKP}}(u,v)$ for all $(u,v)\in\mathbb T_{d\ell}^2$.

\medskip

\ref{enum:SWrealGKP}: This is a direct consequence of Definition~\ref{def:ZGWigner} and, e.g., Lemma~\ref{lem:logicalparity}.

\medskip

\ref{enum:SWstGKP}: By Lemma~\ref{lem:logicalparity}, $\Tr[\hat\Delta^{\mathcal C_\mathrm{GKP}}(0,0)]=1$, so $W_{\hat{\mathbb I}}^{\mathcal C_\mathrm{GKP}}(u,v)=1$ for all $(u,v)\in\mathbb T_{d\ell}^2$. This property then follows from \ref{enum:SWtrGKP} with $\hat B=\hat{\mathbb I}$.

\medskip

\ref{enum:SWcovGKP}: Consider a continuous displacement $\hat D (x,p)$. Covariance under these transformations follows essentially by construction:  for all $(u,v)\in\mathbb T_{d\ell}^2$,
\begin{equation}
\begin{aligned}
    \hat D^\dagger(x,p)\hat\Delta^{\mathcal C_\mathrm{GKP}}(u,v)\hat D(x,p) &= \frac1{{2\pi}}\sum_{m,n \in \Z} (-1)^{mn} e^{i\ell (nv - mu)} \hat D(-x,-p) \hat D(\ell n,\ell m) \hat D^\dagger(-x,-p) \\
    &=\frac1{{2\pi}}\sum_{m,n \in \Z} (-1)^{mn} e^{i\ell (nv - mu)} e^{-i( -x\ell m + p\ell n )}  \hat D(\ell n,\ell m) \\
    &= \frac1{{2\pi}}\sum_{m,n \in \Z} (-1)^{mn} e^{i\ell (n(v-p) - m(u-x))}  \hat D(\ell n,\ell m) \\
    &= \hat \Delta^{\mathcal C_\mathrm{GKP}}(u - x, v - p).
\end{aligned}
\end{equation}
where we used the commutation relations \eqref{eq:CVDV_displacement_conjugation} and where arithmetic operations in the last line are modulo $d\ell$. Hence for the Wigner function,
\begin{equation}
\begin{aligned}
    W^{\mathcal C_\mathrm{GKP}}_{ \hat D(x,p) {\hat\rho} \hat D^\dagger(x,p)}(u,v) &= \Tr[ \hat D(x,p) {\hat\rho} \hat D^\dagger(x,p) \hat \Delta^{\mathcal C_\mathrm{GKP}}(u,v)] \\
    &= \Tr[ {\hat\rho} \hat D^\dagger(x,p) \hat \Delta^{\mathcal C_\mathrm{GKP}}(u,v) \hat D(x,p)] \\
    &= \Tr[ {\hat\rho} \hat \Delta^{\mathcal C_\mathrm{GKP}}(u-x,v-p)] \\
    &= W^{\mathcal C_\mathrm{GKP}}_{ {\hat\rho} }(u-x,v-p),
\end{aligned}
\end{equation}
where arithmetic operations in the last line are modulo $d\ell$.

In particular, the logical displacements $\hat{\bar{D}}(a,b)$ \eqref{encoded_DV_dis_via_CV_dis} give 
\begin{equation}\label{logical_wigner_covariance}
    W^{\mathcal C_\mathrm{GKP}}_{ \hat{\bar{D}}(a,b) {\hat\rho} \hat{\bar{D}}^\dagger(a,b)}(u,v) =  W^{\mathcal C_\mathrm{GKP}}_{\hat\rho}(u - a\ell, v - b\ell),
\end{equation}
for all $(a,b)\in\Z_d$, while the stabilizer actions are trivial,
\begin{equation}\label{trivial_stabilizer_action}
    W^{\mathcal C_\mathrm{GKP}}_{ \hat{\bar{D}}(a,b) {\hat\rho} \hat{\bar{D}}^\dagger(a,b)}(u,v) =  W^{\mathcal C_\mathrm{GKP}}_{\hat\rho}(u,v).
\end{equation}

\medskip

\ref{enum:SWtrGKP}: Firstly, note that $\Tr[\mathcal E(\hat A)\mathcal E(\hat B)]=\Tr[\mathcal E(\hat A)\hat B]=\Tr[\hat A\mathcal E(\hat B)]$ is a direct consequence of the fact that $\hat{P}_{(s,t)}$ are orthogonal projectors for all $\mathbb{T}_\ell^2$, together with the linearity and cyclicity of the trace. Now we compute
\begin{equation}
    \begin{aligned}
        \int_{\mathbb{T}^2_{d\ell}}\frac1ddudv\,W^{\mathcal C_\mathrm{GKP}}_{\hat A}(u,v)\hat{\Delta}^{\mathcal C_\mathrm{GKP}}(u,v)&=\frac1d\int_{\mathbb{T}^2_{\ell}}dsdt \sum_{a,b\in\Z_d}W^\mathrm{DV}_{\hat{\underline A}(s,t)}(a,b)\hat D(a\ell+s,b\ell+t)\hat{\Delta}^{\mathcal C_\mathrm{GKP}}(0,0)\hat D^\dag(a\ell+s,b\ell+t)\\
        &=\int_{\mathbb{T}^2_{\ell}}dsdt\hat D(s,t)\left[\frac1d\sum_{a,b\in\Z_d}W^\mathrm{DV}_{\hat{\underline A}(s,t)}(a,b)\hat{\bar{\Pi}}_d(a,b)\right]\hat D^\dag(s,t),
    \end{aligned}
\end{equation}
where we used Lemma~\ref{lem:logicalparity} in the second line.
Applying the GKP-encoding map to the DV reconstruction formula in Eq.~(\ref{eq:reconstruction}) yields
\begin{equation}
    \frac1d\sum_{a,b\in\Z_d}W^\mathrm{DV}_{\hat{\underline A}(s,t)}(a,b)\hat{\bar{\Pi}}_d(a,b)=\hat{ A}(s,t),
\end{equation}
for all $(s,t)\in\mathbb T_\ell^2$, so the previous equation rewrites as
\begin{equation}
    \begin{aligned}
        \int_{\mathbb{T}^2_{d\ell}}\frac1ddudv\,W^{\mathcal C_\mathrm{GKP}}_{\hat A}(u,v)\hat{\Delta}^{\mathcal C_\mathrm{GKP}}(u,v)&=\int_{\mathbb{T}^2_{\ell}}dsdt\hat D(s,t)\hat A(s,t)\hat D^\dag(s,t)\\
        &=\int_{\mathbb{T}^2_{\ell}}dsdt\hat D(s,t)\hat P(0,0)\hat D^\dag(s,t)\hat A\hat D(s,t)\hat P(0,0)\hat D^\dag(s,t)\\
        &=\mathcal E(\hat A).
    \end{aligned}
\end{equation}
This shows Eq.~(\ref{eq:mapE}) in the main text and concludes the proof by linearity using Definition~\ref{def:ZGWigner}.

\section{Zak measurement via characteristic function}
\label{appendix_gkp_dim1}

The Zak transform of a pure state $\psi$ is a quasi-periodic function on $\R^2$ \eqref{Zak_quasi-periodicity}, however the ``cross-Zak transform'' of two states $\psi$ and $\phi$, 
\begin{equation}
    [Z_\alpha\psi]^*(k,q) [Z_\alpha\phi](k,q),
\end{equation}
is exactly periodic on the rectangle $[0,\alpha)\times [0, \frac{2\pi}{\alpha})$.  This includes in particular the Zak measurement distribution $|[Z_\alpha\psi](k,q)|^2$.  Hence the Fourier series can be computed,
\begin{equation}
    |[Z_\alpha\psi](k,q)|^2 = \sum_{n,m} c_{nm} e^{i2\pi \frac{k}{(\frac{2\pi}{\alpha})} n} e^{i2\pi \frac{q}{\alpha} m},
\end{equation}
with coefficients
\begin{equation}\label{zak_squared_fourier_coeffs}
    c_{nm} = \frac{1}{2\pi} \int_0^\alpha dq \int_0^{\frac{2\pi}{\alpha}} dk |[Z_\alpha\psi](k,q)|^2 e^{-i2\pi \frac{k}{(\frac{2\pi}{\alpha})} n} e^{-i2\pi \frac{q}{\alpha} m}.
\end{equation}
On the other hand it follows from the quasi-periodicity relations \eqref{Zak_quasi-periodicity} that the following holds for phase-space displacements along the rectangle,
\begin{equation}
    [Z_\alpha e^{i \frac{2\pi}{\alpha} \hat x} e^{-i \alpha \hat p}\psi](k,q) = e^{i \frac{2\pi}{\alpha} q} e^{-i\alpha k} [Z_\alpha\psi](k,q).
\end{equation}
This leads to
\begin{equation}
    \begin{aligned}
        |[Z_\alpha\psi](k,q)|^2 e^{-i2\pi \frac{k}{(\frac{2\pi}{\alpha})} n} e^{-i2\pi \frac{q}{\alpha} m}&=  [Z_\alpha\psi]^*(k,q) [Z\psi](k,q) e^{-i \alpha k n} e^{-i\frac{2\pi}{\alpha} q m} \\
        &= [Z_\alpha\psi]^*(k,q) [Z e^{-i \frac{2\pi}{\alpha} m \hat x} e^{-i \alpha n \hat p} \psi](k,q),
    \end{aligned}
\end{equation}
making the coefficients
\begin{equation}
    c_{nm} = \frac{1}{2\pi} \int_0^\alpha dq \int_0^{\frac{2\pi}{\alpha}} dk [Z_\alpha\psi]^*(k,q) [Z_\alpha e^{-i \frac{2\pi}{\alpha} m \hat x} e^{-i \alpha n \hat p} \psi](k,q).
\end{equation}
This is the inner product in the Zak representation,
\begin{equation}
    c_{nm} = \frac{1}{2\pi} ( Z\psi, Z e^{-i \frac{2\pi}{\alpha} m \hat x} e^{-i \alpha n \hat p} \psi )_{L^2(\mathbb{T}^2)},
\end{equation}
which via \eqref{zak_inner_iso} can be related to the $L^2(\mathbb{R})$ inner product,
\begin{equation}
    c_{nm} = \frac{1}{2\pi} ( \psi, e^{-i \frac{2\pi}{\alpha} m \hat x} e^{-i \alpha n \hat p} \psi )_{L^2(\R)}.
\end{equation}
Using the relations for the CV displacement operator \eqref{eq:CVDV_displacement_operators} and the CV characteristic function \eqref{eq:CVDV_char_wig_def} the coefficients further become
\begin{equation}
\begin{aligned}
    c_{nm} &= \frac{1}{2\pi} (-1)^{nm} ( \psi, \hat D(\alpha n, - \frac{2\pi}{\alpha}m) \psi )_{L^2(\R^2)} = \frac{1}{2\pi} (-1)^{nm} \chi_\psi (\alpha n, - \frac{2\pi}{\alpha}m).
\end{aligned} 
\end{equation}
The Zak probability distribution $|[Z_\alpha\psi](k,q)|^2$ becomes 
\begin{equation}
    \begin{aligned}
        &\quad \frac{1}{2\pi} \sum_{n,m} (-1)^{nm} \chi_\psi (\alpha n, - \frac{2\pi}{\alpha}m) e^{i\alpha k n} e^{i2\pi \frac{q}{\alpha} m} = \frac{1}{2\pi} \sum_{n,m} (-1)^{nm} \chi_\psi (\alpha n, \frac{2\pi}{\alpha}m) e^{i\alpha k n} e^{-i2\pi \frac{q}{\alpha} m}.
    \end{aligned}
\end{equation}

\section{Double discrete marginal is equivalent to syndrome measurement distribution}
\label{appendix_double_marginalization}

The double discrete marginal of the kernel is
\begin{equation}
\begin{aligned}
    \frac1d\sum_{j,k=0}^{d-1} \hat \Delta^{\mathcal{C}_{\text{GKP}}}(u-j\ell,v-k\ell) &= \frac1{2\pi d}\sum_{j,k=0}^{d-1} \sum_{m,n \in \Z} (-1)^{nm} e^{i\ell(nv - mu)} e^{-i\ell^2 j n} e^{-i\ell^2 k m} \hat D(\ell n, \ell m) \\
    &=\frac1{2\pi d}\sum_{m,n \in \Z} (-1)^{nm} e^{i\ell(nv - mu)} \left( \sum_{j=0}^{d-1} \omega^{- j n} \right) \left( \sum_{k=0}^{d-1} \omega^{- k m} \right) \hat D(\ell n, \ell m) \\
    &=\frac{d}{2\pi}\sum_{m',n' \in \Z} (-1)^{d^2 n'm'} e^{id\ell(n'v - m'u)} \hat D(d\ell n', d\ell m') \\
    &= \frac{d}{2\pi}\sum_{n',m' \in \Z} (-1)^{n'm'} e^{id\ell(n'v - m'u)} e^{i \frac{d^2 \ell^2 n'm'}{2}} \hat{\bar{X}}^{d n'} \hat{\bar{Z}}^{dm'} \\
    &= \frac{d}{2\pi} \sum_{m',n' \in \Z} e^{id\ell(n'v - m'u)} \hat{\bar{X}}^{d n'} \hat{\bar{Z}}^{dm'}.
\end{aligned}
\end{equation}
Because our kernel and the code space projectors are related by the same displacements, it is sufficient to compare the undisplaced case
\begin{equation}\label{eq:undisplaced_double_marg_kernel}
    \sum_{j,k=0}^{d-1} \hat \Delta^{\mathcal{C}_{\text{GKP}}}(-j\ell,-k\ell) = \frac{d}{2\pi} \sum_{m,n \in \Z} \hat{\bar{X}}^{d n} \hat{\bar{Z}}^{dm} = \frac{d}{2\pi} \sum_{m,n \in \Z} e^{-i d\ell n \hat p} e^{i d\ell m \hat x}
\end{equation}
with the undisplaced code space projector $\hat P_{(0,0)} = \sum_{j=0}^{d-1} \ketbra{\bar j}{\bar j}$. Similar to the derivation in section \ref{app:prooflogicalparity}, we obtain
\begin{equation}
    \frac1d\sum_{j,k=0}^{d-1} \hat \Delta^{\mathcal{C}_{\text{GKP}}}(-j\ell,-k\ell) | x \rangle = \begin{cases}
        \frac1\ell\sum_{n\in\mathbb Z}\ket{nd\ell+x}, & x \in \ell\Z \\
        0, & \text{otherwise},
    \end{cases}
\end{equation}
which is identical to the action of $\hat P_{(0,0)}$ in position basis.

\section{Derivation of the Zak--Gross Wigner function for thermal states}
\label{appendix_thermal}

We seek to compute 
\begin{equation}
    W_{\hat \rho}^{\mathcal{C}_{\text{GKP}}}(u,v) = \Tr[ \hat \Delta^{\mathcal{C}_{\text{GKP}}}(u,v) \hat \rho ] = \frac{1}{Z(\beta)}\Tr[ \hat \Delta^{\mathcal{C}_{\text{GKP}}}(u,v) e^{-\beta \hat H} ],
\end{equation}
where $\hat H = \frac{1}{2}(\hat x^2 + \hat p^2)$ is the oscillator Hamiltonian and $Z(\beta) = \frac{1}{2 \sinh(\frac{\beta}{2})}$ is the partition function \cite{Feynman_1998}.  We proceed entirely in the position basis 
\begin{equation}\label{eq:thermal_to_solve}
    W_{\hat \rho}^{\mathcal{C}_{\text{GKP}}}(u,v) = \int dx' \langle x' | \bar\Delta_d(u,v) \frac{e^{-\beta \hat{H}}}{Z(\beta)} | x' \rangle = \int \int dx' dx \langle x' | \hat \Delta^{\mathcal{C}_{\text{GKP}}}(u,v) | x \rangle \langle x | \frac{e^{-\beta \hat{H}}}{Z(\beta)} | x' \rangle,
\end{equation}
where for this appendix section we will drop the $\hat x$ subscript in the position kets to reduce clutter.  Using the Mehler kernel,
\begin{equation}
    \langle x | e^{-\beta \hat{H}} | x' \rangle = \frac{1}{\sqrt{2\pi \sinh(\beta)}} e^{ - \frac{1}{2\sinh(\beta)} [ (x^2 + x'^2)\cosh(\beta) - 2xx' ] },
\end{equation}
the thermal state matrix elements can be brought to the form
\begin{equation}
    \langle x | \frac{e^{-\beta \hat{H}}}{Z(\beta)} | x' \rangle = \frac{1}{\sqrt{\pi}} \sqrt{\tanh(\frac{\beta}{2})} e^{ -\frac{1}{4}[ (x+x')^2 \tanh(\frac{\beta}{2}) + (x - x')^2 \coth(\frac{\beta}{2}) ] }.
\end{equation}
The matrix elements of the kernel are
\begin{equation}
\begin{aligned}
    \langle x' | \hat \Delta^{\mathcal{C}_{\text{GKP}}}(u,v)  | x \rangle &= \frac1{2\pi} \sum_{n,m \in\mathbb Z} \omega_d^{2^{-1}nm} e^{i\ell(nv-mu)} \langle x' | \hat{\bar{X}}^n \hat{\bar{Z}}^m | x \rangle \\ 
    &= \frac1{2\pi} \sum_{n,m\in\mathbb Z}\omega_d^{2^{-1}nm} e^{i\ell(nv - mu)} e^{im\ell x} \delta_{x', x + n\ell }.
\end{aligned}
\end{equation}
Plug the above into \eqref{eq:thermal_to_solve} and use the delta $\delta_{x', x + n\ell }$ to remove the $x'$ integral,
\begin{equation}
    W_{\hat \rho}^{\mathcal{C}_{\text{GKP}}}(u,v) =  \frac{1}{2\pi\sqrt{\pi}} \sqrt{\tanh(\frac{\beta}{2})} \sum_{n,m\in\mathbb Z}\omega_d^{2^{-1}nm} e^{i\ell(nv-mu)}  \int dx e^{im\ell x} e^{ -\frac{1}{4}[ (2x + n\ell )^2 \tanh(\frac{\beta}{2}) + (n\ell )^2 \coth(\frac{\beta}{2}) ] }.
\end{equation}
The integral can be put into the form of the weighted Gaussian $\int dx e^{irx} e^{-ax^2 + bx + c}$ with
\begin{equation}
    r = m\ell \qquad
    a = \tanh(\frac{\beta}{2}) \qquad
    b = - n \ell \tanh(\frac{\beta}{2}) \qquad 
    c = - \frac{1}{4} (n \ell)^2 [ \tanh(\frac{\beta}{2}) + \coth(\frac{\beta}{2}) ].
\end{equation}
This integral has solution
\begin{equation}
    \int dx e^{irx} e^{-ax^2 + bx + c} = \sqrt{\frac{\pi}{a}} e^{\frac{(b + ir)^2}{4a}+c} = \sqrt{\frac{\pi}{a}} e^{i \frac{br}{2a}} e^{\frac{b^2 - r^2}{4a}+c}
\end{equation}
for $a > 0$.  Here $a = \tanh(\frac{\beta}{2}) > 0$ for all finite temperature $T = \frac{1}{\beta}$.  Inserting and simplifying, we get
\begin{equation}
\begin{aligned}
    W_{\hat \rho}^{\mathcal{C}_{\text{GKP}}}(u,v) &= \frac1{2\pi} \sum_{n,m\in\mathbb Z} \omega^{2^{-1}nm} e^{i\ell(nv-mu)} e^{-i\frac{\pi}{d}nm} e^{- \frac{\pi}{d}(n^2 + m^2) \frac{1}{2}\coth(\frac{\beta}{2})} \\
    &= \frac1{2\pi} \sum_{n,m\in\mathbb Z} (-1)^{nm} e^{i\ell(nv-mu)} e^{- \frac{\pi}{d}(n^2 + m^2) \frac{1}{2}\coth(\frac{\beta}{2})} \\
    &= \frac1{2\pi} \sum_{n,m\in\mathbb Z} (-1)^{nm} e^{i\ell(nv-mu)} e^{- \frac{\pi}{d}(n^2 + m^2) \langle \hat H \rangle_\beta },
\end{aligned}
\end{equation}
where $ \langle \hat H \rangle_\beta = \frac{1}{2}\coth(\frac{\beta}{2})$ is the energy expectation value \cite{Feynman_1998}.

\section{Physical GKP state derivation}
\label{appendix:physical_GKP}

It is convenient here to work in the position basis.  There, the Zak--Gross Wigner function \eqref{def:ZGWigner} of a pure state $\ket{\psi}$ is expressed as
\begin{equation}
    W_{\psi}^{\mathcal{C}_{\text{GKP}}}(u,v) = \frac1{2\pi}\sum_{n,m\in\mathbb Z}\omega_d^{2^{-1}nm} e^{i\ell(nv - mu)} \int dx e^{im\ell x}  \psi^{*}(x+n\ell) \psi(x),
\end{equation}
where $\psi(x) = \langle x | \psi \rangle$.  We have (see \cite[Equation (24)]{Matsuura2020equivalenceGKPcodes})
\begin{equation}
    \psi_{\tilde j}(x) \sim \sum_{n \in \Z} e^{-\frac{1}{2} \kappa^2 (j\ell + d\ell n)^2} e^{- \frac{1}{2} \frac{(x - j \ell - d\ell n)^2}{\sigma^2}} \equiv \sum_{n \in \Z} c_n e^{- \frac{1}{2} \frac{(x - j \ell - d\ell n)^2}{\sigma^2}}.
\end{equation}
Then we have
\begin{equation}
\begin{aligned}
    e^{im\ell x}  \psi^{*}(x+n\ell) \psi(x) &= \sum_{k,k'} e^{im\ell x}  c_k c_{k'} e^{- \frac{1}{2} \frac{(x + n\ell - j \ell - d\ell k)^2}{\sigma^2}}  e^{- \frac{1}{2} \frac{(x - j \ell - d\ell k')^2}{\sigma^2}} \\
    &= \sum_{k,k'} e^{im\ell x}  c_k c_{k'} e^{- \frac{1}{2\sigma^2}\big[ (x + n\ell - j \ell - d\ell k)^2 + (x - j \ell - d\ell k')^2 \big]}
\end{aligned}
\end{equation}
where recall the position wavefunction is real.  This Gaussian integral evaluates to
\begin{equation}
    \int dx e^{im\ell x}  \psi^{*}(x+n\ell) \psi(x) = \sqrt{\pi } \sigma \sum_{k,k'} c_k c_{k'} e^{\left(-\frac{\ell^2}{4 \sigma ^2} \left(-2 i m \sigma ^2 (d k+ d k' + 2 j-n) + ( d k'- d k+n)^2+m^2 \sigma ^4\right)\right)}.
\end{equation}
Absorbing the constant factor of $\sqrt{\pi } \sigma$ into the normalization, the Zak--Gross Wigner function becomes
\begin{equation}
\begin{aligned}
    W_{\ket{\tilde j}}^{\mathcal{C}_{\text{GKP}}}(u,v) &\sim \sum_{n,m,k,k'} e^{i\ell(nv - mu)} e^{i\pi nm} e^{i \frac{nm \ell^2}{2}} \, c_k c_{k'} e^{\left(-\frac{\ell^2}{4 \sigma ^2} \left(-i 2 m \sigma ^2 (d k+ d k' + 2 j-n) + ( d k'- d k+n)^2+m^2 \sigma ^4\right)\right)} \\
    &= \sum_{n,m,k,k'} e^{i\ell(nv - mu)} e^{i\pi nm} c_k c_{k'} e^{\left(-\frac{\ell^2}{4 \sigma ^2} \left(-i 2 m \sigma ^2 (d k+ d k' + 2 j) + ( d k'- d k)^2 + 2n(dk' - dk) + n^2 + m^2 \sigma ^4\right)\right)} \\
    &= \sum_{k,k'} c_k c_{k'} e^{-\frac{\ell^2}{4\sigma^2}(dk' - dk)^2} \sum_{n,m} e^{i\ell(nv - mu)} e^{i\pi nm} e^{\left(-\frac{\ell^2}{4 \sigma ^2} \left(-i 2  m \sigma ^2 (d k+ d k' + 2 j) + 2n(dk' - dk) + n^2 + m^2 \sigma ^4\right)\right)} \\
    &= \sum_{k,k'} c_k c_{k'} e^{-\frac{\pi d}{2\sigma^2}(k' - k)^2} \sum_{n,m} \underbrace{e^{i\ell(nv - mu)} e^{-\frac{\ell^2}{4\sigma^2}( -i 2  m \sigma ^2 (d k+ d k' + 2 j) + 2n(dk' - dk) )}}_{\text{linear in }n,m} \underbrace{e^{i\pi nm} e^{\left(-\frac{\ell^2}{4 \sigma ^2} \left(n^2 + m^2 \sigma ^4\right)\right)}}_{\text{quadratic}}.
\end{aligned}
\end{equation}
The linear factor simplifies,
\begin{equation}\label{eq:pGKP_linear_factors}
\begin{aligned}
    e^{i\ell(nv - mu)} e^{-\frac{\ell^2}{2\sigma^2}( -i  m \sigma ^2 (d k+ d k' + 2 j) + n(dk' - dk) )} &= e^{i\ell(nv - mu)} e^{i\frac{\ell^2}{2}  m (d k+ d k' + 2 j) - \frac{\ell^2}{2\sigma^2} n(dk' - dk) } \\
    &= e^{i 2 \pi \left[ \frac{\ell}{2\pi}(nv - mu) + \frac{\ell^2}{4\pi}  m (d k+ d k' + 2 j) + i \frac{\ell^2}{4\pi\sigma^2} n(dk' - dk)  \right]} \\
    &= e^{i 2 \pi \left[ n \left( \frac{\ell}{2\pi}v  + i \frac{\ell^2}{4\pi\sigma^2} (dk' - dk) \right) + m \left( - \frac{\ell}{2\pi}u + \frac{\ell^2}{4\pi} (d k+ d k' + 2 j) \right)  \right]} \\
    &= e^{i 2 \pi \left[ n \left( \frac{v}{d\ell}  + i \frac{k' - k}{2\sigma^2} \right) + m \left( - \frac{u}{d\ell} + \frac{1}{2} (k+ k' + \frac{2}{d} j) \right)  \right]}.
\end{aligned}
\end{equation}
The quadratic factor,
\begin{equation}\label{eq:pGKP_quadratic_factors}
\begin{aligned}
    e^{i\pi nm} e^{\left(-\frac{\ell^2}{4 \sigma ^2} \left(n^2 + m^2 \sigma ^4\right)\right)} &= e^{i \pi \left[ n m + i \frac{\ell^2}{4\pi \sigma^2}n^2 + i \frac{\ell^2}{4\pi}\sigma^2 m^2 \right]} = e^{i \pi \left[2 \frac{n m}{2} + i \frac{n^2}{2 d \sigma^2} + i \frac{\sigma^2}{2d} m^2 \right]}.
\end{aligned}
\end{equation}
From Eqs.\ \eqref{eq:pGKP_linear_factors} and \eqref{eq:pGKP_quadratic_factors} we see that the $(n,m)$-sum is a two dimensional theta function \eqref{theta_multi}, making the final Zak--Gross Wigner function a summation over such theta functions with weights governed by the Gaussian envelope $\kappa$ through $c_k$:
\begin{equation}
    W_{\ket{\tilde j}}^{\mathcal{C}_{\text{GKP}}}(u,v) \sim \sum_{k,k'} c_k c_{k'} e^{-\frac{\pi d}{2\sigma^2}(k' - k)^2} \theta\left( \frac{v}{d\ell}  + i \frac{k' - k}{2\sigma^2}, - \frac{u}{d\ell} + \frac{1}{2} (k+ k' + \frac{2}{d} j), \bm \tau \right), \qquad \bm \tau = \frac{1}{2}
    \begin{pmatrix}
        i \frac{1}{d\sigma^2} & 1 \\ 1 & i \frac{\sigma^2}{d}
    \end{pmatrix}.
\end{equation}

\end{document}